\documentclass[11pt,a4paper,iop,apj]{emulateapj}
\usepackage{graphicx}
\usepackage{color} 
\begin{document}
\title{The late type eclipsing binaries in the Large Magellanic Cloud: catalogue of fundamental physical parameters
\footnote{This paper includes data gathered with the 6.5m Magellan Clay Telescope at Las Campanas Observatory, Chile.}} 
\author{Dariusz Graczyk\altaffilmark{1,2,3}, Grzegorz Pietrzy{\'n}ski\altaffilmark{4}, Ian B. Thompson\altaffilmark{5}, Wolfgang Gieren\altaffilmark{3,2}, Bogumi{\l} Pilecki\altaffilmark{4}, Piotr Konorski\altaffilmark{6}, Sandro Villanova\altaffilmark{3}, Marek G{\'o}rski\altaffilmark{3}, Ksenia Suchomska\altaffilmark{4}, Paulina Karczmarek\altaffilmark{6}, Kazimierz St{e}pie{\'n}\altaffilmark{6}, Jesper Storm\altaffilmark{7}, M{\'o}nica Taormina\altaffilmark{4}, Zbigniew Ko{\l}aczkowski\altaffilmark{4,12}, Piotr Wielg{\'o}rski\altaffilmark{4}, Weronika Narloch\altaffilmark{4}, Bart{\l}omiej Zgirski\altaffilmark{4}, Alexandre Gallenne\altaffilmark{8}, Jakub Ostrowski\altaffilmark{4}, Rados{\l}aw Smolec\altaffilmark{4}, Andrzej Udalski\altaffilmark{6}, Igor Soszy{\'n}ski\altaffilmark{6}, Pierre Kervella\altaffilmark{9}, Nicolas Nardetto\altaffilmark{10}, Micha{\l} K. Szyma{\'n}ski\altaffilmark{6}, {\L}ukasz Wyrzykowski\altaffilmark{6}, Krzysztof Ulaczyk\altaffilmark{6}, Rados{\l}aw Poleski\altaffilmark{6,11}, Pawe{\l} Pietrukowicz\altaffilmark{6}, Szymon Koz{\l}owski\altaffilmark{6}, Jan Skowron\altaffilmark{6} \\ and Przemys{\l}aw Mr{\'o}z\altaffilmark{6}}
\affil{$^1$Centrum Astronomiczne im. Miko{\l}aja Kopernika, PAN, Rabia{\'n}ska 8, 87-100 Toru{\'n}, Poland; darek@ncac.torun.pl}
\affil{$^2$Millennium Institute of Astrophysics (MAS), Chile} 
\affil{$^3$Universidad de Concepci{\'o}n, Departamento de Astronomia, Casilla 160-C, Concepci{\'o}n, Chile; darek@astro-udec.cl}
\affil{$^4$Centrum Astronomiczne im. Miko{\l}aja Kopernika, PAN, Bartycka 18, 00-716 Warsaw, Poland}
\affil{$^5$Carnegie Observatories, 813 Santa Barbara Street, Pasadena, CA 91101-1292, USA}
\affil{$^6$Warsaw University Observatory, Al. Ujazdowskie 4, 00-478 Warsaw, Poland }
\affil{$^7$Leibniz-Institut f\"{u}r Astrophysik Potsdam, An der Sternwarte 16, 14482 Potsdam, Germany}
\affil{$^8$ European Southern Observatory, Alonso de C\'ordova 3107, Casilla 19001, Santiago, Chile}
\affil{$^{9}$LESIA (UMR 8109), Observatoire de Paris, PSL Research University, CNRS, UPMC, Univ. Paris-Diderot, 5 place Jules Janssen,
92195 Meudon, France}
\affil{$^{10}$Universit\'e C$\hat{\rm o}$te d'Azur, Observatoire de la C$\hat{\rm o}$te d'Azur, CNRS, Laboratoire Lagrange, UMR7293, Nice, France}
\affil{$^{11}$Department of Astronomy, Ohio State University, 140 W. 18th Ave., Columbus, OH  43210, USA}
\affil{$^{12}$Instytut Astronomiczny Uniwersytetu Wroc{\l}awskiego, Kopernika 11, 51-622 Wroc{\l}aw, Poland}

\begin{abstract}
We present a determination of precise fundamental physical parameters of twenty detached,  double-lined, eclipsing binary stars in the Large Magellanic Cloud (LMC) containing G- or early K-type giant stars. Eleven are new systems, the remaining nine are systems already analyzed by our team for which we present updated parameters. The catalogue results from our long-term survey of eclipsing binaries in the Magellanic Clouds suitable for high-precision determination of distances (the Araucaria project). 
The V-band brightnesses of the systems range from 15.4 mag to 17.7 mag and their orbital periods range from 49 days to 773 days. Six systems have favorable geometry showing total eclipses. 
The absolute dimensions of all eclipsing binary components are calculated with a precision of better than 3\% and all systems are suitable for a precise distance determination. 
The measured stellar masses are in the range 1.4 to 4.6 M$_\odot$ and comparison with the MESA isochrones gives ages between 0.1 and 2.1 Gyr. The systems show an age-metallicity relation with no evolution of metallicity for systems older than 0.6 Gy, followed by a rise to a metallicity maximum at age 0.5 Gy, and then a slow metallicity decrease until 0.1 Gy. Two systems have components with very different masses: OGLE LMC-ECL-05430 and OGLE LMC-ECL-18365. Neither system can be fitted by single stellar evolution isochrone, explained by a past mass transfer scenario in the case of ECL-18365 and a gravitational capture or a hierarchical binary merger scenario in the case of ECL-05430.
The longest period system OGLE LMC SC9\_230659 shows a surprising apsidal motion which shifts the apparent position of the eclipses. This is a clear sign of a physical companion to the system, however neither investigation of the spectra nor light curve analysis indicate a third light contribution larger than 2-3\%. In one spectrum of OGLE LMC-ECL-12669 we noted a peculiar dimming of one of the components by 65\% well outside of the eclipses. We interpret this observation as arising from an extremely rare occultation event as a foreground Galactic object covers only one component of an extragalactic eclipsing binary.   
  
\end{abstract} 

\section{Introduction}
The Large Magellanic Cloud (LMC) is a nearby dwarf barred galaxy connected by a hydrogen gas and stellar bridge to the Small Magellanic Cloud (SMC) \citep[e.g.][]{jac16,jac17}. Both are nearby satellites of our Milky Way galaxy and because of this proximity they are ideal environments to study stellar populations and to calibrate many different stellar standard candles. For these reasons, determination of distances to these galaxies has played a central role in our Araucaria Project \citep{gie05} whose main goal is to produce a significantly improved calibration of the extragalactic distance scale. 

To this end, for more than a decade we have been observing late-type eclipsing binary systems in both Magellanic Clouds, resulting in an analysis of nine binaries in the LMC and five binaries in the SMC, leading to a determination of the distances to the galaxies accurate to 2\% \citep[][hereafter P13]{pie13}, \citet{elg16} and 3\% \citep[][hereafter G14]{gra14}, respectively. A by-product of that research was the determination of precise parameters of 28 low-metallicity giant stars. Giant stars in eclipsing binaries are near-perfect laboratories to study late phases of stellar evolution, especially with a potential for the calibration of stellar evolution models and their internal parameters such as convective core-overshooting \citep[e.g.][]{egg17,cla17,ost18}. They are also a precise tool to calibrate important asteroseismic scaling relations \citep{gau16,bro18} and to study chromospheric activity of the late phases of stellar evolution \citep{rat16}. 
Single giant stars in the LMC are also important for the investigation of kinematics and chemodynamical evolution \citep[e.g.][]{har06,son17}.  

Here we present an extended analysis of twenty late-type eclipsing binary stars in the LMC. Nine systems were already analyzed in our previous papers. All these eclipsing binaries are composed of G- and K-type normal giant stars. We will derive their individual distances and use them for a new determination of the distance to the LMC with an accuracy of 1\% in a follow-up paper \citep{pie18}. 

The paper is organized as follows. In Section 2 we describe photometric and spectroscopic observations, Section 3 gives details of our method, Section 4 describes the determination of  physical parameters, and Section 5 gives short notes about individual binaries. Section 6 is devoted to a comparison with evolutionary models and a discussion of the results. 

\section{Observations and Data Reduction}
\label{obs}
\begin{figure}
\hspace*{-0.8cm}
\includegraphics[angle=0,scale=0.315]{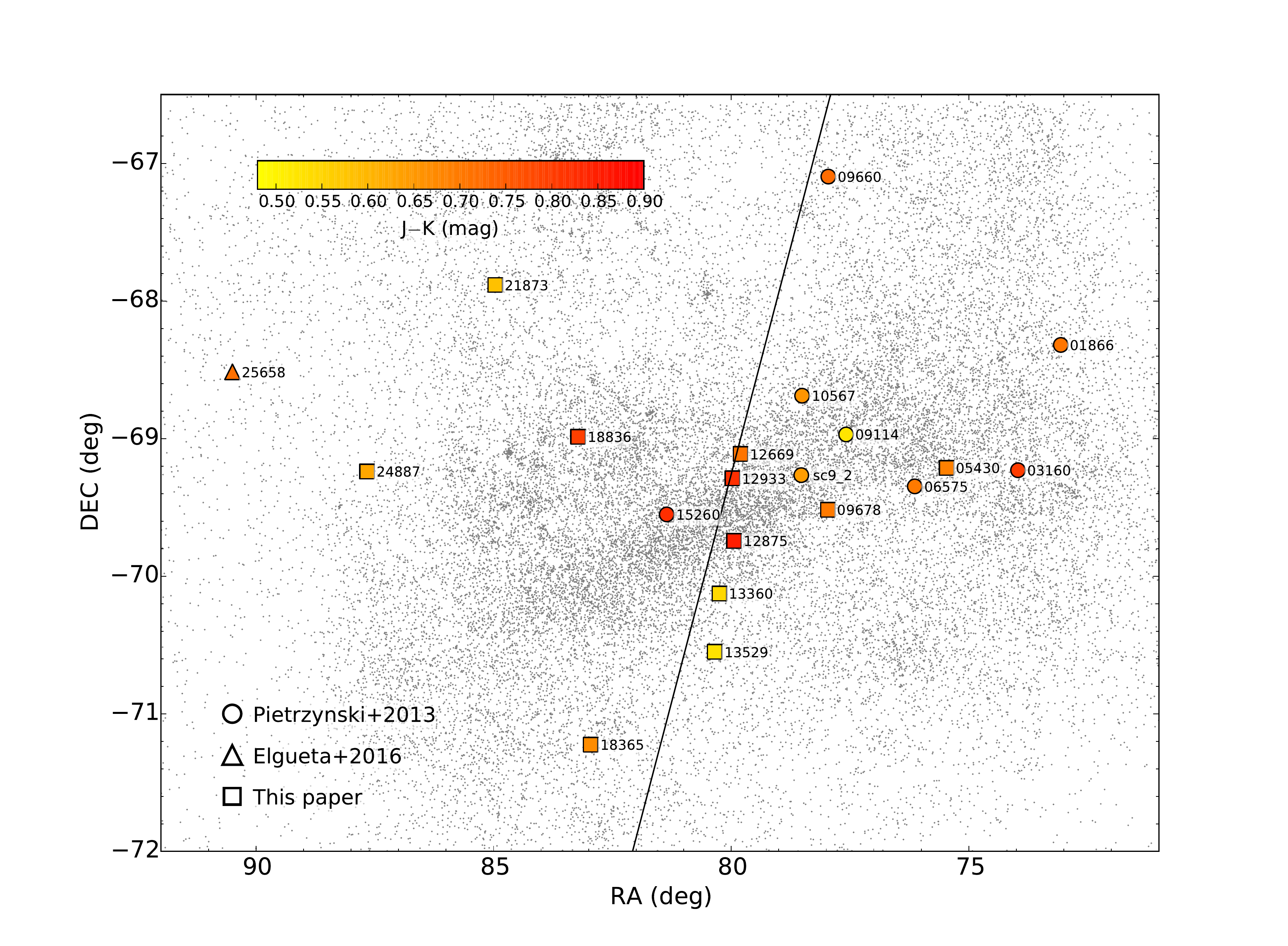}
\caption{The positions of the eclipsing binaries within the LMC. We separately denote eight systems from P13, one system from \cite{elg16}, and the eleven new systems presented in this paper. The colorbar gives their $(J\!-\!K)$ color. In the background the positions (grey dots) of some 44000 eclipsing binaries detected by the OGLE project in the LMC \citep{gra11,paw16} are shown. The sloping line shows the position of the nodes derived from our 20 eclipsing binaries \citep{pie18}. \label{fig0}}
\end{figure}

The basic data for our sample of twenty LMC eclipsing binary systems are given in Table~\ref{tbl:1}. Their positions within the LMC are presented in Fig.~\ref{fig0}. Eleven binaries are new systems for which no analysis has been published. Two systems which are apparent members of open clusters in the LMC are identified in the last column of Table~\ref{tbl:1} using the notation of \cite{sha63}.

Five of our targets were discovered by the OGLE-II survey \citep{uda97,wyr03} and fourteen more were identified as eclipsing binaries by the OGLE-III survey \citep{uda03,gra11}. One system, LMC SC9\_230659, was confirmed to be an eclipsing binary by our internal work. We used optical $V$- and $I$-band photometry in the Johnson-Cousins filters obtained with the Warsaw 1.3 m telescope at Las Campanas Observatory in the course of the second, third and fourth phase of the OGLE project \citep{uda97,uda03,uda15} and published in catalogues of eclipsing binary stars in the LMC \citep{gra11,paw16}. Because of the long orbital periods consecutive epochs were usually taken  on different nights. The raw data were reduced with   difference image analysis \citep{woz00,uda03} and instrumental magnitudes were  converted onto the standard system using Landolt standard  stars. The $I$-band photometry was detrended from long-term instrumental effects and small short-term light modulations for all targets.    

Near-infrared  $J$-, $H$- and $K$-band photometry was collected with the ESO NTT telescope on La Silla, equipped with the SOFI camera. The setup of the instrument, reduction and calibration of the data onto the UKIRT system are described in \cite{pie09}. The photometry was later converted onto the 2MASS system using the  transformation equations given by \cite{car01}. We collected at least 5 epochs of infrared photometry for each of our targets outside of eclipses.

High resolution echelle spectra were collected with the MIKE spectrograph on the Magellan Clay 6.5 m telescope at Las Campanas, with the HARPS spectrograph on the 3.6 m telescope at La Silla,  and with the UVES spectrograph on the 8.4 m VLT Unit 2 telescope at Paranal. We used a $5\times0.7$ arc sec slit with MIKE giving a resolution of about 40000. In the case of HARPS we used the EGGS mode giving a resolution of about R$\sim$80000. UVES was operated in the high-resolution mode with R$\sim$80000. The typical S/N at $\sim$5500 \AA~was about 15 and 4 for the MIKE-Red  and HARPS spectra, respectively, and about 8 at $\sim$4700 \AA~for the MIKE-Blue  spectra. 

In order to determine radial velocities of the components of the binaries we employed the Broadening Function (BF) formalism introduced by \citet{ruc92,ruc99} and implemented in the RaveSpan software \citep{pil13,pil17}. We used numerous metallic lines in the wavelength regions 4125-4230, 4245-4320, 4350-4840, 4880-5000, 5350-5850, 5920-6250, 6300-6390, and 6600-6800 \AA~to construct the BH profiles. As templates we used synthetic spectra with [Fe/H]=$-0.3$ from the library computed by \cite{col05}. The templates were chosen to match the atmospheric properties of the stars in a grid of $T_{\rm eff}$ and $\log g$. We also calculated instrumental shifts between different spectrographs based on residual radial velocities from our best models and took them into account when they were larger than 0.65$\times${rms}. Those shifts are changing from one system to another because the orbital phase coverage of each spectrograph is different, e.g. in some cases orbital quadratures were covered by different spectrographs. Also, the S/N of the analysed spectra and temperature differences between the components (the blue shift)  influenced the measured shifts. In other words these are not true instrumental shifts but $a~posteriori$ corrections to minimize resulting residuals. The average shift in respect to HARPS is:  $-0.52\pm0.40$ km s$^{-1}$ for MIKE (based on 17 systems) and +1.16  km s$^{-1}$ for UVES (based on one system). The resulting velocities are presented in Table~\ref{tbl:spec}.
 
\begin{deluxetable*}{@{}lcclcccccccc@{}}
\tabletypesize{\scriptsize}
\tablecaption{The target stars \label{tbl:1}}
\tablewidth{0pt}
\tablehead{
\colhead{OGLE ID} & \colhead{RA} & \colhead{Dec} & \colhead{$V$} & \colhead{$V-I$} & \colhead{$V-K$} & \colhead{$J-K$} & \colhead{$P_{\rm obs}$}  & \colhead{$T_0$} & \colhead{New} & \colhead{Ref} & \colhead{Cluster}\\
\colhead{} & \colhead{h:m:s} & \colhead{deg:m:s} & \colhead{mag} & \colhead{mag} & \colhead{mag} &\colhead{mag}& \colhead{d} & \colhead{HJD-2450000} & & &\colhead{memb.}
}
\startdata
LMC-ECL-01866 & 04:52:15.28 &$-$68:19:10.3 &  16.126 & 1.165 &2.693&0.708&251.3&3844.3& No& 1,3 &\\
LMC-ECL-03160 & 04:55:51.48 &$-$69:13:48.0 & 17.126 & 1.305 & 2.985 &0.801&150.1&2226.0& No& 1,3 &\\
LMC-ECL-05430 & 05:01:51.74 & $-$69:12:48.8 & 16.611 &1.204 &2.708  &0.690 &505.2 &4923.9& Yes &3 &SL181\\
LMC-ECL-06575 & 05:04:32.87 & $-$69:20:51.0 & 15.811 & 1.195 & 2.704 & 0.699&190.0&2194.6&No&1,3&\\
LMC-ECL-09114\tablenotemark{a} & 05:10:19.64 & $-$68:58:12.2 & 16.711 & 1.033&2.309  &0.524 &214.4 &1141.1 & No & 1,3&\\
\\
LMC-ECL-09660 & 05:11:49.45 &$-$67:05:45.2  & 16.269 & 1.186&2.661 &0.722 &167.8 &3815.6& No & 1,3&\\
LMC-ECL-09678 & 05:11:51.76 & $-$69:31:01.1 & 17.228 &1.203 &2.694  &0.699 &114.5 &3951.3& Yes &3&\\
LMC-ECL-10567 & 05:14:01.89 & $-$68:41:18.2 & 16.430 & 1.091&2.563  &0.654 &118.0 &2168.4& No& 1,3& SL327\\
LMC SC9\_230659\tablenotemark{b,c}& 05:14:06.04 & $-$69:15:56.9 & 16.630 &1.092 &2.597 &0.642 &772.6 &4990.7& No &1,3&\\
LMC-ECL-12669 & 05:19:12.80 & $-$69:06:44.4 & 17.205 &1.197 &2.720  &0.709 &749.6 &5570.7& Yes &3&\\
\\
LMC-ECL-12875 & 05:19:45.39 & $-$69:44:38.5 & 17.320 &1.394 &3.243  &0.850 &152.8 &1845.3& Yes &3&\\
LMC-ECL-12933 & 05:19:53.69 & $-$69:17:20.4 & 17.379 &1.372 &3.143  &0.822 &125.4 &3771.8& Yes &3&\\
LMC-ECL-13360 & 05:20:59.46 & $-$70:07:35.2 & 15.822 &1.072 &2.304  &0.544 &262.4 &2177.1& Yes &3&\\
LMC-ECL-13529 & 05:21:23.34 & $-$70:33:00.0 & 17.362 &1.012 &2.215  &0.530 &$\;\;\;\,$49.47 &3931.2& Yes &3&\\
LMC-ECL-15260 & 05:25:25.66 & $-$69:33:04.5 & 16.924 & 1.358 & 3.105 & 0.824 & 157.5 & 1168.0 & No & 1,3&\\
\\
LMC-ECL-18365 & 05:31:49.56 & $-$71:13:28.3 & 16.669 &1.172 &2.614  &0.669 &$\;\;\;\,$78.54 &2154.7& Yes &3&\\
LMC-ECL-18836 & 05:32:53.06 & $-$68:59:12.3 & 17.432 &1.286 &2.954  &0.795 &182.6 &3621.8&Yes &3 &\\
LMC-ECL-21873 & 05:39:51.19 & $-$67:53:00.5 & 16.792 &1.056 &2.335  &0.582 &144.2 &2127.2& Yes &3&\\
LMC-ECL-24887 & 05:50:39.02 & $-$69:14:20.7 & 17.676 &1.160 &2.522  &0.626 &$\;\;\;\,$94.84 &3939.1& Yes &3&\\
LMC-ECL-25658 & 06:01:58.77 & $-$68:30:55.1 & 16.997 &1.165 &2.640  &0.715 &192.8 & 3891.8 & No & 2,3& 
\enddata
\tablenotetext{a}{OGLE-III temporary name OGLE-051019.64-685812.3 was used by \cite{pie09} }
\tablenotetext{b}{OGLE-II internal ID is given; OGLE-III and -IV internal IDs are LMC111.2 62181 and LMC503.24 153, respectively.} 
\tablenotetext{c}{The name LMC-ECL-26122 was used by \cite{pie13}, however, this name  was assigned to another star in the latest release of the OGLE catalogue of eclipsing binary stars \citep{paw16}.}

\tablecomments{Columns give coordinates, observed magnitudes, colors, orbital periods and epochs of primary minimum.  Identification numbers are from the OGLE-III catalogue of variable stars \citep{gra11}. Observed NIR $J$, $K$  magnitudes
are expressed in the 2MASS photometric system.
{\bf Reference} to photometric parameters: 1 - \cite{pie13}, 2 - \cite{elg16}, 3 - this paper}
\end{deluxetable*}

\begin{deluxetable*}{@{}lrccccl}
\tabletypesize{\scriptsize}
\tablecaption{The radial velocity
 measurements \label{tbl:spec}}
\tablewidth{0pt}
\tablehead{
\colhead{Name} & \colhead{HJD} & \colhead{RV1} & \colhead{Err1} & \colhead{RV2} & \colhead{Err2} &\colhead{Instrument}  \\
\colhead{} & \colhead{HJD-2450000} & \colhead{km s$^{-1}$} & \colhead{km s$^{-1}$} &  \colhead{km s$^{-1}$} & \colhead{km s$^{-1}$} & \colhead{}
}
\startdata
OGLE LMC-ECL-01866  & 4784.79993 & 322.340 & 0.450 & 265.297 & 0.150 & MIKE-Red \\ 
OGLE LMC-ECL-01866  & 4808.68996 & 334.992 & 0.450 & 252.089 & 0.150 & HARPS \\ 
OGLE LMC-ECL-01866  & 4810.85189 & 334.794 & 0.450 & 252.502 & 0.150 & HARPS \\ 
OGLE LMC-ECL-01866  & 4816.76038 & 333.537 & 0.450 & 254.163 & 0.150 & MIKE \\ 
OGLE LMC-ECL-01866  & 4888.68153 & 274.190 & 0.450 & 313.191 & 0.150 & HARPS \\ 
OGLE LMC-ECL-01866  & 5187.82593 & 269.593 & 0.450 & 318.279 & 0.150 & HARPS \\ 
OGLE LMC-ECL-01866  & 5430.78591 & 268.765 & 0.450 & 318.703 & 0.150 & HARPS 
\enddata
\tablecomments{This table is available entirety in electronic format in the online journal. A portion is shown here for guidance regarding its form and content.}
\end{deluxetable*}

\section{Modeling approach}
The light and radial velocity curves were analyzed using the Wilson-Devinney (hereafter WD) program, version 2007 \citep{wil71,wil79,wil90,van07}, equipped with a Python wrapper written by P. Konorski.  The code simultaneously solves light and radial velocity curves. The surfaces of the component stars are represented as  Roche potentials and the radiative treatment is based on model stellar atmospheres by \cite{kur93}. For the analysis we follow the methodology described in \cite{gra12,gra14} and the Supplementary Information section of P13. We adopted the nomenclature in which the star which is eclipsed during the primary eclipse (the deeper one) is called the primary and its companion is called the secondary; the primary is not always the brighter, more massive, nor hotter star of the two components. 

\subsection{Initial Parameter Choice}
\label{par}

\begin{deluxetable}{lccc}
\tabletypesize{\scriptsize}
\tablecaption{The spectroscopic light ratios at 5500 \AA \label{tbl:spec2}}
\tablewidth{0pt}
\tablehead{
\colhead{OGLE ID} & \colhead{Line intensity } & \colhead{Correction} & \colhead{Light ratio} \\
\colhead{} & \colhead{$I_2/I_1$} & \colhead{$k_{21}$} & \colhead{$L_2/L_1$} 
}
\startdata
LMC-ECL-01866 &$1.61\pm0.12$&$0.70 \pm 0.02$&$1.13 \pm 0.10$\\
LMC-ECL-03160 & $3.39 \pm 0.15$ & $0.81 \pm 0.03$&$2.75 \pm 0.16$\\
LMC-ECL-05430 & $1.49 \pm 0.02$ & $1.01 \pm 0.03$ &$1.51 \pm 0.05$\\
LMC-ECL-06575 &$0.93\pm0.03$&$0.89 \pm 0.03$&$0.83 \pm 0.04$\\
LMC-ECL-09114&$0.526\pm0.019$&$1.12 \pm 0.03$&$0.590 \pm 0.026$\\
\\
LMC-ECL-09660&$2.12\pm0.07$&$ 0.75 \pm 0.03$&$1.59 \pm 0.08$\\
LMC-ECL-09678& $3.74 \pm 0.18$ & $0.64 \pm 0.02$ &$2.39\pm0.14$\\
LMC-ECL-10567& $1.63\pm0.06$&$0.89 \pm 0.02$&$1.45 \pm0.06$\\
LMC SC9\_230659&$0.490 \pm 0.020$&$1.00\pm0.02$&$0.490\pm0.022$\\
LMC-ECL-12669& $1.08 \pm 0.02$ & $1.05 \pm 0.03$&$1.13\pm0.04$\\
\\
LMC-ECL-12875 & $4.37 \pm 0.36$ & $0.80 \pm 0.02$& $3.50\pm0.30$\\
LMC-ECL-12933 & $2.90 \pm 0.07$&$0.84 \pm 0.02$&$2.44 \pm 0.09$\\
LMC-ECL-13360 & $1.38\pm0.04$&$0.71\pm0.02$&$0.99\pm0.04$\\
LMC-ECL-13529 &$0.87 \pm 0.03$&$1.09 \pm 0.03$&$0.80 \pm 0.05$\\
LMC-ECL-15260 & $2.41 \pm 0.10$& $0.86 \pm 0.03$ & $2.07 \pm 0.11$\\
\\
LMC-ECL-18365 & $0.196\pm0.013$&$1.04\pm0.03$&$0.204 \pm 0.015$\\ 
LMC-ECL-18836 &$2.76\pm0.10$&$0.66\pm 0.02$&$1.82 \pm 0.08 $\\
LMC-ECL-21873&$1.55\pm0.06$&$0.71 \pm 0.02$&$1.11 \pm 0.05$\\ 
LMC-ECL-24887 & $1.13\pm0.03$&$0.97 \pm 0.03$&$1.10 \pm 0.04$ \\
LMC-ECL-25658 & $1.48 \pm 0.08$ &$0.94\pm0.03$&$1.40\pm0.09$
\enddata
\tablecomments{Line intensity - the relative strengths of the absorption lines of the secondary with respect to those of the primary.}
\end{deluxetable}

\begin{deluxetable*}{@{}lcccccccccc}
\tabletypesize{\scriptsize}
\tablecaption{Atmospheric parameters\label{tbl:atmo}}
\tablewidth{0pt}
\tablehead{
\colhead{OGLE ID} &\colhead{}&\colhead{$T_{\rm a,eff}$}& \colhead{$\log{g}$}& \colhead{$[$Fe/H$]$}& \colhead{$v_{t}$}&\colhead{$v_{mt}$}&\colhead{$T_{\rm s,eff}$}&\colhead{$T_{\rm c,eff}$}&\colhead{$T_{\rm b,eff}$}&\colhead{$[$Fe/H$]_{\rm C}$}\\
& &\colhead{K} & \colhead{cgs} & \colhead{dex} &\colhead{km $s^{-1}$} &\colhead{km $s^{-1}$}&\colhead{K}&\colhead{K}&\colhead{K}&\colhead{dex}
}
\startdata
LMC-ECL-01866 &p&5300\tablenotemark{a}& 2.1\tablenotemark{a}&$-$0.21& 2.0\tablenotemark{a}&4.9&5210&5325&5340&$-$0.21\\
			  &s 	&  4450 & 1.65 &  $-$0.70  & 1.9&3.8&4515&4505&4505&$-$0.67\\   
LMC-ECL-03160 &p	& 5140  &2.25 &  $-$0.14 &  1.7&3.9& 4980&4865&4875&$-$0.29\\
                            &s& 4420 & 1.55  & $-$0.82 &  1.8&3.7&4415&4475&4480&$-$0.80\\
LMC-ECL-05430 &p	&4630  &  1.85  &  $-$0.48  &  1.5&3.9&4730&4700&4735&$-$0.42\\
                            &s& 4645  &  1.89  &  $-$0.41  &  1.4&4.1&4775&4735&4775&$-$0.33\\
LMC-ECL-06575 &p	& 4915  &1.94  & $-$0.47  & 2.7&4.6&5055 &4865&4880&$-$0.47\\
                            &s&4655 & 1.75 &  $-$0.44  & 1.7&4.1&4680&4635&4645&$-$0.45\\
LMC-ECL-09114  &p  &   5300    &    2.18   &  $-$0.28   &2.0&4.8&5165&5200&5230&$-$0.33\\
                            &s& 5820& 2.95& $-$0.16&2.2&4.8&5325&5320&5355&$-$0.47\\
LMC-ECL-09660  &p	& 5320 & 2.26  & $-$0.37 &  1.7&4.7&5200&5235&5260&$-$0.42\\
                            &s& 4655 & 1.92  & $-$0.50  & 1.7&4.2&4945&4615&4635&$-$0.48\\
LMC-ECL-09678  &p	&  5350  &  2.68  &  $+$0.11  &  0.9&4.2&4895&5205&5275&$+$0.03\\
                            &s& 4710  &  1.95  &  $-$0.50  &  1.6&3.9&4770&4615&4655&$-$0.50\\
LMC-ECL-10567  &p	&  4840 & 1.85 &  $-$0.96 &  1.7&4.4&5290&5060&5065&$-$0.80\\
                            &s&4680 & 1.80 &  $-$0.67 &  2.2&4.0&4800&4690&4700&$-$0.65\\
LMC SC9\_230659  &p 	&  5060 & 1.96  & $-$0.25   &1.9&4.5&5045&4970&4910&$-$0.29\\
                            &s& 5140 & 2.25 &  $-$0.05 &  1.8&4.2&5085&4970&4910&$-$0.13\\
LMC-ECL-12669   &p & 4430  &  1.69  &  $-$0.45  &  1.1&3.6&4600&4620&4730&$-$0.31\\
                            &s& 4820  &  2.16  &  $-$0.22  &  1.5&3.7&4560&4770&4805&$-$0.29\\
LMC-ECL-12875  &p	&   	5260  &  2.78  &  $-$0.03  &  3.4&3.7&4985&4790&4760&$-$0.32\\
                            &s& 4540  &  2.04  &  $-$0.41  &  1.9&3.6&4620&4305&4310&$-$0.52\\
LMC-ECL-12933  &p	&  5175  &  2.82  &  $-$0.07  &  1.9&3.9&5065&4690&4685&$-$0.26\\
                            &s&4695  &  2.27  &  $-$0.26  &  1.2&3.7&4520&4345&4320&$-$0.42\\
LMC-ECL-13360  &p	&  5300  &  0.90  &  $-$0.43  &  1.8&5.5&5500&5480&5535&$-$0.29\\
                            &s& 5195  &  1.64  &  $-$0.23  &  2.3&4.9&5150&4980&5025&$-$0.31\\
LMC-ECL-13529  &p	&  5255  &  2.50  &  $-$0.28  &  0.0&4.5&5320&5280&5315&$-$0.25\\
                            &s& 5390  &  2.96  &  $-$0.01  &  0.1&4.4&5185&5210&5250&$-$0.10\\
LMC-ECL-15260   &p &  5150 & 3.50 &  $-$0.42 &  2.4 &3.9&4735&4675&4695&$-$0.66\\
                            &s& 4550 & 2.50 &  $-$0.52  & 1.8&3.7&4360&4370&4375&$-$0.61\\
LMC-ECL-18365  &p	&  5275  &  3.03  &  $-$0.36  &  0.0&4.4&4970&4830&4850&$-$0.62\\
                            &s& 4950\tablenotemark{a}& 2.2\tablenotemark{a}&  $-$0.25  &  1.5&3.9&4905&4925&4945&$-$0.25\\
LMC-ECL-18836  &p	& 5445  &  3.03  &  $-$0.06  &  0.3&4.3&5050&5155&5105&$-$0.26\\
                            &s& 4630  &  1.93  &  $-$0.44  &  1.5&3.7&4705&4540&4570&$-$0.46\\
LMC-ECL-21873  &p	&  5175  &  1.93  &  $-$0.39  &  0.7&4.6&5265&5260&5305&$-$0.33\\
                            &s& 5020  &  2.16  &  $-$0.27  &  1.6&4.3&5185&5000&5030&$-$0.25\\
LMC-ECL-24887  &p	& 5200\tablenotemark{a} & 2.39& $-$0.20	&1.8&4.1&5090&5130&5165&$-$0.25\\
                            &s& 5150\tablenotemark{a} &2.45 & $-$0.13&2.1&4.1&5140&5015&5050&$-$0.19\\
LMC-ECL-25658   &p & 4775  &  2.08  &  $-$0.42  &  1.3&3.9&4855&4820&4820&$-$0.35\\
                            &s& 4790  &  2.21  &  $-$0.55  &  1.4&3.9&4770&4690&4695&$-$0.57
 \enddata
 \tablecomments{$T_{\rm a,eff}$ is spectroscopic temperature (Sec.~\ref{atmo}), $v_t$ is the microturbulent velocity, $v_{mt}$ is the macroturbulent velocity, $T_{\rm s,eff}$ - is temperature from \cite{kov06} calibration (Sec.~\ref{sec:temp}), $T_{\rm c,eff}$ is a color temperature (Sec.~\ref{sec:temp}) and $T_{\rm b,eff}$ is a temperature derived from the bolometric flux scaling (Sec.~\ref{sec:temp}), [Fe/H]$_C$ is a corrected metallicity used through the paper.}
 \tablenotetext{a}{Fixed value}
 \end{deluxetable*}
 
The orbital periods and epochs of primary minima were adopted from \cite{gra11}. In case of LMC SC9\_230659 the initial period was found using a combination of period finding programs as described in \cite{gra10}. The phased light curves were inspected visually to verify the photometric stability of the systems - a lack of spots, flares, the O'Connell effect \citep{oco51} - at the precision level of the OGLE photometry. The initial orbital parameters (radial velocity semiamplitudes, $\gamma$, etc.) were usually found with the RaveSpan code.

The initial effective temperatures of the components of each system were estimated following a procedure described in G14 but assuming [Fe/H]$\,=-0.4$. This procedure gave temperatures close to the final ones to within about 250 K. 

We estimated rotational velocities of the components from the Broadening Function (BF) profile analysis using the RaveSpan code. The BF profile contains the rotational broadening convolved with broadenings caused by the macroturbulent velocity $v_{\rm mt}$ and the instrumental profile  broadening $v_{\rm ip}$. We removed both effects using the equation for the total "macrobroadening" \citep[][their Sec.~4.2.1]{tak08}:

\begin{equation}
\label{eqn:mac }
v^2\!\!\!_{\rm M}=v^2\!\!\!_{\rm ip} + v^2\!\!\!_{\rm rt} + v^2\!\!\!_{\rm mt}\,.
\end{equation}
We calculated the macroturbulent velocity using several relations \citep{hek07,tak08,mas08}: 
\begin{eqnarray}
 v_{\rm mt} & = & 0.00195\, T_{\rm eff}  -3.953  \\
 v_{\rm mt} & = &  4.3 - 0.67 \log{g}\\
 \log{v_{\rm mt}} & = & 3.50 \log{T_{\rm eff}} + 0.25 \log{L/L_\odot} - 12.97
\end{eqnarray}
assuming in the last equation $v_{\rm mt}=0.5\,\zeta_{\rm RT}$, where $\zeta_{\rm RT}$ is the Radial
-Tangential macroturbulence dispersion \citep{gray05}.  As a final estimate of $v_{\rm mt}$ we adopted the average from the above three equations. The instrumental broadening of the MIKE spectrograph was calculated from the equation  \citep[][footnote 10]{tak08}:
\begin{equation}
v_{\rm ip} = \frac{300000}{2 \sqrt{2} R}
\end{equation}
 where $R$ is the resolving power of MIKE. Whenever rotation was consistent with synchronous rotation, as it is for most components, we set the rotation parameter $F=1.0$. In the few systems where the components rotate super-synchronously we set the parameter $F$ accordingly. The albedo parameter was set to 0.5 and the gravity brightening to 0.32, both values appropriate for a cool, convective atmosphere. The limb darkening coefficients were calculated internally by the WD code (setting LD=$-2$) according to the logarithmic law \citep{kli70} during each iteration of the Differential Correction (DC) subroutine using tabulated data computed by \cite{van93}.         

As free parameters of the WD model we chose the orbital period $P_{\rm obs}$, the semimajor axis $a$, the orbital eccentricity $e$, the argument of periastron $\omega$, the epoch of the primary spectroscopic conjunction $T_0$, the systemic radial velocities $\gamma_{1,2}$, the orbital inclination $i$, the primary or the secondary star average surface temperature $T_{1,2}$, the modified surface potential of both components ($\Omega_1$ and $\Omega_2$), the mass ratio $q$, and the relative monochromatic luminosity of the primary star in two bands ($L1_V$ and $L1_I$). We usually set the temperature of the more luminous component as constant during the analysis, varying the temperature of the companion star. In  cases where proximity effects are more clearly visible between eclipses we also adjusted the albedo parameter $A$ for  one or both components. 

Initial input parameters were used to obtain a preliminary solution which was used to determine the spectroscopic light ratio (Sec.~\ref{sec:spec}), update reddening estimates (Sec.~\ref{red}), and for the renormalization of the disentangled spectra (Sec.~\ref{atmo}).

\subsection{Spectroscopic Light Ratio}
\label{sec:spec}
It is well known that for detached eclipsing binaries showing partial eclipses a strong degeneracy usually exists between the radii of the components: many light curve solutions exist having different individual radii $r_1$ and $r_2$ but a similar sum of their radii. When there are proximity effects visible in the light curve, they can be used to break this degeneracy. However, the best method is to utilize additional information obtained from the composite spectra of the systems, the so called spectroscopic light ratio. For equal temperature systems the ratio of the strength of the absorption lines for the components is a direct indication of the true light ratio of the components: a brighter component more strongly dilutes the lines of the fainter companion star and thus in the lines of the primary appear deeper and stronger. In the case of unequal temperature components (like most in our sample) it is important to account for the growing equivalent width of metallic absorption lines with decreasing temperature for a given chemical composition. To account for both effects, we first measure the line intensity ratio from the strength of the broadening function profiles by using a properly matched template spectrum. The templates were calculated from a synthetic spectra library \citep{col05} for the temperature and surface gravity of the components. We used a mask corresponding to a wavelength range of 5300-5800 ${\rm \AA}$, i.e. an approximate Johnson $V$-band response curve. The line intensity ratio is approximately equal to the ratio of the equivalent widths in the individual compontents. The line intensity ratios $I_2/I_1$ are given in Table~\ref{tbl:spec2}. 

To convert $I_2/I_1$  into light ratios we calculated the corrections $k_{21}$ in the following way. Using preliminary temperatures and surface gravities, and assuming [Fe/H]$\,=-0.4$ (mean metallicity of our sample -- see Section~\ref{atmo}), we synthesized spectra corresponding to both components using the Spectrum v2.76 \citep{gray94} software and ATLAS9 models from \cite{cas04}. During the spectral synthesis we also accounted for the different projected $v\sin{i}$ velocities of the components  and their macroturbulence velocities $v_{\rm mt}$. The resulting spectra were re-normalized over 5250-5800 ${\rm \AA}$ and we  measured the integrated equivalent width $EW_t$ of all lines within that wavelength range. We then  computed the ratios $k_{21}=EW_2/EW_1$ which are efffectively corrections for the different temperatures of the components of a given system. The final spectroscopic light ratios corresponding to true V-band light ratios were computed simply as the product $k_{21}\cdot I_2/I_1$, and these are also  given in Table~\ref{tbl:spec2}.  

\subsection{Spectral disentangling and atmospheric parameters analysis}
\label{atmo}
We used the  method outlined by \cite{gon06} to disentangle the individual spectra of the binary components. The method is an iterative procedure in which a spectrum of one component is alternately used to calculate a spectrum of the other one. The method works in the real domain. For the renormalization of the disentangled spectra we follow the methodology described in G14. The disentangling was done with the RaveSpan code. We used MIKE spectra and in two cases also UVES spectra to obtain the disentangled spectra. The resulting $S/N$ ratios vary from as low as 8 (LMC-ECL-24887) to about 45 for the brightest components.

Atmospheric parameters and the iron content were obtained from the equivalent widths (EWs) of the iron spectral lines in the wavelength range  5800 ${\rm \AA}$ to 6800 ${\rm \AA}$. See \cite{mar08} for a more detailed explanation of the method  used to measure the EWs and \cite{vil10} for a description of  the line list that was used. We adopted $\log{\epsilon}(Fe)=7.50$ as the solar iron abundance. The local thermodynamic equilibrium  program MOOG \citep{sne73} was used for the abundance analysis together with Kurucz atmosheric models \citep{kur70}. T$_{\rm eff}$ values were derived by requiring excitation equilibrium for the FeI lines, surface gravities by requiring ionization equilibrium between Fe~I and Fe~II, and microturbulent velocities by imposing no trend in the relation between Fe~I abundances and EWs. Usually between 70 and 75 Fe~I lines and 5-8 Fe~II lines were used depending on the quality of the spectrum. The resulting  effective temperatures $T_{\rm a,eff}$, surface gravities $\log g$, microturbulence velocities $v_t$, and [Fe/H] metallicities  are given in Table~\ref{tbl:atmo}. The mean errors of the estimated parameters are 75 K, 0.4 dex, 0.3 km s$^{1}$, and 0.15 dex, respectively. We calculated two sets of solutions, one with $\log g$ as a free parameter and one with $\log g$ fixed to the dynamical values from the WD model. We then averaged the results. 

During this analysis we noted a significant degeneracy between T$_{\rm eff}$ and [Fe/H], especially for lower S/N spectra. This probably arises from misplacement of the continuum level in those spectra: some systematic shifts of the continuum arise from unrecognized weak lines and from the renormalization of such low S/N spectra. To account for this correlation, we recalculated solutions for some low S/N spectra by fixing T$_{\rm eff}$ to values resulting from the color-temperature relations (see Sec.~\ref{red} for references). We measured a mean metallicity - temperature shift of 0.07 dex per 100 K. We used this value to adjust all measured [Fe/H] using differences between $T_{\rm a,eff}$ and the final temperatures from our analysis (see Sec.~\ref{fin}). The corrected metallicities [Fe/H]$_{\rm C}$ are reported in the final column of Table~\ref{tbl:atmo}.

\begin{deluxetable}{@{}lcccc@{}}
\tabletypesize{\scriptsize}
\tablecaption{Color Excess E($B\!-\!V$) \label{tbl-5}}
\tablewidth{0pt}
\tablehead{\colhead{OGLE ID} &\colhead{Na I D1} &\colhead{Red. map} & \colhead{Atmos.} &\colhead{Adopted}\\
\colhead{} &\colhead{(mag)} &\colhead{(mag)} & \colhead{(mag)} &\colhead{(mag)}
}
\startdata
LMC-ECL-01866 & 0.134 & 0.118 & 0.094 & $0.115\pm0.016$ \\
LMC-ECL-03160 & 0.109 & 0.125 & 0.151 & $0.128\pm0.020$ \\
LMC-ECL-05430 & 0.108 & 0.118 & 0.120 & $0.115\pm0.012$ \\
LMC-ECL-06575 & 0.105 & 0.096 & 0.134 & $0.112\pm0.016$ \\
LMC-ECL-09114 & 0.091 & 0.114 & 0.169 & $0.125\pm0.028$ \\
LMC-ECL-09660 & 0.069 & 0.125 & 0.149 & $0.114\pm0.029$ \\
LMC-ECL-09678 & 0.093 & 0.143 & 0.116 & $0.117\pm0.023$ \\
LMC-ECL-10567 & 0.097 & 0.111 & 0.119 & $0.109\pm0.012$ \\
LMC-SC9\_230659 & 0.126 & 0.136 & 0.183 & $0.148\pm0.027$ \\
LMC-ECL-12669 & 0.122 & 0.119 & 0.074 & $0.105\pm0.025$ \\
LMC-ECL-12875 & 0.122 & 0.096 & 0.210 & $0.142\pm0.049$ \\
LMC-ECL-12933 & 0.124 & 0.086 & 0.186 & $0.132\pm0.041$ \\
LMC-ECL-13360 & 0.112 & 0.150 & 0.159 & $0.140\pm0.022$ \\
LMC-ECL-13529 & 0.100 & 0.132 & 0.132 & $0.121\pm0.020$ \\
LMC-ECL-15260 & 0.106 & 0.093 & 0.175 & $0.125\pm0.032$ \\
LMC-ECL-18365 & 0.101 & 0.139 & 0.149 & $0.130\pm0.024$ \\
LMC-ECL-18836 & 0.167 & 0.201 & 0.189 & $0.186\pm0.019$ \\
LMC-ECL-21873 & 0.115 & 0.122 & 0.127 & $0.121\pm0.012$ \\
LMC-ECL-24887 & 0.139 & 0.172 & 0.202 & $0.171\pm0.028$ \\
LMC-ECL-25658 & 0.065 & 0.096 & 0.106 & $0.089\pm0.022$ 
\enddata
\end{deluxetable}
 
  \subsection{Reddening}
 \label{red}
The interstellar extinction in the direction to each of our target stars was derived in three ways following G14. First we used a calibration  between the equivalent width of the interstellar  absorption Na I D1 line (5890.0 \AA) and the reddening \citep{mun97}. The calibration works best for relatively small values of reddening, E($B\!-\!V$)$\,<0.4$ mag. The method uses an empirical relation between gas and dust content calibrated for our Galaxy. We can expect deviations from this relation for the LMC, however, since on average more than half of the total reddening comes from foreground galactic extinction. As a result any possible systematic offset is small and is included in the statistical error of final reddening. We calculated the reddening separately for the galactic  absorption component(s) and the LMC absorption component of Na I D1 line. Then we add them and the second column of Table~\ref{tbl-5} gives appriopriate values.

The second method is based on the Magellanic Cloud reddening maps published by \cite{has11}. We calculate an average E($V\!-\!I$) value from all of the reddening estimates within a 2 arc min radius of our stars and divided this average by a factor of 1.3 to get a ($B\!-\!V$) color excess for each system. We then added $\Delta$E($B\!-\!V$)$\,=0.064$ mag, the mean foreground Galactic reddening in the direction of the LMC as derived from extinction maps published by \cite{sch98} with the renormalization of \cite{sch11}. The third column of Table~\ref{tbl-5} gives the resulting reddening to each eclipsing binary.   

The third method uses the effective temperatures we derived from the atmospheric analysis described in Section~\ref{atmo}. We estimated the intrinsic ($V\!-\!I$) and ($V\!-\!K$) colors of each component from the effective temperature - ($V\!-\!I$) and ($V\!-\!K$) color calibrations \citep{ben98, hou00, ram05, gon09, wor11}. These colors were compared with the observed colors of the components obtained from the preliminary solution to directly derive E($V\!-\!I$) and E($V\!-\!K$) color excesses, and then combined to derive a E($V\!-\!B$) color excess. The reddening to each system was calculated as the mean value of the two components with the exception of the system LMC-ECL-01866 where we have at our disposal a reddening estimate from only the secondary component.

\begin{figure}
\hspace*{-0.75cm}
\includegraphics[angle=0,scale=0.33]{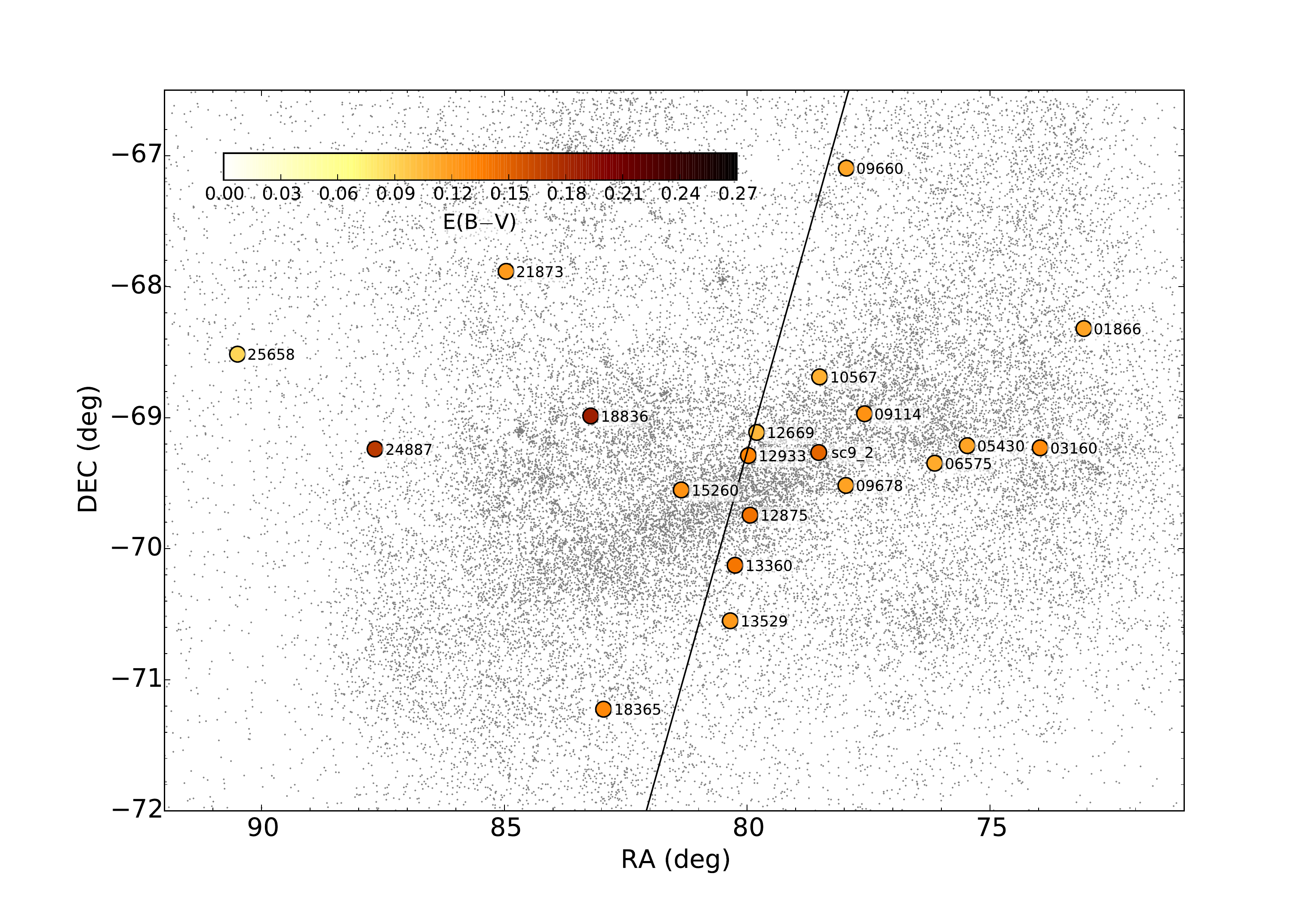}
\caption{The reddenings of the eclipsing binary stars. \label{figebv}}
\end{figure}

Each of the three methods has a precision of approximately 0.03 - 0.04 mag. We calculated an average reddening for each system from the three estimates, and used this new reddening estimate to update the temperature scale of the components. We then calculated new models and repeated the reddening derivation using the third method. The fourth column of Table~\ref{tbl-5} gives the reddening estimates after two such iterations. The fifth column presents the adopted E($B\!-\!V$) for each system - see also Fig.~\ref{figebv}. The reddenings were used to calculate bolometric flux scaled temperatures and to determine the intrinsic colors of the components. 

\subsection{Effective temperatures}
\label{sec:temp}
The effective temperature was also calculated by using calibrations given by \cite{kov06} which utilize absorption depth ratios of specific metallic lines. Because of the low S/N ratio of the spectra and heavy blending of the spectral lines, we used multicomponent gaussian deblending of lines. The formal precision of such derived temperatures are of order of a few Kelvins but the absolute precision is about 70-100 K. Those temperatures $T_{\rm s,eff}$ are given in Table~\ref{tbl:atmo}. In most cases the temperatures from the \cite{kov06} calibrations  correspond well to temperatures derived from atmospheric analysis. Even better agreement is found with temperatures derived from color calibrations. 

The combined temperatures from the both spectroscopic methods were used to derive reddenings to individual systems (Sec.~\ref{red}, third method) and to update the temperature scale of the WD solutions.

\subsection{Fitting  Light and Radial Velocity Curves} 
\label{fit}
 The light and radial velocity curves were solved simultaneously following  the methodology of data fitting described by G14, however with some modifications. The first modification was the inclusion of the spectroscopic light ratios into the modeling. We searched for best solutions giving a model V-band light ratio in agreement with the spectroscopic light ratio (Table~\ref{tbl:spec2}) to within the estimated errors. For a number of systems where total eclipses are observed, we obtained  very good agreement without additional constraints. However, for partial eclipse systems  we usually had to enforce solutions by fixing one of the modified Roche lobe potentials $\Omega$, which approximately corresponds to fixing one of the fractional radii $r$, during the fitting procedure. 
A second modification was to allow for negative third light $l_3$ in our models. This was motivated by the fact that for many systems we observed that some improvement of the fit can be obtained by simply allowing for negative $l_3$ in $V$-band. Usually a small value of $l_3$ in a range of $-0.04$ to $-0.02$ was found with $l_3$ in $I$-band being simultaneously consistent with zero. Finally we allowed for different systemic velocities of the components of the binary, caused by the combined influence of  convective blueshift and  gravitational redshift on the stellar surface.  For some systems the difference of the systemic velocities is comparable or even larger than the rms of the solution.

For a few systems showing noticeable out-of-eclipse proximity effects, we fitted an additional  free parameter, the albedo $A$. While $A$ was usually found to be close to the expected value for a convective atmosphere (i.e. 0.5), in two systems LMC-ECL-10567 and LMC-ECL-15260 the albedo of the secondary component is significantly smaller,  $A_2=0.36\pm0.02$ and $A_2=0.16\pm0.05$, respectively. 
 
\begin{turnpage}
\begin{deluxetable*}{@{}rlllllll | lllllcc}
\tabletypesize{\scriptsize}
\tablecaption{Model parameters from the Wilson-Devinney code \label{tbl-3}}
\tablewidth{0pt}
\tablehead{
\colhead{OGLE ID} &\multicolumn{7}{c}{Orbital parameters} & \multicolumn{6}{c}{Photometric parameters} &\colhead{Type} \\
\colhead{LMC-ECL-} &\colhead{$P_{\rm obs}$} &\colhead{$q=\frac{M_2}{M_1}$}&\colhead{$a$} &\colhead{$\gamma$}&\colhead{$e$} & \colhead{$\omega$}&\colhead{$K$} & \colhead{$i$} & \colhead{$T_{\rm eff}$\tablenotemark{a}} & \colhead{$\Omega$} &\colhead{$r$}&\colhead{$\left(\frac{L2}{L1}\right)_V$} & \colhead{$l_{3,V}$} &\colhead{of} \\
\colhead{} &\colhead{(days)} &\colhead{}&\colhead{($R_\odot$)}& \colhead{$\left(\frac{\rm km}{\rm s}\right)$}&\colhead{} & \colhead{(deg)} &\colhead{$\left(\frac{\rm km}{\rm s}\right)$}& \colhead{(deg)} & \colhead{(K)} & \colhead{} &\colhead{} & \colhead{$\left(\frac{L2}{L1}\right)_I$} & \colhead{$l_{3,I}$} &\colhead{Sol.} \vspace{0.08cm}
}
\startdata
01866$\;$ p&251.247(3)&0.997(4)&322.3(8)&293.67(12)&0.241(1)&15.6(4)&33.20(14)&83.9(4)&5350(14)&13.34(12)&0.0832(16)&1.13(4)&0.073(22) & 3\\
           s&&&&293.40(4)&&&33.30(5)&&4500&8.15(5)&0.1463(15)&1.51(3)&0.075(22) & \\
03160$\;$ p&150.1541(17)&1.006(6)&182.2(5)&267.55(12)&0&--&30.47(14)&82.02(15)& 4852(6)&11.70(13)&0.0936(15)&2.80(5)&$-$0.042(21) & 3\\          
           s&&&&267.77(10)&&&30.31(11)&&4450&5.92(3)&0.2056(12)&3.30(5)&0 &\\
05430$\;$ p& 505.183(5)&1.242(4)&487.7(8)&257.79(5)&0.192(3)&97.7(1)&27.54(6)&87.65(3)&4730&18.37(14)&0.0595(7)&1.51(4)&0 &1 \\
           s&&&&257.82(6)&&&22.18(7)&&4770(5)&18.59(13)&0.0711(6)&1.49(3)&0 &\\ 
06575$\;$ p&189.9943(9)&0.957(3)&280.0(5)&274.31(9)&0&--&36.10(11)&81.81(14)& 4870&7.34(4)&0.1570(15)&0.832(32)&$-$0.023(15) & 3\\
             s&&&&274.28(6)&&&37.71(8)&&4631(3)&6.77(3)&0.1671(15)&0.909(29)&0 &\\   
09114$\;$ p&214.3655(10)&0.970(5)&281.5(8)&272.16(11)& 0.036(4) &97.3(9)&32.73(12)&88.7(5)&5250&11.70(5)&0.0936(12)&0.594(31)&0.004(15) & 3\\    
            s&&&&271.92(21)&&&33.74(10)&&5413(5)&15.61(12)&0.0668(13)&0.567(25)&0.016(15) & \\   \\            
09660$\;$ p&167.7964(13) &1.005(3)&232.4(3)&286.13(5)&0.050(2)&208(2)&35.14(6)&87.46(25)&5264(5)&10.88(9)&0.1019(9)&1.66(2)&$-$0.028(17) & 3\\
             s&&&&286.38(6)&&&34.96(7)&&4620&6.33(3)&0.1912(11)&2.07(3)&0 &\\  
09678$\;$ p&114.4497(5)&1.062(8)&169.1(6)&246.03(18)&0&--&38.11(20)&81.77(12)&5252(36)&13.33(13)&0.0815(9)&2.45(3)&$-$0.050(25)& 4 \\    
             s&&&&246.38(17)&&&35.87(20)&&4673&6.86(5)&0.1811(15)&3.17(3)&0 & \\  
10567$\;$ p&117.9747(7)&0.955(5)&189.1(5)&264.99(12)&0&--&39.39(14)&83.7(3)&5059(7)&8.65(7)&0.1302(15)&1.41(3)&0.056(25) & 3\\
            s&&&&265.16(13)&&&41.23(16)&&4680&5.97(3)&0.1939(12)&1.61(3)&0.007(26) &\\     
230659$\;$ p&772.638(11)&0.953(6)&679(2)&266.39(7)&0.417(2)&301.9(3)&23.86(8)&88.45(6)&4990&22.29(9)&0.0484(3)&0.500(14)&0 & 1\\
            s&&&&266.34(11)&&&25.04(12)&&4982(12)&29.4(3)&0.0345(4)&0.501(9)&0& \\      
12669$\;$ p&749.632(6)&1.064(10) &542(3)&247.81(11)&0.491(3)&77.3(2)&21.66(14)&89.11(5)&4700&24.50(28)&0.0446(6)&1.11(4)&$-$0.005(23) & 4\\
             s&&&&247.76(11)&&&20.35(14)&&4724(40)&26.65(33)&0.0431(5)&1.01(3)&0 &\\     \\  
12875$\;$ p&152.8498(9)&1.014(8)&186.0(7)&266.82(15)&0&--&30.79(18)&83.25(11)& 4802(31)&12.91(8)&0.0841(6)&3.36(2)&0 &2\\
                 s&&&&267.02(12)&&&30.33(14)&&4300&5.65(3)&0.2197(15)&4.32(2)&0 &\\   
12933$\;$ p& 125.3946(12)&0.999(7)&152.6(6)&271.38(11)&0&--&30.57(14)&83.49(28)&4873(8)&9.81(10)&0.1136(14)&2.57(5)&$-$0.040(20)&3\\
                 s&&&&271.67(14)&&&30.61(18)&&4470&5.22(3)&0.2388(18)&3.00(4)&0& \\ 
13360$\;$ p& 262.4386(18)&1.028(4)&345.3(7)&263.08(7)&0&--&33.59(9)&84.66(8)&5550&12.39(8)&0.0883(11)&0.99(5)&$-$0.040(18)&3\\
                 s&&&&263.61(8)&&&32.68(9)&&5031(5)&9.99(6)&0.1144(10)&1.15(4)&0 &\\ 
13529$\;$ p& 49.46703(14)&0.984(3)&101.1(2)&259.94(12)&0&--&50.54(13)&80.18(18)&5360&6.94(6)&0.1685(21)&0.82(4)&0&3\\
                 s&&&&260.15(11)&&&51.38(12)&&5285(10)&7.25(7)&0.1581(22)&0.83(4)&$-$0.020(25)& \\
15260$\;$ p&157.4692(8)&0.981(6)&174.5(7)&276.67(9)&0&--&27.56(11)&83.2(9)&4791(5)&8.50(4)&0.1332(24)&2.06(15)&$-$0.028(11)&3\\
                 s&&&&276.65(11)&&&28.10(14)&&4450&5.10(3)&0.2421(52)&2.38(16)&0 &\\ \\
18365$\;$ p&78.5436(6)&0.313(2)&140.6(4)&274.90(13)&0&--&21.61(14)&89.23(14)&4890&4.11(2)&0.2656(15)&0.206(7)&0&1\\    
                 s&&&&274.52(17)&&&68.93(20)&&4993(9)&4.11(2)&0.1135(9)&0.199(5)&0 &\\       
18836$\;$ p&182.5526(13)&0.974(7)&241.2(9)&254.75(19)&0.107(7)&293.8(9)&33.07(20)&85.57(10)&5116(47)&16.20(21)&0.0662(10)&1.80(5)&0&2\\
                 s&&&&254.68(13)&&&33.95(15)&&4600&8.74(7)&0.1281(13)&2.35(5)&0& \\ 
21873$\;$ p&144.1882(8)&1.037(5)&211.3(5)&296.79(12)&0.008(3)&292(8)&37.48(13)&83.49(11)&5300&11.47(9)&0.0960(10)&1.11(3)&$-$0.028(23)&3\\
                 s&&&&297.07(11)&&&36.16(11)&&5026(8)&9.87(8)&0.1169(11)&1.20(3)&0 &\\
24887$\;$ p& 94.8444(5)&1.083(11)&156.6(8)&289.62(23)&0&--&43.33(28)&86.01(7)&5150&10.61(14)&0.1050(16)&1.08(3)&0&1\\
                 s&&&&289.65(27)&&&40.00(32)&&5070(9)&10.47(15)&0.1140(18)&1.11(3)&0 &\\
25658$\;$ p&192.7893(14)&1.000(5)&231.3(6)&257.16(15)&0.373(4)&263.9(1)&32.70(13)&89.79(18)&4860&12.41(8)&0.0926(6)&1.405(9)&$-$0.015(8)&3\\
                 s&&&&257.67(16)&&&32.71(14)&&4723(6)&9.97(5)&0.1195(7)&1.477(7)&0 &            
\enddata 
\tablecomments{Simultaneous solutions of $V$-band and $I$-band light curves together with radial velocity curves of both components. Quoted uncertainties are the standard errors from the Differential Corrections subroutine combined with errors from
Monte Carlo simulations with the JKTEBOP code ver.~34 (see Section 4 for details).
The last column gives the type of the solution (see Section 4 for details).}
\tablenotetext{a}{The effective temperatures used in the fitting. The values without uncertainties are fixed values.}
\end{deluxetable*}
\end{turnpage}

\section{Adopted solutions and final parameters}
\label{fin}

After the final reddening estimate we set the temperature of the more luminous component to be the average of the spectroscopic and color temperatures, and we then recalculated all of the models. The consistency of the temperature scale  of each eclipsing binary was checked by computing the distance to each system resulting from scaling  the bolometric flux observed at Earth. To calculate the bolometric corrections we used an average from several calibrations \citep{cas10,mas06,alo99,flo96}. 

We computed four types of models for every system: 1) setting limb darkening coefficients according to a logarithmic law (LB$=-2$) and the third light $l_3=0$; 2) adjusting coefficients of a linear law of limb darkening (LB=+1) for both stars and both light curves and $l_3=0$; 3) adjusting the third light $l_3$ in two bands with a logarithmic law of limb darkening (LB$=-2$); and 4) adjusting both the third light and linear coefficients of limb darkening together giving six more free parameters. We compared the models according to their reduced $\chi^2_{\rm r}$ and chose the one with the lowest value. Models with very high ($>1.1$) or very low ($<0.1$) limb darkening coefficients were excluded even if they produced a better formal fit with lower $\chi^2_{\rm r}$. Table~\ref{tbl-3} lists the parameters of the best model for each system. Individual model solutions to the radial velocity and the $I$-band light curves are presented in the Appendix in Figures:~\ref{fig1} -- \ref{fig10}. 

The quoted uncertainties were calculated by combining the formal errors reported by the Differential Correction routine of the Wilson-Devinney code and errors from the Monte-Carlo simulations with the JKTEBOP code ver. 34 \citep{pop81,sou04,sou13}. We typically used the larger error of the two but in some cases we used the average of the two errors. For the Monte Carlo simulations with  JKTEBOP we ran 10000 models, simultaneously solving light and radial velocity curves. For the JKTEBOP calculations we used the same set of free parameters as for the WD code. Usually very good agreement was obtained between model parameters of the WD code and the JKTEBOP code. However, for some systems having larger proximity effects (i.e. having more deformed components) we fixed some parameters like the orbital inclination, the radii ratio, or the eccentricity in order to obtain a consistent solution with  the WD model.

\begin{figure}
\hspace*{-0.5cm}
\includegraphics[angle=0,scale=0.5]{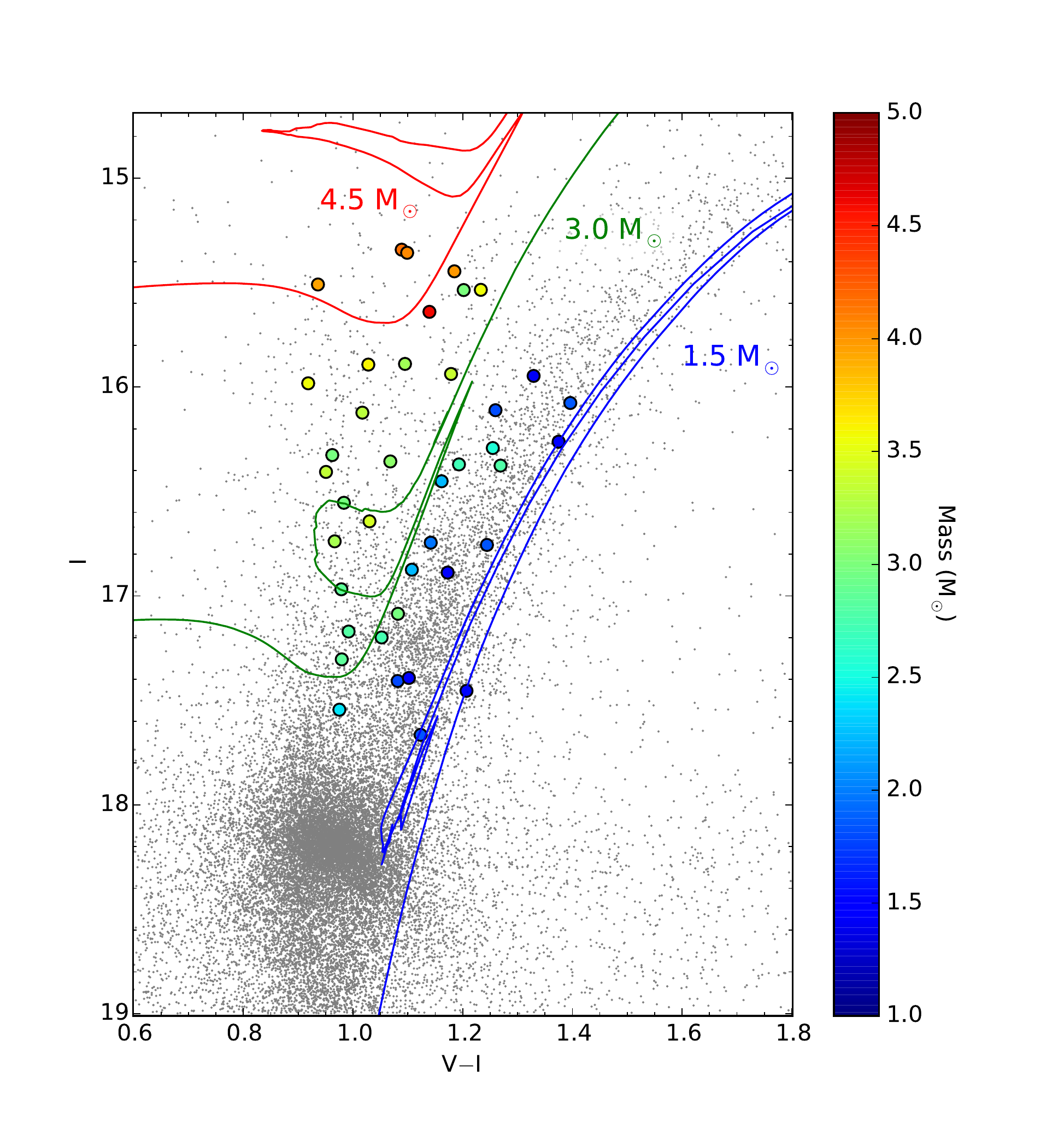}
\caption{Color-magnitude diagram of the central part of the LMC from the OGLE-III  lmc100.2 field (grey dots). The superimposed circles are the observed positions of the 40 giants in our sample. The positions were shifted to account for differential reddening. The mean reddening of the lmc100.2 field was assumed to be $E(B\!-\!V)=0.1$ mag. The stars are color-coded by their mass. \label{figVI}}
\end{figure}

Absolute physical parameters for all components are given in Table~\ref{tbl-6}. The masses and radii of the stars were calculated by adopting astrophysical constants from IAU resolution B3 \citep{mam15}. Fig~\ref{figVI} shows the positions of individual stars in the color-magnitude diagram of the central part of the LMC. Superimposed are evolutionary tracks for solar composition stars interpolated from the MIST tracks \citep{cho16}, assuming [Fe/H]=$-0.4$ dex and $E(B-V)=0.1$ mag.     

Spectral types for the components were indirectly derived from surface temperatures and gravities using the calibration of \cite{alo99}. Final temperatures are averages of both spectroscopic temperatures, a color temperature derived from several calibrations, and the temperature resulting from bolometric flux scaling. Because of this, the final temperature ratios are somewhat different from the temperature ratios obtained from the WD models. Our procedure leads to a slight revision of the parameters for nine previously published systems \citep[P13,][]{elg16}. The precision of the mass and radii determinations is better, sometimes by a factor of three. However the temperature errors quoted in Table~\ref{tbl-6} are sometimes larger than those reported in P13. This reflects the fact that temperatures determined with different methods  in some cases show a significant spread (see Tab.~\ref{tbl:atmo}). In our opinion the uncertainties in effective temperatures given in Table~\ref{tbl-6} are a better estimate of the  real errors on the absolute temperature scale. 
The metallicity of each system was calculated as a flux weighted average of the metallicity of both components. In this way we assumed the same  composition for the components in any particular system.    

\section{Notes on individual systems}
\subsection{OGLE LMC-ECL-01866}
Although the system consists of two almost equal mass giants, their other physical parameters are much different. In P13 we derived precise parameters of the system for the first time. Here we present an update on the parameters, especially the radius of the hotter and smaller primary component was revised by 1.5$\sigma$. Our revised  metallicity estimate is larger. Most of the changes are caused by detection of a third light contribution of about $7\%$ in the OGLE light curves (3.5$\sigma$ detection). The inferred extinction corrected third light color is $(V-I)_3=1.10$ for a Galactic foreground object or $(V-I)_3=1.03$ for a LMC object. In the first case the color corresponds to a halo main sequence dwarf of spectral type K3.5 which lies at a distance of 2.5 kpc from the Sun. In the second case the only possibility is that the third light comes from a K1 giant with an effective temperature of 4600 K and a luminosity of about 100 $L_\odot$. Such the giant could not co-evolve with the system (see Fig.~\ref{fig11} in Appendix) and would be a star unrelated to the system, lying along the line-of-sight. This interpretation seems quite unlikely. Interestingly we cannot detect any third light in our spectra. 

The system has a significant eccentricity and the rotation of the primary component is super-synchronous, while the secondary (the larger star) rotates pseudo-synchronously. The fast rotation of the primary can be explained by  angular momentum conservation of the star during its relatively rapid shrinking while descending on the red giant branch. 

\subsection{OGLE LMC-ECL-03160}
The system contains  components with very similar masses but very different temperatures, radii, and luminosities. The updated physical parameters are in full agreement with those presented by P13. The system has a circular orbit and both stars rotate synchronously. 

\subsection{OGLE LMC-ECL-05430}
The system contains components with quite discrepant masses, and it is unlikely that the components evolved together in the system with present day masses. A discussion of its evolutionary status is given Section~\ref{ev05430}. LMC-ECL-05430 is an eccentric system with the secondary rotating pseudo-synchronously, i.e. with  tidal locking at periastron, while the primary rotates slightly faster. The system shows small out-of-eclipse light variations which have  timescales shorter than the orbital period. 

\subsection{OGLE LMC-ECL-06575}
This is the most massive system in our sample. The system shows partial, shallow eclipses and the updated stellar radii are significantly different (by 2.5$\sigma$) from those presented by P13. We attribute this change to the inclusion of spectroscopic light ratios in our analysis which led to a significant change in the derived ratio of the radii but not the sum of the radii. In fact, the sum of the absolute radii $R_1+R_2=90.7 \pm 0.3$ $R_\odot$ is consistent with the previous value,  $89.1 \pm 1.2$ $R_\odot$ (P13) at a 1.2$\sigma$ level. The remaining fundamental parameters (masses, temperatures, metallicity) are fully consistent with the previous results. The primary rotates slightly faster than synchronous rotation, while the secondary slightly slower. This results from a contraction of the primary on the horizontal loop and an expansion of the secondary which is ascending  the red giant branch (see Fig.~\ref{fig11} in the Appendix), while the system was rotationally synchronized. 

\subsection{OGLE LMC-ECL-09114}
This system was the first LMC eclipsing binary analyzed by our team \citep{pie09}, which led to the first distance determination to the LMC from late type eclipsing binaries. In our subsequent paper (P13) we presented updated physical parameters of the system. Using OGLE-IV photometry and our dedicated photometric $V$- and $I$-band observations we found that there are total eclipses in this system,  leading  to a revision of some parameters. In particular, the radius of the hotter and smaller  secondary component was revised by almost 3$\sigma$. In the present work the derived physical parameters are in full agreement with the parameters reported by P13. We have slightly improved the mass and radii determinations, however, we could not improve the precision of the estimate of the  effective temperature. Both components rotate  faster than synchronous rotation by a factor of 1.5.   

The analysis of the OGLE light curves showed possible very red third light in this system. The detection is only at a $1\sigma$ level but nevertheless may significantly influence the intrinsic infrared colors of the system. The estimated intrinsic third light color is $(V-I)=2.45$. It is very unlikely that this could be a companion to the system, a more plausible explanation is that it is a foreground halo M dwarf blending with the system. The $(V-I)$ color corresponds to a spectral type of M3 and an intrinsic $(V-K)=4.65$ \citep{pec13}.    
The inferred distance to the dwarf is 14 kpc and the estimated light contribution in K-band is $l_{3K}=0.041$. 

\begin{turnpage}
\begin{deluxetable*}{@{\hspace{-12pt}}rcccllllllllllll}
\tabletypesize{\scriptsize}
\tablecaption{Physical Properties of Giants in the LMC \label{tbl-6}}
\tablewidth{0pt}
\tablehead{
\colhead{OGLE ID} &\colhead{Spectral}&\colhead{V$_0$}& \colhead{(V$-$K)$_0$}&\colhead{Mass}& \colhead{Radius}&\colhead{$\log{g}$}& \colhead{$T_{\rm eff}$}&\colhead{Lumin.} &\colhead{$M_{\rm bol}$}&\colhead{$M_V$}&\colhead{$\left[{\rm Fe/H}\right]$}&\colhead{$v\sin{i}$}&\colhead{$v_{\rm sync}$}&\colhead{$v_{\rm psync}$}&\colhead{Age}\\
\colhead{LMC-ECL-}&\colhead{Type} & \colhead{(mag)} &\colhead{(mag)} & \colhead{M$_\sun$} &\colhead{R$_\sun$} & \colhead{(cgs)} &\colhead{(K)} &\colhead{L$_\sun$}&\colhead{(mag)}&\colhead{(mag)}&\colhead{(dex)}&\colhead{km s$^{-1}$}&\colhead{km s$^{-1}$}&\colhead{km s$^{-1}$}&\colhead{Gyr}
}
\startdata
01866$\;$ p &G3III & 16.672& 1.844& 3.560(20) & 26.79(52) & 2.134(17) & 5300(80) &  510(37) & $-$2.02(8) & $-$1.86(8) & $-$0.49(17)		&  16.2(1.0) & 5.4(1) & 9.1(2) & 0.21(1)\\
                 s &K1.5III &16.541 & 2.694& 3.550(31) & 47.11(50) & 1.642(9) & 4495(60) &  816(47) & $-$2.53(6) & $-$1.99(6)&			 	&  8.2(9) & 9.5(1) &16.0(2) &\\
03160$\;$ p &G7III & 18.179& 2.269& 1.792(16) & 17.03(28) & 2.229(14) & 4930(100) &   154(14) & $-$0.72(10) & $-$0.33(10)&$-$0.68(18) 	&   6.4(6) & 5.75(9) & -- & 1.17(7) \\
                 s & K2III& 17.061& 2.736& 1.802(18) & 37.42(24) & 1.548(5) & 4450(70) &   495(32) & $-$1.99(7) &$-$1.45(7) & 				&  13.9(4) &12.6(1) &--\\
   05430$\;$ p & G9III & 17.252& 2.417& 2.717(17) & 28.99(36) & 1.948(11) & 4710(70) &   373(24) & $-$1.68(7)& $-$1.30(7)& $-0.37(10)$	 &  6.4(4) & 2.91(4) & 4.37(6) & 0.25(2)\\
                    s & G9III & 16.807& 2.373 & 3.374(18) & 34.64(28) & 1.887(7) & 4760(65) &   555(32) & $-$2.11(6)&$-$1.74(6)& 				&  5.7(4) & 3.47(3) & 5.22(5)&0.43(4)\\
  06575$\;$ p &G6II-III & 16.121& 2.268& 4.167(22) & 43.93(43) & 1.773(8) & 4920(80) &  1020(69) & $-$2.77(7) &$-$2.39(7) & $-$0.46(10)	&  13.7(8) &11.7(1) &--&0.15(1)\\
                  s & K0III& 16.320& 2.531& 3.989(26) & 46.75(43) & 1.699(8) & 4645(60) &   918(50) & $-$2.66(6) & $-$2.19(6)& 				 &   9.3(8) &12.5(1) &--&\\
 09114$\;$ p & G4III&16.834 &1.968 & 3.304(23) & 26.33(34) & 2.116(11) & 5230(60) &   467(25) & $-$1.92(6) & $-$1.68(6)& $-$0.38(12) 	&   9.9(4) & 6.22(8) & 6.7(1)&0.28(3)\\
                  s &G2III & 17.399& 1.841& 3.205(25) & 18.79(37) & 2.396(17) & 5425(110) &   275(25) & $-$1.35(10) & $-$1.11(10)& 			&  10.1(8) & 4.44(9) & 4.8(1)&\\ \\        
  09660$\;$ p & G3III&16.979 &1.914 & 2.981(13) & 23.66(21) & 2.165(8) & 5250(65) &   383(20) & $-$1.71(6) & $-$1.51(6)& $-$0.46(10)		&   7.7(8) & 7.14(6) & 7.91(7)&0.35(3)\\
                   s & K0III&16.427 & 2.543& 2.997(12) & 44.40(26) & 1.620(5) & 4685(95) &   856(70) & $-$2.58(9) & $-$2.06(9)& 				&  14.9(6) &13.4(1) &14.8(1)&\\                                     
   09678$\;$ p & G4III & 18.211& 1.923 & 2.400(29) & 13.77(16) & 2.540(10) & 5230(80) &   128(8) & $-$0.52(7)& $-$0.32(7)&$-$0.38(16)	 &  5.7(1.1) & 6.09(7) & --&0.58(5)\\
                    s & G9III & 17.236& 2.512 & 2.549(31) & 30.60(28) & 1.873(8) & 4705(90) &   413(33) & $-$1.79(9)& $-$1.30(9)&				& 10.5(4) &13.5(1) &-- &\\     
  10567$\;$ p & G6III &17.109 &2.084 & 3.333(29) & 24.60(29) & 2.179(10) & 5065(100) &   359(30) & $-$1.64(9) & $-$1.37(9)& $-$0.70(10)	&  11.5(5) &10.6(1) &--&0.25(2)\\
                    s &G9III & 16.737& 2.467& 3.184(26) & 36.64(25) & 1.813( 6) & 4715(75) &   598(39) & $-$2.19(7) & $-$1.74(7)& 				&  17.2(5) &15.7(1) &--&\\                                    
 230659$\;$ p & G6III & 16.610&2.187 & 3.598(38) & 32.83(22) & 1.962(5) & 5000(70) &   607(35) & $-$2.21(10) & $-$1.87(10)&$-$0.24(11) 	&  7.3(8) & 2.15(1) & 5.76(5)&0.24(2)\\
            	   s & G6III & 17.363&2.190 & 3.429(30) & 23.40(31) & 2.235(11) & 5030(100) &   316(26)& $-$1.50(11) & $-$1.11(11)&			&  6.4(8) & 1.53(2) & 4.11(6) &\\  
  12669$\;$  p & K0.5III& 17.688& 2.505& 1.843(29) & 24.17(34) & 1.937(12) & 4630(85) &   242(19) & $-$1.21(9) & $-$0.80(9)&$-$0.30(14)	&  2.9(1.1) & 1.63(2) & 5.49(9)&1.13(8) \\ 
                    s & G9III& 17.579& 2.359& 1.962(30) & 23.36(30) & 1.994(11) & 4715(95) &   243(21) & $-$1.21(10) &$-$0.90(10)& 			 &  3.8(1.0) & 1.58(2) & 5.31(8)&\\	   \\
   12875$\;$ p & G8III & 18.479&2.350 & 1.831(20) & 15.62(13) & 2.313(7) & 4845(100) &   121(10) & $-$0.46(9) &$-$0.04(9) & $-$0.48(15)	 &  4.9(1.8) & 5.18(4) & --&1.20(5)\\
                    s & K2.5III& 17.163&2.965 & 1.858(23) & 40.82(32) & 1.486(6) & 4385(110) &   555(56) & $-$2.11(11) &$-$1.35(11) & 			& 15.0(8) &13.5(1) &--&\\ 
   12933$\;$ p & G7III& 18.351& 2.424& 1.516(19) & 17.32(22) & 2.142(11) & 4900(200) &   156(26) & $-$0.73(18) &$-$0.17(18) & $-$0.38(13)&  6.8(1.9) & 7.00(9) & --&1.97(19)\\
                    s & K2III& 17.327& 2.890& 1.514(17) & 36.41(31) & 1.496(7) & 4470(150) &   477(64) & $-$1.95(15) & $-$1.19(15)&			& 15.7(1.0) &14.7(1) &--&\\  
   13360$\;$ p &G1II-III &16.138 & 1.682& 3.950(24) & 30.46(38) & 2.067(11) & 5495(90) &   762(54) & $-$2.46(8) &$-$2.37(8) &$-$0.30(10) 	& 13.6(1.2) & 5.88(7) & --&0.18(1) \\
                    s &G4II-III &16.143 & 2.112& 4.060(24) & 39.46(35) & 1.854(8) & 5085(80) &   938(61) & $-$2.68(7) & $-$2.36(7)&			& 11.0(7) & 7.62(7) & -- &\\      
   13529$\;$ p &G4III & 17.635& 1.853& 2.857(16) & 17.03(21) & 2.432(11) & 5295(75) &   205(13) & $-$1.03(7) & $-$0.88(7)&$-$0.18(14)	& 19.5(6) &17.4(2) &--&0.42(3)\\
                    s & G4III& 17.855& 1.914& 2.810(16) & 15.98(22) & 2.480(12) & 5260(90) &   176(13) & $-$0.86(8) & $-$0.66(8)&				& 19.3(1.6) &16.4(2) &-- &\\ 
  15260$\;$ p &G8III &17.750 &2.470 & 1.449(18) & 23.22(43) & 1.868(16) & 4810(130) &   260(30) & $-$1.29(12) &$-$0.75(12) &$-$0.63(12) 	&   8.6(6) & 7.5(1) & --&2.0(2)\\
                   s & K2III&16.966 & 2.878& 1.422(16) & 42.20(92) & 1.340(19) & 4420(85) &   612(54) & $-$2.22(10) & $-$1.53(10)& 		 	&  15.7(8) &13.6(3) &--&\\ \\               
   18365$\;$ p &G8III & 16.469& 2.272& 4.596(37) & 37.30(23) & 1.957(5) & 4900(120) &   723(71) & $-$2.40(11) & $-$2.03(6)&$-$0.56(17)	& 21.0(8) &24.1(2) &--&0.11(1)\\
                    s &G7III & 18.187& 2.174& 1.441(15) & 15.94(13) & 2.192(7) & 4940(70) &   136(8) & $-$0.59(6) & $-$0.31(6)& 				& 12.1(1.2) &10.3(1) &--&2.1(2)\\     
   18836$\;$ p & G5III&17.973 &2.003 & 2.858(31) & 15.95(25) & 2.489(13) & 5155(100) &   162(14) & $-$0.77(9) & $-$0.53(9)&$-$0.40(10)	& 10.1(7) & 4.43(7) & 5.5(1) &0.36(2)\\
                    s & K1III& 17.335& 2.625& 2.784(36) & 30.87(33) & 1.904(9) & 4605(80) &   386(28) & $-$1.72(8) &$-$1.17(8) & 				& 10.2(4) & 8.56(9) &10.7(2)&\\ 
  21873$\;$ p & G4III& 17.229&1.869 & 2.984(21) & 20.26(22) & 2.300(9) & 5265(75) &   284(17) & $-$1.38(7) &$-$1.23(7) &$-$0.28(12)		 &  8.8(4) & 7.12(8) & 7.23(9) &0.34(3)\\
                    s & G6III& 17.113&2.107 & 3.093(24) & 24.67(24) & 2.144(8) & 5055(80) &   358(24) & $-$1.63(7) &$-$1.35(7) & 				& 12.3(5) & 8.67(8) & 8.8(1)&\\  
   24887$\;$ p & G5III&17.942 &2.013 & 2.747(47) & 16.43(26) & 2.446(14) & 5130(80) &  168(12) & $-$0.82(8) &$-$0.60(8) &$-$0.22(12) 	& 11.1(7) & 8.8(1) & --&0.42(3)\\
                    s & G6III& 17.857 &2.084 & 2.976(45) & 17.83(30) & 2.409(14) & 5070(80) &  189(14) & $-$0.94(8) &$-$0.67(8) &				 & 12.0(6) & 9.5(2) & -- &\\ 
   25658$\;$ p & G8III&17.674 & 2.309& 2.231(24) & 21.40(15) & 2.126(6) & 4840(70) &   226(14) & $-$1.14(6) & $-$0.78(6)&$-$0.48(13)		&  7.0(9) & 5.62(4) &13.3(2)&0.77(6) \\
                    s & G9III& 17.305& 2.452& 2.230(23) & 27.61(19) & 1.904(6) & 4720(75) &   341(22) & $-$1.58(7) & $-$1.15(7)&  				 & 17.4(8) & 7.25(5) &17.1(2) &                                                                                                                                     
 \enddata
 \tablecomments{Absolute dimensions were calculated assuming nominal solar gravitational constant $\mathcal{GM}_\sun=1.3271244\!\cdot\!10^{20}$ m$^3$ s$^{-2}$, nominal solar radius $\mathcal{R}_\sun=695 700$ km, solar effective temperature $\mathcal{T}_{\rm eff,\sun}=5772$ K \citep{mam15}, and the solar bolometric absolute magnitude $M_{bol,\sun}=+4.75$ .
The magnitudes and colors are extinction corrected values. The $V-K$ color is expressed in the 2MASS system.}
\end{deluxetable*}
\end{turnpage}

\subsection{OGLE LMC-ECL-09660}
The presence of total eclipses in this system makes a determination of the radii more robust and indeed the radii we derived are in full agreement with the previous results presented by P13. The remaining parameters are in full agreement with P13 with the exception of the temperature of the primary which was revised downward by $1\sigma$. The orbit of the system has a low eccentricity and the rotation of both components is tidally locked at periastron. The eccentricity is at odds with the circularization time of the system (Section~\ref{circula}) and the  relative sizes of the stars. The non-circular orbit may result from the influence of a third body in the system. There is no evidence for third light in the spectra nor in the OGLE light curves. 

\subsection{OGLE LMC-ECL-09678}
The system shows partial and shallow eclipses. The components have much different radii and surface temperatures. The primary component rotates synchronously being on the horizontal loop, while the larger companion star is actually expanding quickly on the asymptotic giant branch. As a result its rotation is slower than synchronous.  

\subsection{OGLE LMC-ECL-10567}
The updated physical parameters of this system are in perfect agreement with the parameters published by P13. The rotation of the components is similar to that in the system LMC-ECL-06575, i.e. it results from the combined effects of a prior synchronization of rotation and the angular momentum conservation of rapidly contracting/expanding components during thier current evolutionary stage. As a result, the shrinking primary which is on the horizontal loop rotates slightly super-synchronously, while the expanding secondary which is ascending the red giant branch rotates slightly sub-synchronously (see Fig.~\ref{fig12} in the Appendix). We also confirm the blue third light in the system reported by P13. 

\subsection{OGLE LMC SC9\_230659}
During the reanalysis of the data we noted that this system shows a shift of the secondary eclipse with respect to the primary eclipse. We adopt notation from P13 in which the primary star is occulted during the {\it shallower} eclipse when the components are 
much wider separated than during the secondary {\it deeper} minimum. In order to verify  possible apsidal motion in this system we determined times of minima from $V$-, $I$-band OGLE photometry and from $V$-, $R$-band MACHO photometry \citep{alc99,fac07}. The times of minima are presented in Table~\ref{tbl-oc}.  We fitted a linear ephemeris to both primary and secondary minima: 
\begin{eqnarray*}
\label{ }
{\rm T1(HJD}) & = & 2450355.23(19) + 772.746(48){\rm\, \times\, E}  \\
{\rm T2(HJD}) & = & 2450858.66(4) \;\,+ 772.430(9){\rm \,\times \,E}
\end{eqnarray*}
The secondary eclipses follow a linear ephemeris, however the primary eclipses follow an ephemeris with a significantly different period and with a much larger scatter. For the purpose of obtaining a consistent WD solution we added as a free parameter the first time derivative of the argument of periastron $d\omega/dt$. Inclusion of this additional parameter significantly improved the residuals of the solution during eclipses. We estimated an apsidal motion  of 0.0025 rad per orbital period. This is a surprisingly fast motion given the long orbital period system. Because the components are  well detached the only possible origin of the apsidal motion is an interaction with a third body orbiting the system. We carefully examined the spectra of the system by removing a contribution from both components and analyzing the residual spectra. Using templates with temperatures between 3000 K and 7000 K we tried to find any signature of the third spectrum. This search was unsuccessful, and  we put an upper limit for $l_3$  of 3\% in the visual region of the spectrum. 

We also examined the light curves in order to search for a  third light contribution. First we solved only for the $I$-band light curve, such solutions are fully consistent with $l_3=0$. We then added the $V$-band light curve, but because of the poor $V$-band coverage of the primary eclipse  a strong correlation between the photometric parameters prevents a determination of $l_3$. The third light contribution of between 0 and 8\% is possible in the $V$-band. The inclusion of MACHO photometry did not help because of much larger scatter in the light curve compared to OGLE photometry (by a factor of 3) and a large number of outliers. We conclude that a third light contribution to the  $V$-band is lower than 3\% and lower than 0.5\% in the $I$-band. If this is the explanation of the apsidal motion then the perturbing object is most probably a main sequence star of spectral type later than A2 with a mass smaller than 2 $M_\odot$ in a sufficiently tight outer orbit.

\begin{deluxetable}{@{}rcccc@{}}
\tabletypesize{\scriptsize}
\tablecaption{Times of minima of LMC SC9\_230659 \label{tbl-oc}}
\tablewidth{0pt}
\tablehead{\colhead{Epoch} &\colhead{HJD} &\colhead{Error} & \colhead{Type} &\colhead{Source\tablenotemark{a}}\\
\colhead{} &\colhead{+2400000} &\colhead{(days)} & \colhead{} &\colhead{}
}
\startdata
$-$2 & 49313.95 & 0.10 &  Sec & M \\
$-$1 & 49582.89 & 0.09 &  Pri &  M \\
$-$1 & 50086.22 &  0.12 & Sec& M\\
0  &    50355.31 & 0.11 &  Pri & M \\
0  &   50858.54  & 0.04  &  Sec & O, M\\     
1   &  51127.60  &0.09 & Pri & O, M\\
1 &   51630.96  & 0.03  & Sec & O\\     
3  &  52673.35 & 0.04 & Pri & O\\
4   &  53445.90 & 0.23  & Pri & O\\
4  &  53948.44  & 0.18  & Sec & O\\
5  &  54720.78  & 0.12 & Sec & O\\
6   & 55493.31 & 0.02  & Sec & O\\
7  &  56265.70  & 0.08& Sec & O\\
8  &  56537.46 & 0.12  & Pri & O\\
8 &   57038.06  & 0.07   &  Sec & O  
\enddata
\tablenotetext{a}{O - OGLE, M - MACHO}
\end{deluxetable}

\subsection{OGLE LMC-ECL-12669}
\begin{figure*}
\includegraphics[angle=0,scale=.7]{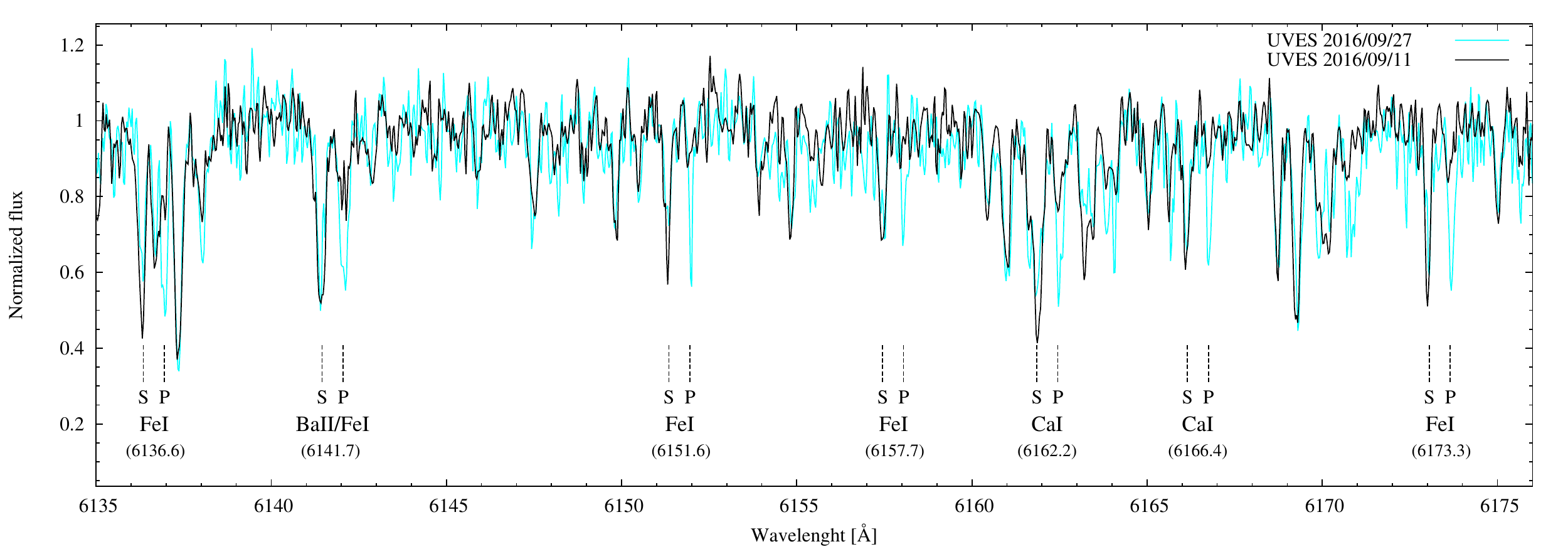}
\caption{Part of two UVES spectra of OGLE LMC-ECL-12669 taken in 2016 within two weeks. Spectra were shifted to the system's rest frame. Both were taken well outside of eclipses at  orbital phases $\sim$0.76 and $\sim$0.78. 
The position of the stronger absorption lines are noted for the primary (P) and the secondary (S) star. The spectrum taken on September 11th shows a peculiar weakening of the absorption lines from the primary. \label{fig_uves}}
\end{figure*}

This system has the second longest orbital period in the sample. The components of the system are very similar physically and they have almost the same radii and luminosities. The rotation of the components is neither synchronous nor pseudo-synchronous, but instead is in between. In the spectra we observe that the strengths of the absorption lines are also similar. However, in one 
spectrum taken with the UVES spectrograph on 11th September 2016~UT we noted that the lines of the primary component are conspicuously weak. The spectrum has an integration time of 20 minutes, S/N$\sim$20 at 6150 ${\rm \AA}$, and was taken well outside of eclipses at an orbital phase of 0.76. We compare the spectrum with another UVES spectrum taken 16 days later having S/N$\sim$18 in Fig~\ref{fig_uves}. The equivalent widths of the spectral lines from the primary are diminished  by 65\% on average while the lines of the companion star are slightly stronger (by about 5\%). This can be explained by a flux decrease of the primary by 65\% during the exposure, resulting in both weaker lines of the primary and a smaller dilution of the lines of the companion in the spectrum. If the secondary star had its flux not decreased during the exposure we would expect its lines to be stronger by about 20\%. The radial velocity of the primary is in full agreement with the expected value from the orbital solution while the radial velocity of the secondary is 1.5 km s$^{-1}$ lower than expected from the model. The question arises what may be the reason for such a remarkable flux decrease; it is equivalent to a full obscuration of the light from the primary for about 15 minutes. The angular diameters of the components are 4.5$\mu{\rm as}$ and 4.3$\mu{\rm as}$, while the projected angular separation on 11th September was 49.7 $\mu{\rm as}$.  The most compelling explanation is an occultation of the system by a distant Solar System body or a Milky Way foreground body. To estimate the size and projected angular velocity of such an object we assume that the body caused the total occultation of the primary and marginally affected the secondary. This implies that an object with an angular diameter of about 50 $\mu{\rm as}$ would cross at most 50 $\mu{\rm as}$ within 15 minutes giving an upper limit for its annual mean motion of about 2$^{''}$. For the obscuring object such estimate an supports an interpretation for the occulting body to be a fast proper motion, low luminosity object like a brown dwarf or a free-floating giant planet lying within about 25 pc of the sun, but outside of our Solar System. The foreground body would cause a shallow, partial eclipse of the secondary affecting its radial velocity during the event and making its absorption lines less strong than expected.

\subsection{OGLE LMC-ECL-12875}
The system shows total eclipses and slightly tidally deformed components. There is some inconsistency between the spectroscopically determined effective temperatures ($T_1=5123$ K and $T_2=4580$) and the photometrically determined temperatures ($T_1=4775$ K and $T_2=4310$ K) -- see Tab.~\ref{tbl:atmo}. The reason for this discrepancy is unclear. One possibility is that the system has a red companion influencing its infrared colors, however we could not detect any third light in the system. Another possibility is that the disentangled spectra used to determine the atmospheric parameters were not of sufficient quality to give accurate results: the spectrum of the primary is of very low S/N while the lines in the secondary spectrum are quite strongly blended because of rotation. We  discarded the outlying  temperature from the atmospheric analysis, retaining the three other estimates and adopted as final temperatures the weighted average.    

\subsection{OGLE LMC-ECL-12933}
The system contains  components with very similar masses but very different temperatures, radii, and luminosities. The system is very similar  to the Galactic system HD 187669 \citep{hel15}, having similar masses with the larger and cooler component being about twice the size of the hotter component. The components of ECL-12933 are larger and hotter than components of HD 187669 which can be explained by their lower metal abundance. The system also shares many similarities with LMC-ECL-12875: almost the same light curve, with total eclipses and relatively prominent proximity effects, similar luminosities, sizes, and temperatures of the components. Both stars show also a similar inconsistency of the temperatures. In the case of LMC-ECL-12933 the discrepancy is even more prominent: the spectroscopic temperatures are $T_1=5120$ K and $T_2=4610$ H and the photometric temperatures are  $T_1=4690$ K and $T_2=4335$ K -- see Tab.~\ref{tbl:atmo}. We also could not find the source of the discrepancy, there is no trace of the red third light in our data. As final temperatures we used weighted averages of all four temperature determinations. 

\subsection{OGLE LMC-ECL-13360}
The system has relatively deep, partial eclipses with almost no proximity effect visible in the light curves. There is small variability out of eclipse with an amplitude of about 0.006 mag which can be attributed to the presence of spots on one of stars. Both components are on the horizontal loop (see Fig.~\ref{fig12}) and, surprisingly, both show significant super-synchronous rotation. 

\subsection{OGLE LMC-ECL-13529}
The system consists of two almost identical stars in a circular orbit. The eclipses are partial and shallow, and the light curves have relatively large scatter. The components lie on bluest part of the horizontal loop (see Fig.~\ref{fig12}) and both rotate slightly super-synchronously. 

\subsection{OGLE LMC-ECL-15260}
This is the lowest mass system in our sample and was  analyzed by P13. The light curves show a circular orbit with wide and deep eclipses and also with relatively prominent proximity effects. Our updated masses and radii of components are in perfect agreement with those published by P13. The final values for the  effective temperatures and metallicities are higher by about $1\sigma$. Both components rotate slightly super-synchronously, with the primary actually shrinking on the second red giant branch and the secondary shrinking on the first red giant branch -- see Fig.~\ref{fig13}. 

\begin{figure}
\hspace*{-0.5cm}
\includegraphics[angle=0,scale=0.42]{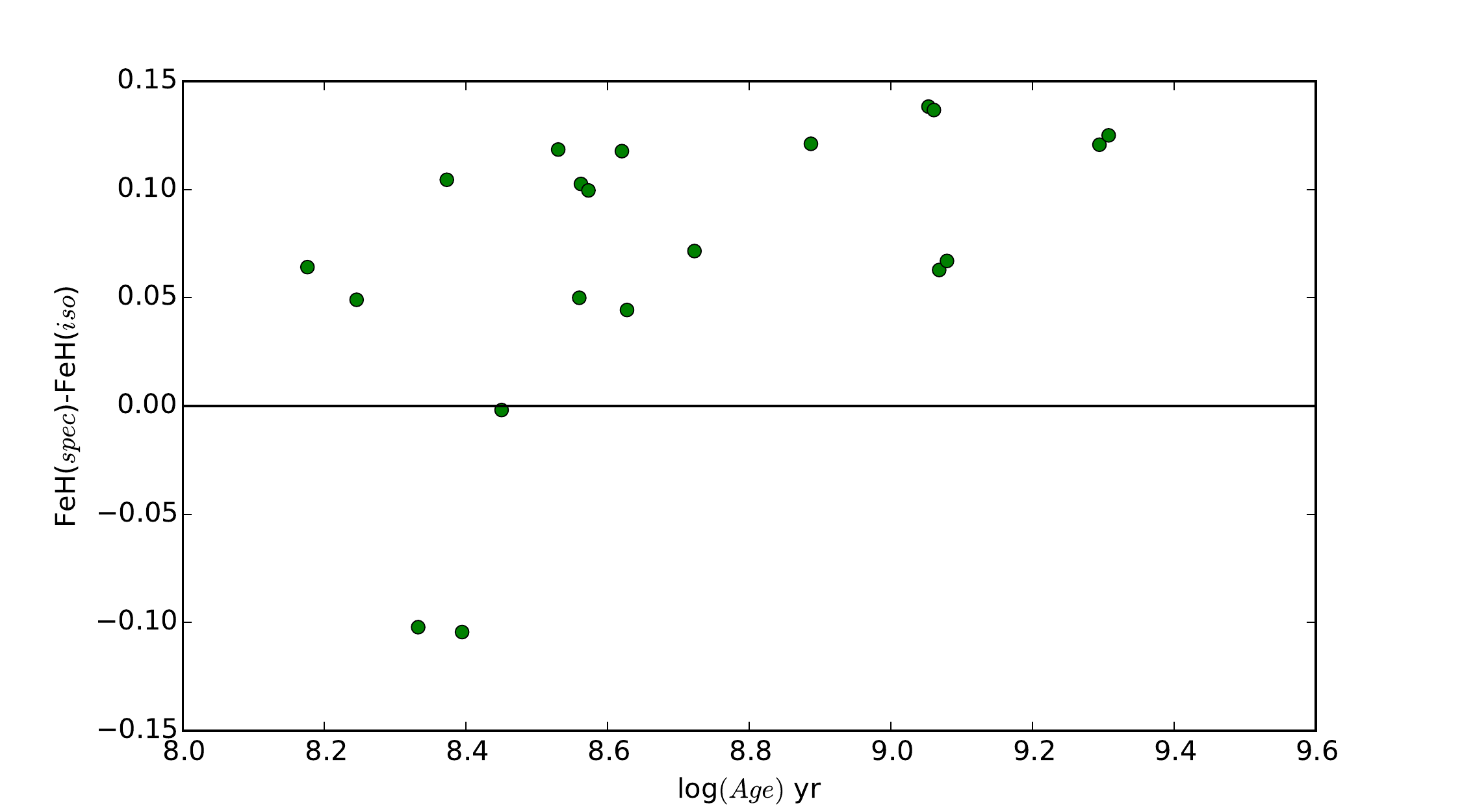}
\caption{The difference betwen the corrected spectroscopic metallicity [Fe/H](spec) and the isochronal metallicity [Fe/H](iso) as a function of systems' age. \label{figFeHm}}
\end{figure}

\begin{figure*}
\begin{minipage}[th]{0.5\linewidth}
\includegraphics[angle=0,scale=.70]{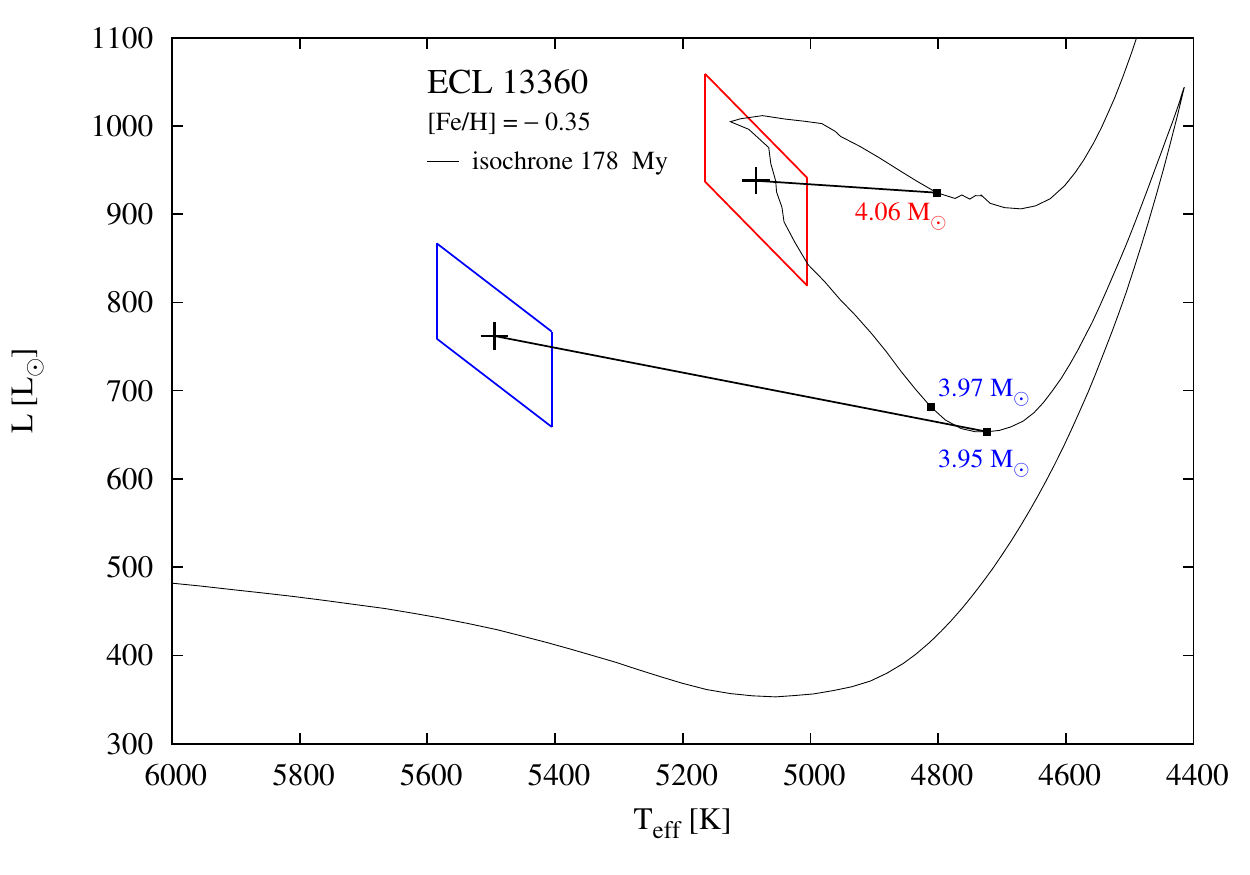} 
\end{minipage}\hfill 
\begin{minipage}[th]{0.5\linewidth}
\includegraphics[angle=0,scale=.70]{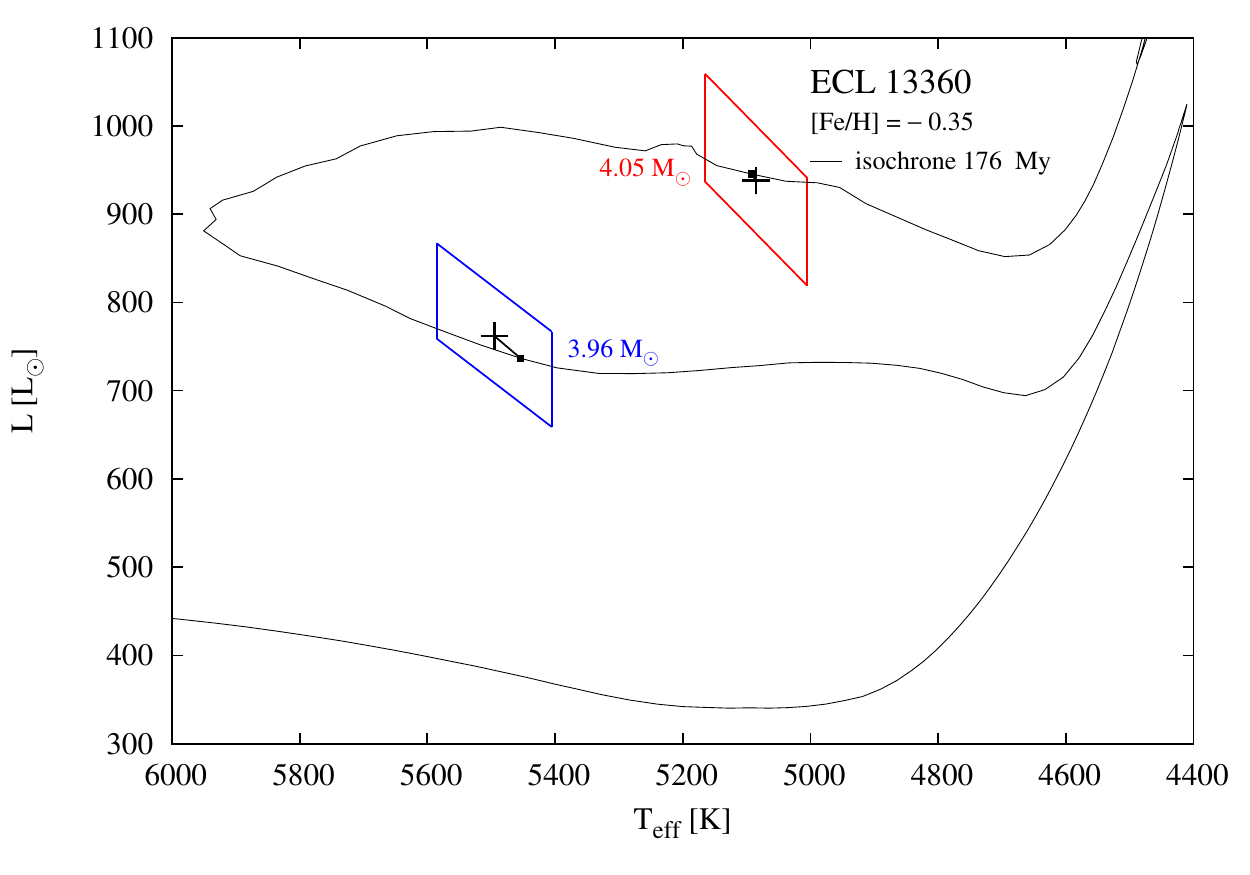}
\end{minipage}\hfill 
\caption{Comparison of the effects of rotating (left panel) and non-rotating MIST models (right panel) in case of the components of OGLE LMC-ECL-13360 . \label{fig13360}}
\end{figure*}

\subsection{OGLE LMC-ECL-18365}
The system shows shallow and total eclipses, with the components having similar surface temperatures but having very different masses and radii. The components have very different evolutionary ages, by a factor of almost 20 -- see Fig.~\ref{fig13}. A discussion of the evolutionary stage of the system is given in Section~\ref{ev18365}. 

\subsection{OGLE LMC-ECL-18836}
The system shows partial eclipses with a mildly eccentric orbit. The secondary component lies very close to the tip of the red giant branch, just before the helium core ignition. Its rotation is tidally locked at periastron -- pseudo-synchronism. The primary rotates much faster, most probably because of rapid and substantial shrinking during its evolution on the horizontal loop. The system has a possible very red third light visible in the OGLE light curves with $(V-I)=1.9$ mag. The detection is of low significance and we  adopted a final solution with the third light $l3=0$. 

\subsection{OGLE LMC-ECL-21873}
The system has a very low eccentricity and has the shortest orbital period from all eccentric systems in our sample. The system shows partial eclipses and almost no proximity effects. Both components rotate slightly super-synchronously. The circularization time of the system is significantly shorter than the life of their components as red giant stars. This may signify that there is an external third body perturbing the inner binary system. However we could not identify any third light in our data. 

\subsection{OGLE LMC-ECL-24887}
The light curves have deep and partial eclipses of almost the same depth. It is the faintest system optically in our sample, and this system has the highest rms residuals of the light curve solution. The system is circular and both components rotate super-synchronously. 

\subsection{OGLE LMC-ECL-25658}
The system was analyzed  by \cite{elg16} where its fundamental physical parameters were derived with  high precision. The system has components of similar temperatures with deep and total eclipses while moving on an eccentric orbit. Our revised physical parameters  are only very slightly different from the previous values and all differences, with the exception of the metallicity, are much smaller than the uncertainties of the parameters. Both components seem to be very evolved and in the phase of shell helium burning after evolution on the horizontal loop -- see Fig.~\ref{fig13}. There is out-of-eclipse semi-periodic variability visible in the OGLE $I_{\mathrm C}$-band light curve with an amplitude of about 0.01 mag and with a period of about 75 days. This period is very close to the rotation period of the secondary star.   

\section{Discussion}
 \subsection{Evolutionary status}
 \label{evolut}
 
\begin{figure}
\hspace{-0.2cm}
\includegraphics[angle=0,scale=0.47]{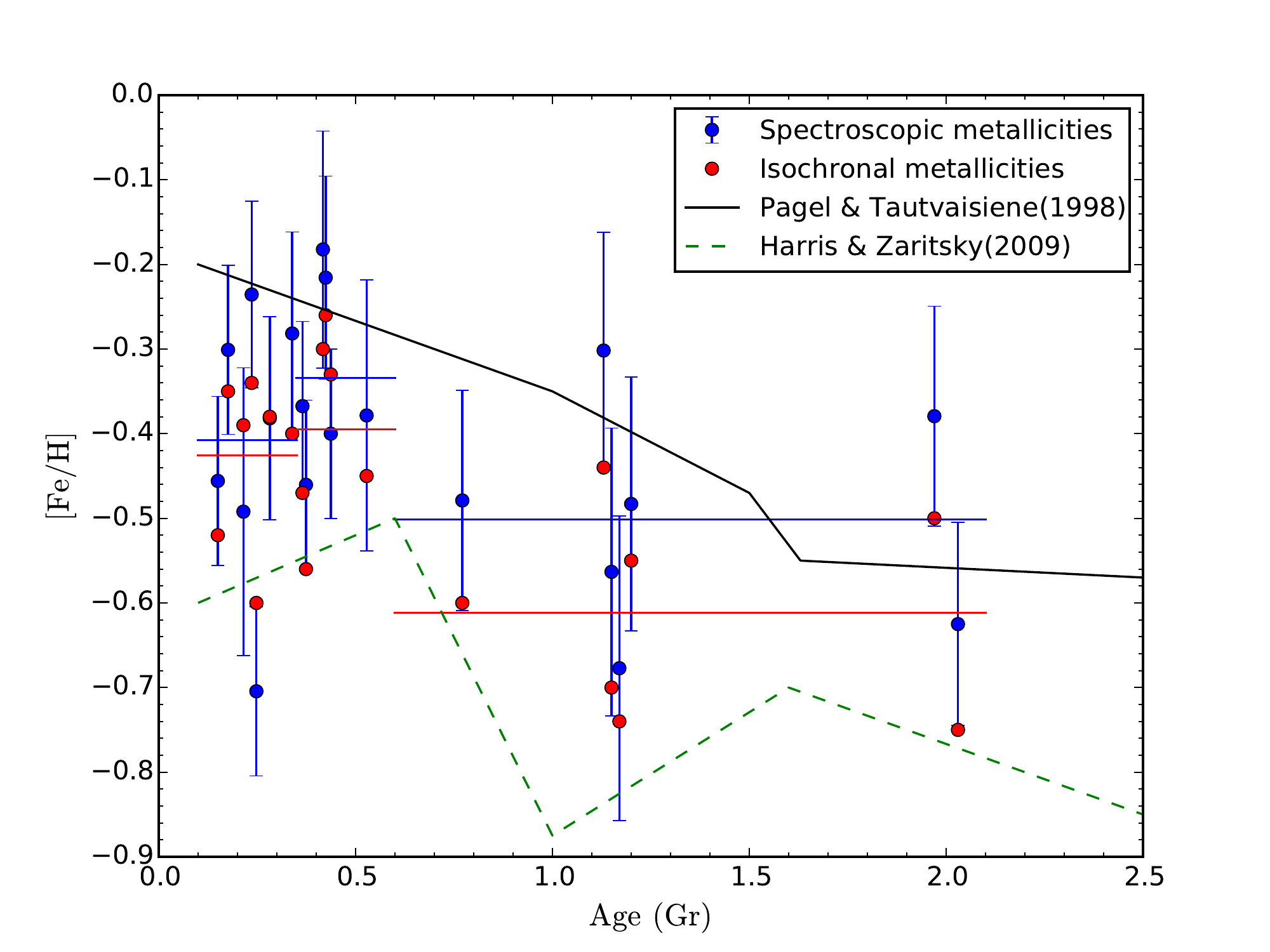}
\caption{Age-metallicity relation for the LMC giant stars. Horizontal lines are mean metallicity values for age ranges of 0.1-0.35 Gyr, 0.35-0.6 Gyr and 0.6-2.1 Gyr. \label{AgeMet}}
\end{figure}

In order to investigate the evolutionary status of the systems we made a comparison of the observed properties of the components with model predictions from the MESA Isochronal and Stellar Tracks code \citep[MIST]{cho16}\footnote{http://waps.cfa.harvard.edu/MIST/} based on Modules for Experiments in Stellar Astrophysics \citep[MESA]{pax11}, an open-source 1D stellar evolution package. The MIST tracks are calculated with a homogenous set of model parameters reported in Table 1 of \cite{cho16}. We fitted isochrones to the positions of the components of a particular system on the Hertzsprung-Russell diagram. The fits were made manually by searching for isochrones which  simultaneously fit the positions of both stars to within $1\sigma$  for the luminosities, the surface temperatures, and the masses. We looked for isochrones within a $1\sigma$ range of the spectroscopic metallicities. In most cases, to obtain the best fits we had to adjust the metallicities, usually by making them smaller  by about $0.6\sigma$ on average. We call such metallicities the isochronal metallicities and they are always within $1\sigma$ of the spectroscopic values. We note that for some  less massive and older systems, to obtain the best fit a decrease in the metallicity of $2\sigma$ or even larger was required. However we refrained from fitting such models because other secondary evolutionary parameters (like core and atmosphere overshooting, mixing length scale, etc.) are expected to play a role. A comparison of the spectroscopic and  isochronal metallicities is shown in Fig.~\ref{figFeHm}.
      
In the case of more massive systems, non-rotating models give significantly better fits than models with some amount of initial rotation ($\nu_{\rm ZAMS}/\nu_{\rm crit} = 0.4$). This is illustrated for the system LMC-ECL-13360 in Fig.~\ref{fig13360}. For non-rotating models the horizontal loop extends sufficiently to the blue to allow a fit of  both stars with a single metallicity. For rotating models the horizontal loop shrinks considerably and  it is impossible to find a good fit, even with a lower metallicity. In the case of less massive systems the difference between non-rotating and rotating models is much smaller and is not significant, but for consistency we also fit non-rotating models  for lower mass binaries. 

In most of the systems we found an isochrone which fits the position of both components - see Fig.~\ref{fig11}, \ref{fig12} and \ref{fig13} in the Appendix. A few systems (LMC-ECL-01866,  LMC-ECL-10567, LMC-ECL-15260 and LMC-ECL-18836) show reversed luminosity ratios i.e. the less massive component is the more luminous one. Investigation of the position and isochrones on the HR diagrams for these systems shows that these are systems where the more massive component shows significant contraction while on the blue loop after ignition of helium in the core while the less massive component is still on the descending or ascending red giant branch. However, the two systems LMC-ECL-05430 and LMC-ECL-18365 apparently have components of discrepant evolutionary age, as there is no single isochrone, even with allowing for $3\sigma$ errors, which could fit the positions of their components. In the system LMC-ECL-01866 a single isochrone is marginally consistent with the position of the components at a $2.5\sigma$ level. 

The evolutionary status of seven systems from our sample was analysed more in more detail by \cite{ost18} but they used the physical parameters from P13. They conclude that in many cases a single isochrone (with internal model parameters having the same values for both components) gives a poor fit to position of stars and the fitting of individual evolutionary tracks was necessary. The evolutionary status and the ages of the systems derived by \cite{ost18} are, however, in perfect agreement with our results for six systems. For the seventh system LMC-ECL-03160 they found that both components are on the RGB before core helium ignition (with an age of 0.97 Gyr) while we found that both components are already after the helium flash (with an age of 1.17 Gyr,  see Fig.~\ref{fig11}). We attribute this difference in the evolutionary stages mostly to different metallicites used for the computations. We add that the metallicities obtained by \cite{ost18} from fitting evolutionary tracks  are also on average larger than the spectroscopically measured values, in  agreement with our findings.  

\subsubsection{OGLE LMC-ECL-05430}
\label{ev05430}
The system is well detached and the components have quite different masses. However, on the HR diagram the components seem to be at a similar evolutionary stage (both being on the asymptotic red giant branch or just after ignition of helium in the core) but with significantly discrepant ages - see Fig.~\ref{fig11}. The eccentricity of the system and large separation between the components makes mass transfer an unlikely explanation for the discrepancy. A stellar capture or a merger of an inner-system close binary star are possible scenarios. The stellar capture scenario is supported by the fact that the system is an apparent member of the cluster SL181 and thus resides in a dense stellar environment where the probability of stellar encounters is enhanced.  

\subsubsection{OGLE LMC-ECL-18365}
\label{ev18365}
This system contains the most massive star (the primary) in our sample and its apparent age seems much less than its companion. The orbit is circular and tight ($r_1+r_2=0.38$) suggesting that past mass transfer occurred from the then more massive secondary star to the primary while ascending the red giant branch. The size of the secondary would reach about 70 $R_\odot$ just before  helium core ignition,  i.e. the star would fill its Roche lobe and thus start mass transfer to the donor star. The primary was rejuvenated this way and it now has the properties of a star just at the beginning of red giant branch ascension.  

\subsection{Age-metallicity relation}
The ages of the systems derived in Section~\ref{evolut} were used to construct an age - metallicity relation for the LMC during the last 2 Gyr. Ages and metallicities and their uncertainties were adopted from Table~\ref{tbl-6}. We used both sets of metallicities: the spectroscopic and the isochronal values. The results are shown in Fig.~\ref{AgeMet}. The metallicities, especially the spectroscopic values, show a large  scatter. However, it is possible to draw some general conclusions.  On average the metallicity of intermediate age stars (older than about 0.6 Gyr) is significantly smaller than the metallicity of the younger population. From 2 Gyr to about 0.6 Gyr the metallicity shows a very large scatter and it is consistent with a flat relation (no evolution) at about $\sim -0.55$ dex. We can notice some metallicity increase at around 0.5 Gyr to [Fe/H] $\sim -0.35$ dex followed by a  possible decrease to a value of about $-0.4$ dex at 0.1 Gyr. This general picture is similar to results obtained from the analysis of field red giant stars in the LMC \citep{har09} where the maximum  metallicity coincides with the peak of star formation in the LMC at about 0.5 Gyr, with a slow evolution of metallicity before that time. The absolute value of the metallicity is however  significantly lower, which can be attributed  in  part to the different way of deriving spectroscopic metallicities by \cite{har09}. The chemical enrichment history we derive is  also marginally consistent with the analytic model of chemical evolution in the LMC by \cite{pag98}  which predicts a smooth metallicity evolution, but larger metallicities. Clearly a model between those of \cite{har09} and \cite{pag98} would probably be the most appropriate, which is in  agreement with a recent study based on LMC stellar clusters by \cite{pal15}. 

\subsection{Circularization of orbits} 
\label{circula}
\begin{figure}
\hspace*{-0.5cm}
\includegraphics[angle=0,scale=0.36]{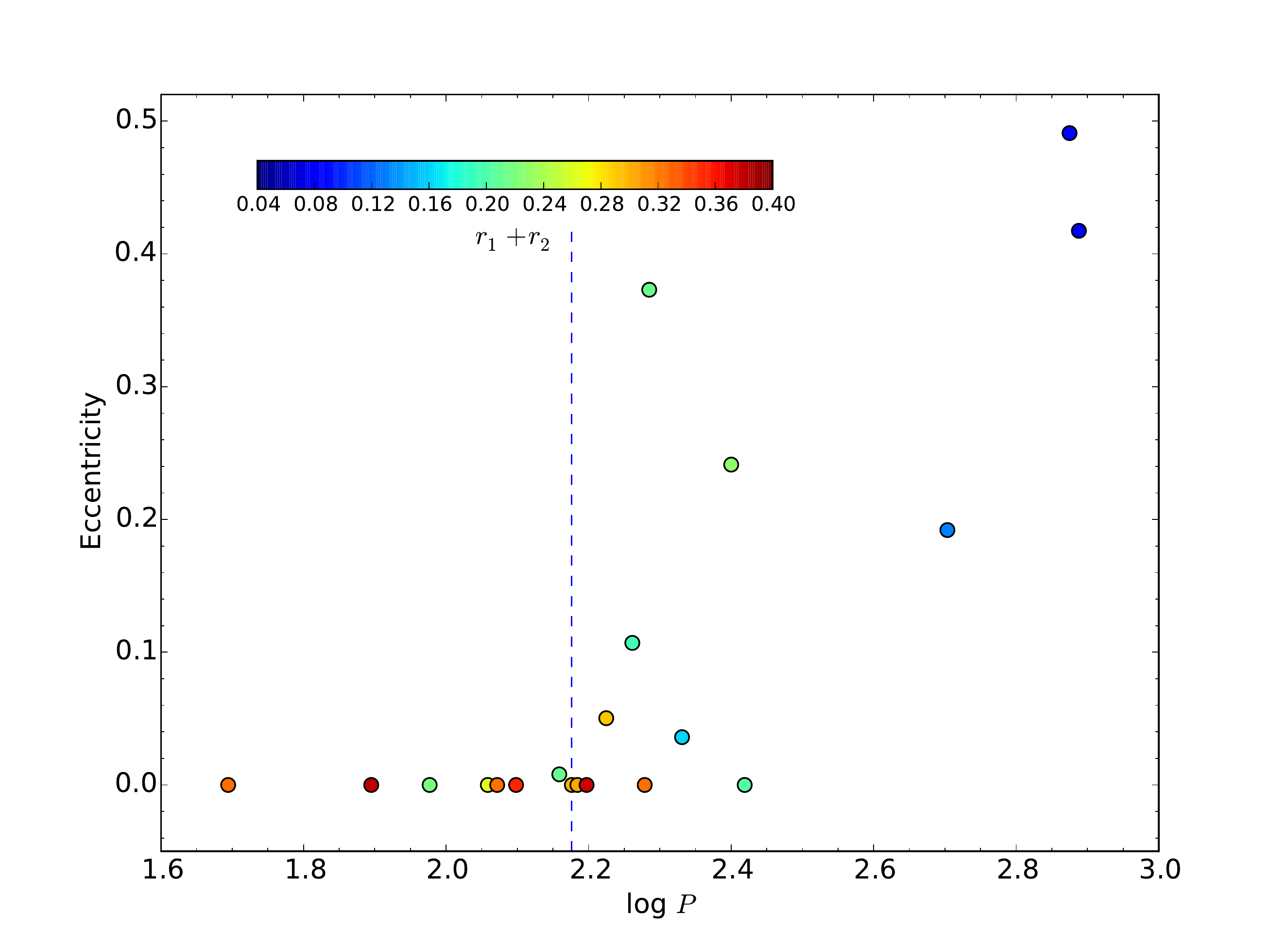}
\caption{The orbital period - eccentricity diagram for 20 LMC systems. The sum of the relative radii is color-coded. The vertical dashed line corresponds to an orbital period of 150 days.\label{figPecc}}
\end{figure}

The circularization of the orbits of eclipsing binaries in the LMC was studied by \cite{fac08}. Their study was confined mostly to shorter period systems ($P_{orb}<20$ d). Our sample contains significantly evolved giant stars on long period orbits. For some systems the radii of the components  are large enough that tidal forces relatively quickly circularize the orbit and synchronize  rotation during the first or second red giant branch ascension. The systems we investigated follow the empirical rule \citep{may84} that all binary stars containing red giant components with orbital periods below 150 days have circular orbits -- see Fig.~\ref{figPecc}. Only one eccentric system LMC-ECL-21873 has a slightly shorter orbital period than 150 days, but its orbit is very close to circular. 

\begin{figure}
\hspace*{0cm}
\includegraphics[angle=0,scale=0.55]{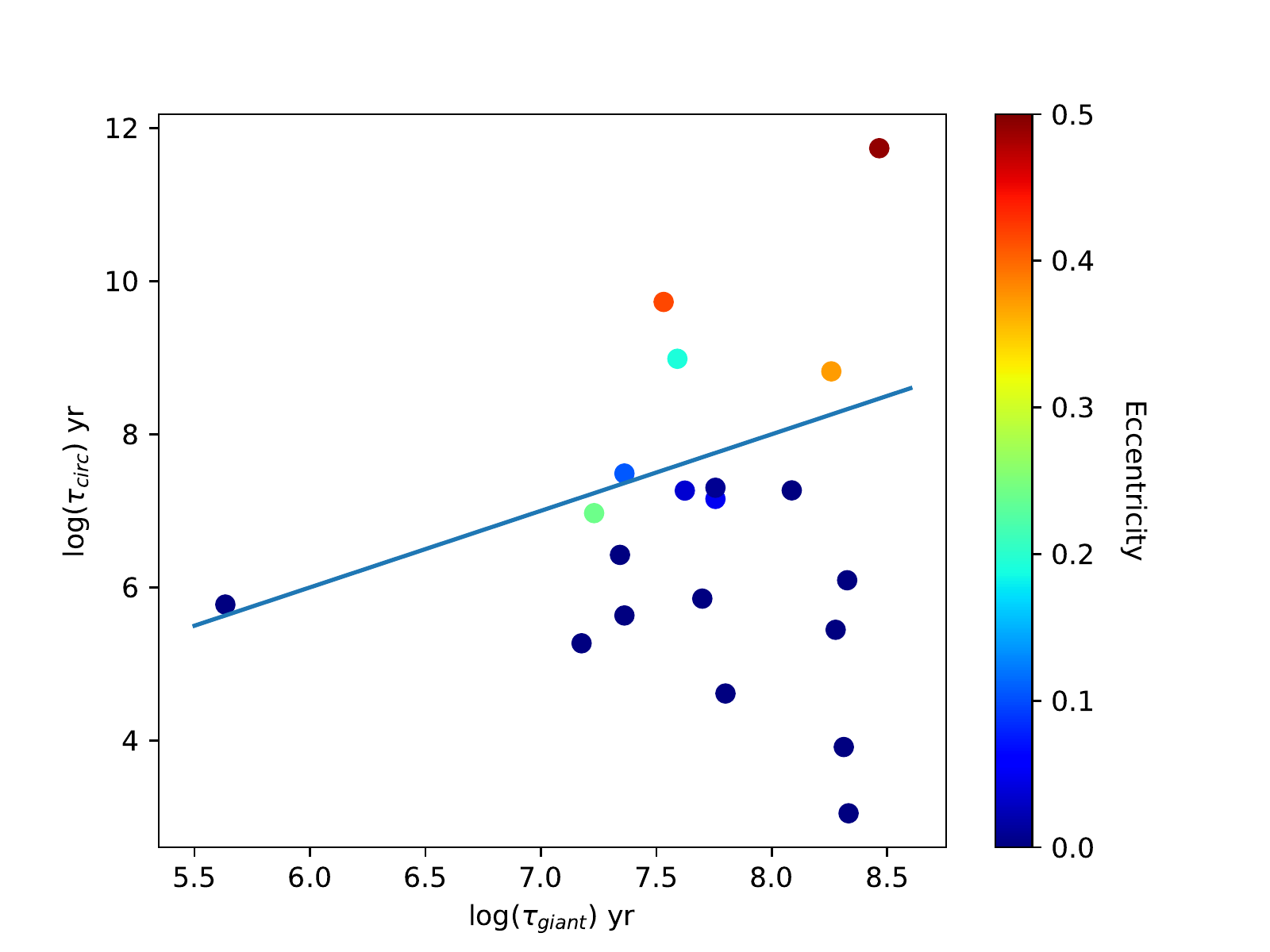}
\caption{ The red giant phase duration -  circularization time diagram for 20 LMC systems. The straight line corresponds to a 1:1 relation. \label{figCirc}}
\end{figure}

For a more quanitative analysis we estimated the times of circularization following the prescription given in \cite{such15}, who used Zahn's formalism \citep{zah89}. However, instead of using the actual radii of the components we used weighted effective radii of the components during their lifetime  as red giants. The effective radii were calculated by utilizing the MIST evolutionary tracks corresponding to masses and chemical compositions of the components (Section~\ref{evolut}). We arbitrarily set a lower limit for a red giant radius at 6.5 $R_\odot$. The times of circularization $\tau_{circ}$ are compared with an actual  red giant phase duration $\tau_{giant}$ of the more massive component in Fig.~\ref{figCirc}. As one would expect almost all circular systems are well below the 1:1 relation i.e. their more massive components spend more time as red giants than the circularization time of a  given system. An exception is  LMC-ECL-18365 in which the more massive component seems to be significantly younger than the mean age of the system - an affect of the probable past mass transfer in the system. Few low eccentricity systems are below the 1:1 relation, however, they are probably in a stage of rapid orbit circularization. One higher eccentricity system (LMC-ECL-05430) is also below the line. This may suggest that this is a triple system with an undetected companion which prevents the circularization of the orbit or  reflects the systems' past complex evolution.

\acknowledgments
The research leading to these results  has received
funding from the European Research Council (ERC) under the European
Union's Horizon 2020 research and innovation program (grant agreement
No 695099).

We are grateful for financial support from Polish National Science
Center grant OPUS 2013/09/B/ST9/01551. 

Support from the BASAL Centro
de Astrof{\'i}sica y Tecnolog{\'i}as Afines (CATA) PFB-06/2007, the
Millenium Institute of Astrophysics (MAS) of the Iniciativa Cientifica
Milenio del Ministerio de Economia, Fomento y Turismo de Chile, project
IC120009.

The OGLE project has received funding from the National Science Centre,
Poland, grant MAESTRO 2014/14/A/ST9/00121 to AU.
S.V. gratefully acknowledges the support provided by Fondecyt reg.~n.~1170518.
The authors acknowledge the support of the French Agence Nationale de la Recherche (ANR), under grant ANR-15-CE31-0012-01 (project UnlockCepheids).

Based on observations made with ESO 3.6m and NTT telescopes in La Silla under programme 074.D-0318, 074.D-0505, 082.D- 0499, 083.D-0549, 084.D-0591, 086.D-0078, 091,D-0469(A), 0100.D-0339(A), 098.D-0263(A,B), 097.D-0400(A). We also acknowledge generous allocation of CNTAC time and time allocated by the Carnegie Observatories TAC.
 

{}    

\begin{appendix}

\section{Light and radial velocity curves solutions for individual systems.}
\begin{figure*}
\begin{minipage}[th]{\linewidth}
\includegraphics[angle=0,scale=0.43]{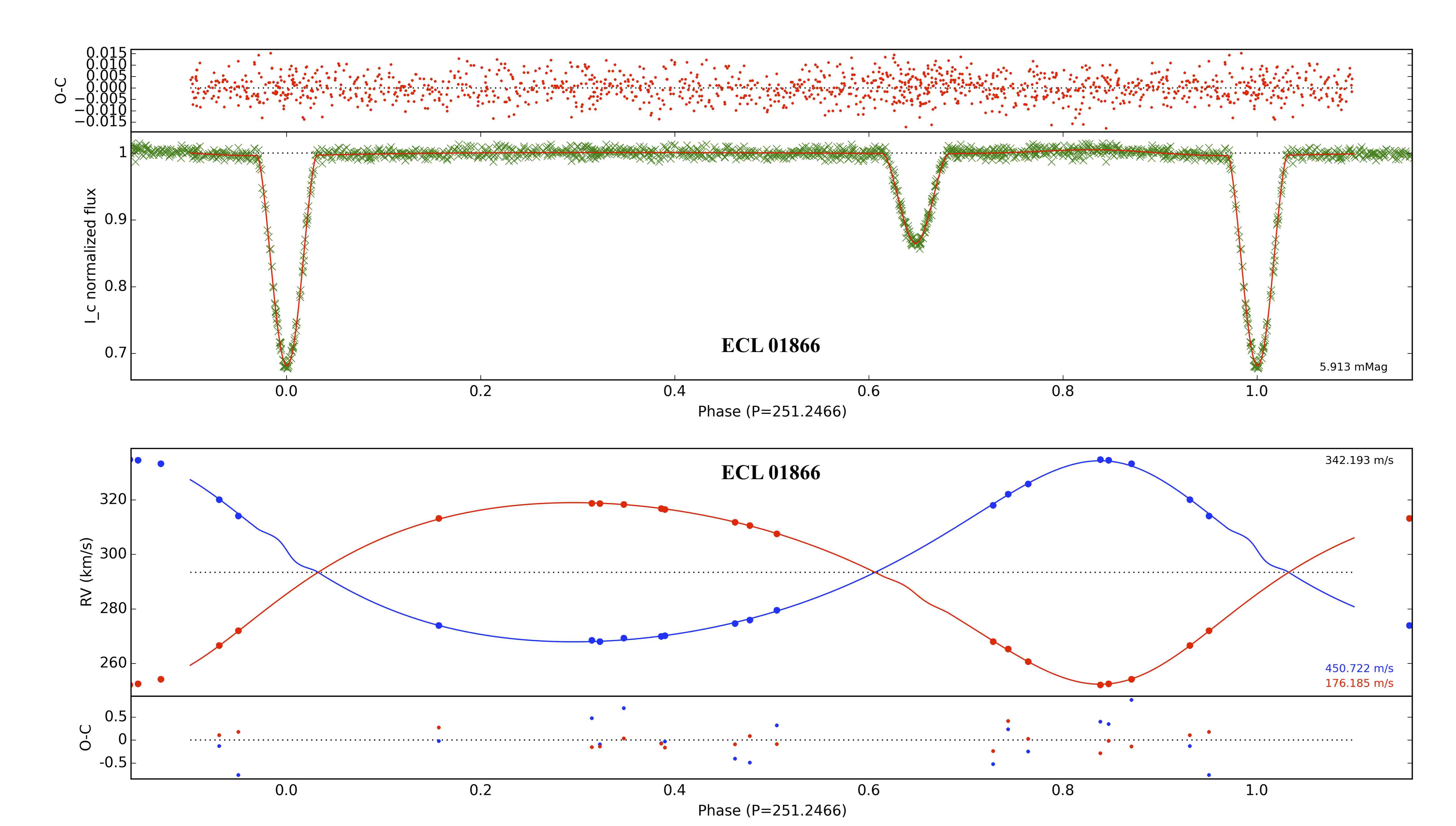}
\mbox{}\\
\includegraphics[angle=0,scale=0.43]{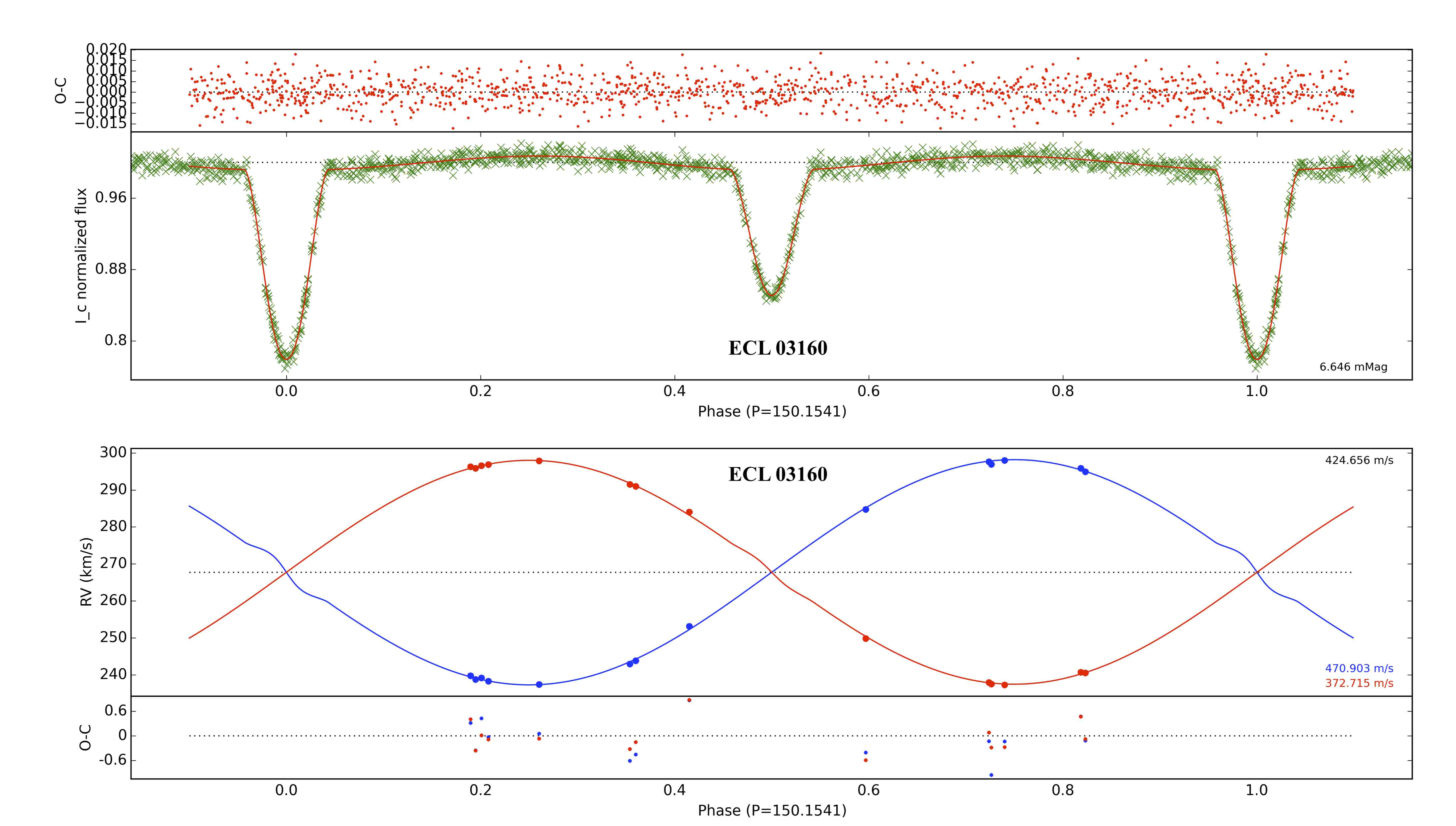} 
\end{minipage}\hfill 
\caption{ The OGLE $I_{\mathrm C}$-band light curve and the radial velocity solutions for  OGLE-LMC-ECL-01866 (two upper panels) and OGLE-LMC-ECL-03160 (lower two panels). \label{fig1}}
\end{figure*}

\begin{figure*}
\begin{minipage}[th]{\linewidth}
\includegraphics[angle=0,scale=0.43]{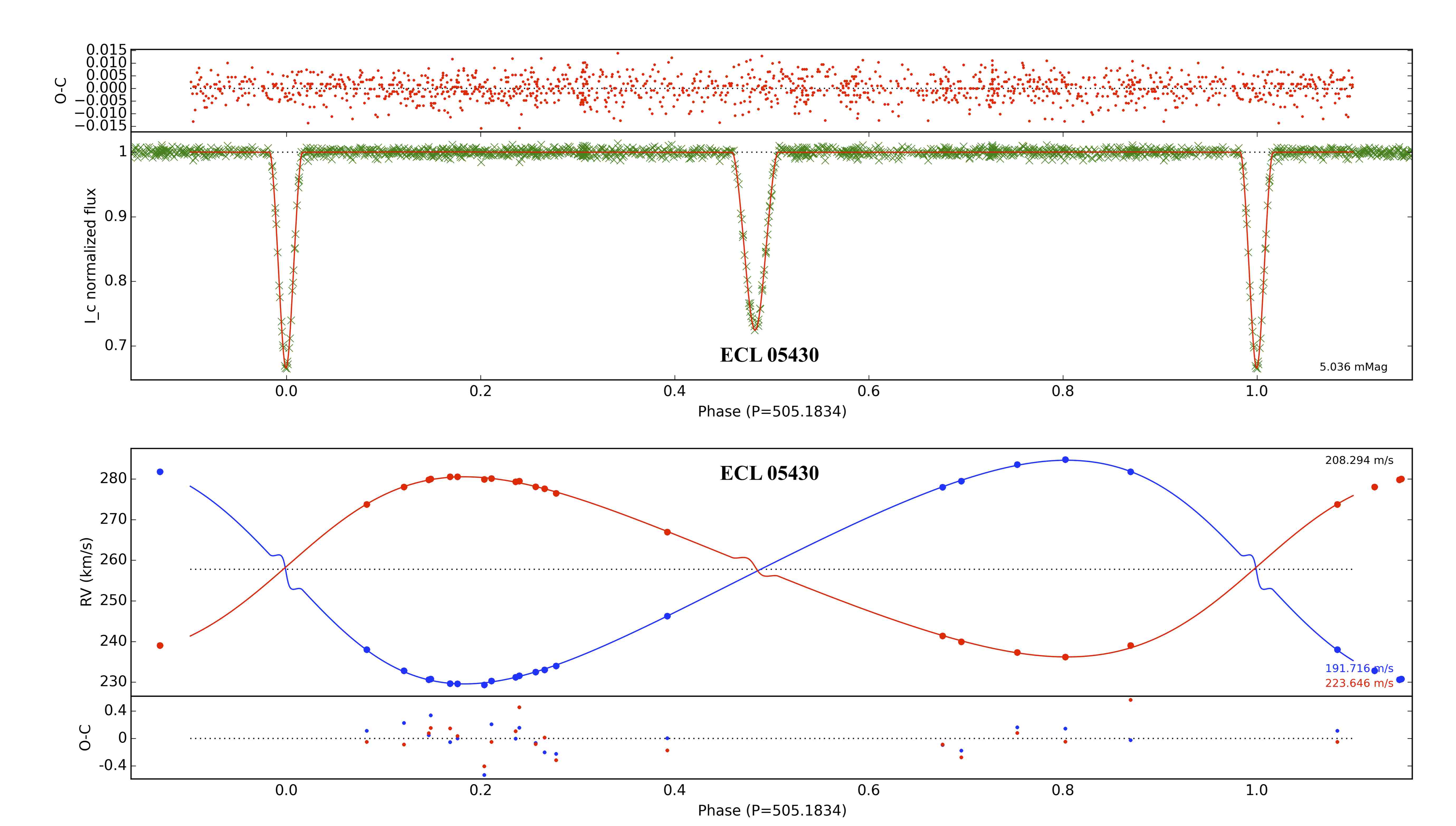}
\\ 
\includegraphics[angle=0,scale=0.43]{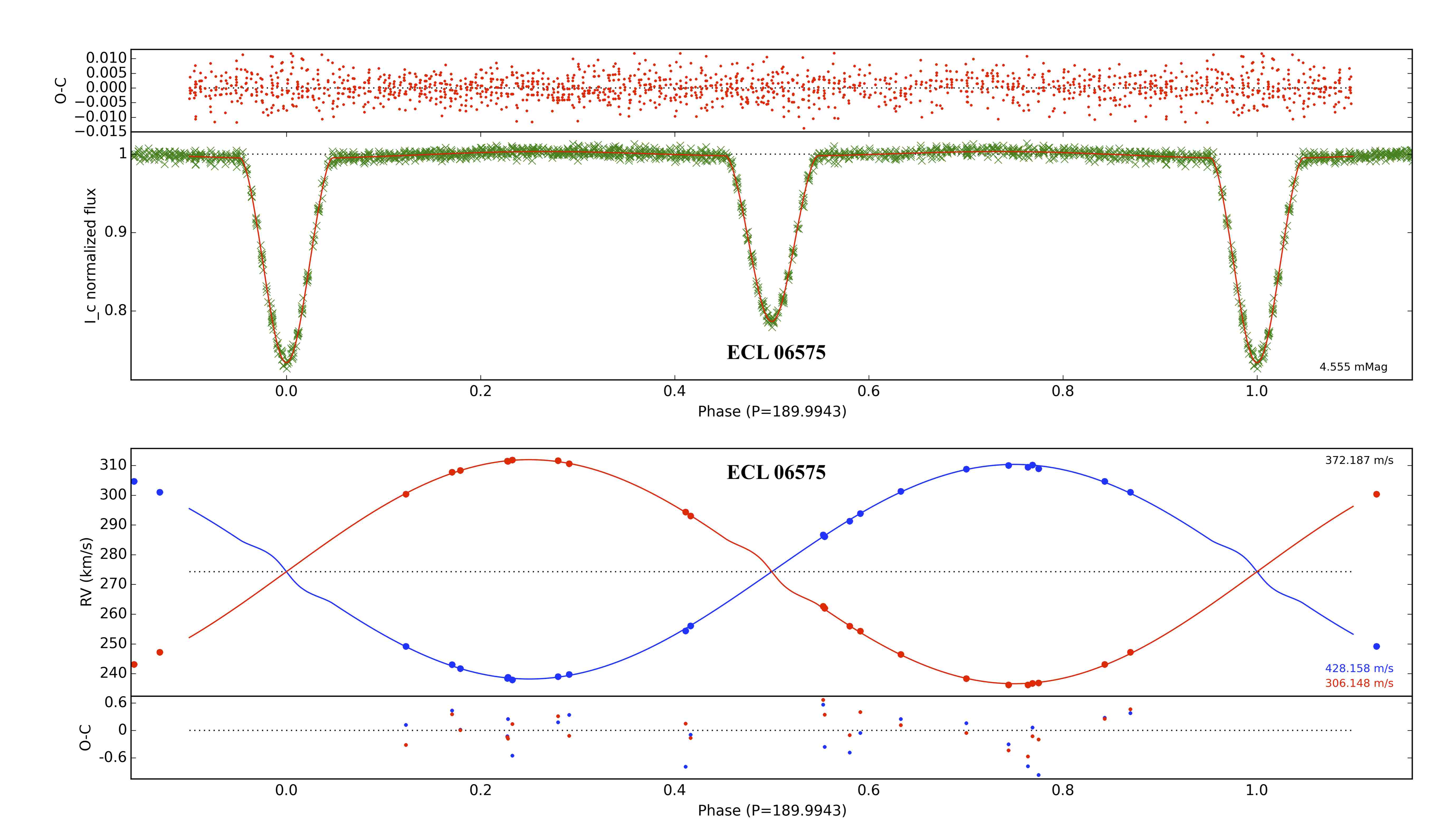} 
\end{minipage}\hfill 
\caption{ The OGLE $I_{\mathrm C}$-band light curve and the radial velocity solutions for OGLE-LMC-ECL-05430 (two upper panels) and OGLE-LMC-ECL-06575 (lower two panels).  \label{fig2}}
\end{figure*}

\begin{figure*}
\begin{minipage}[th]{\linewidth}
\includegraphics[angle=0,scale=0.43]{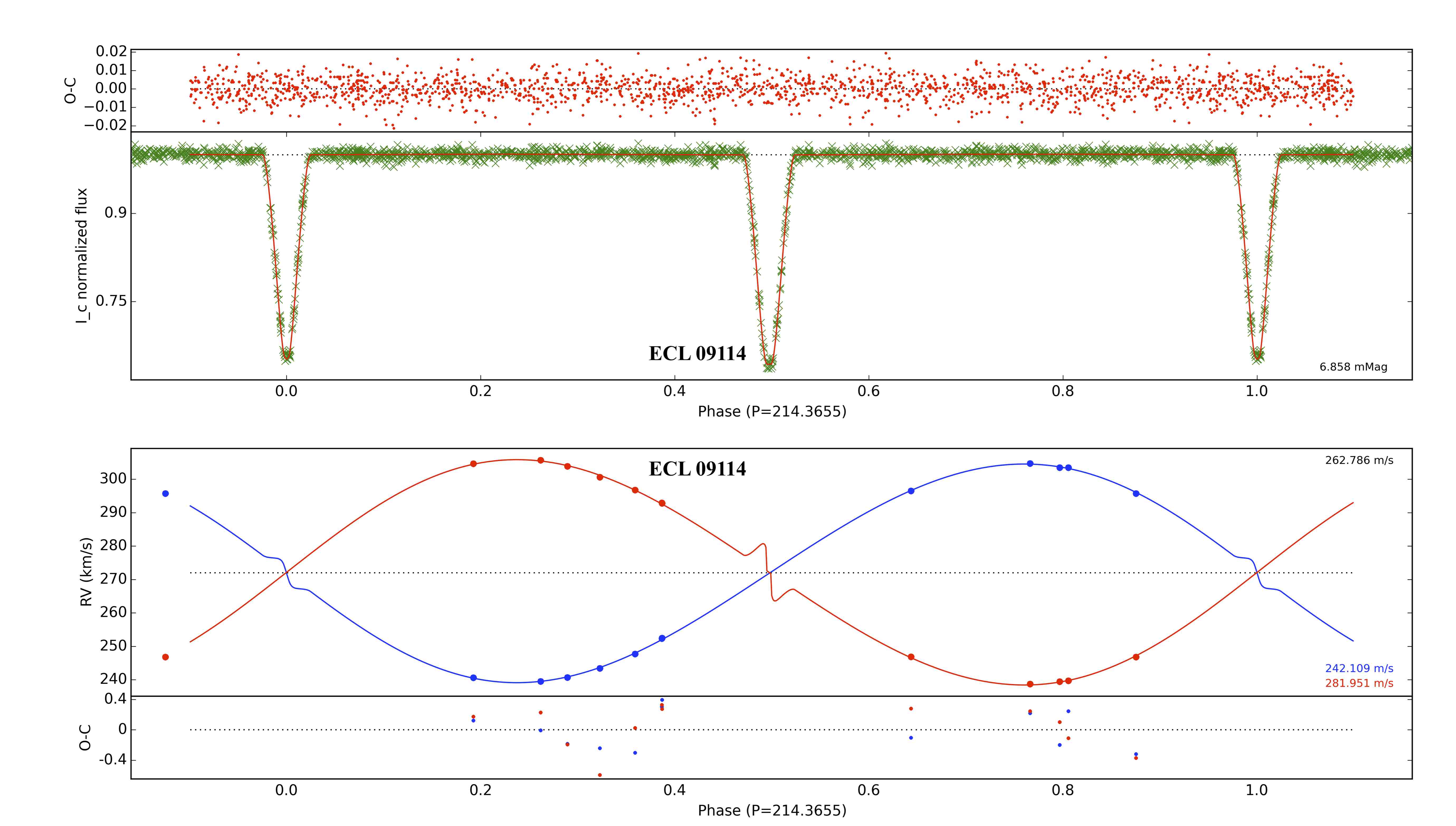}
\\ 
\includegraphics[angle=0,scale=0.43]{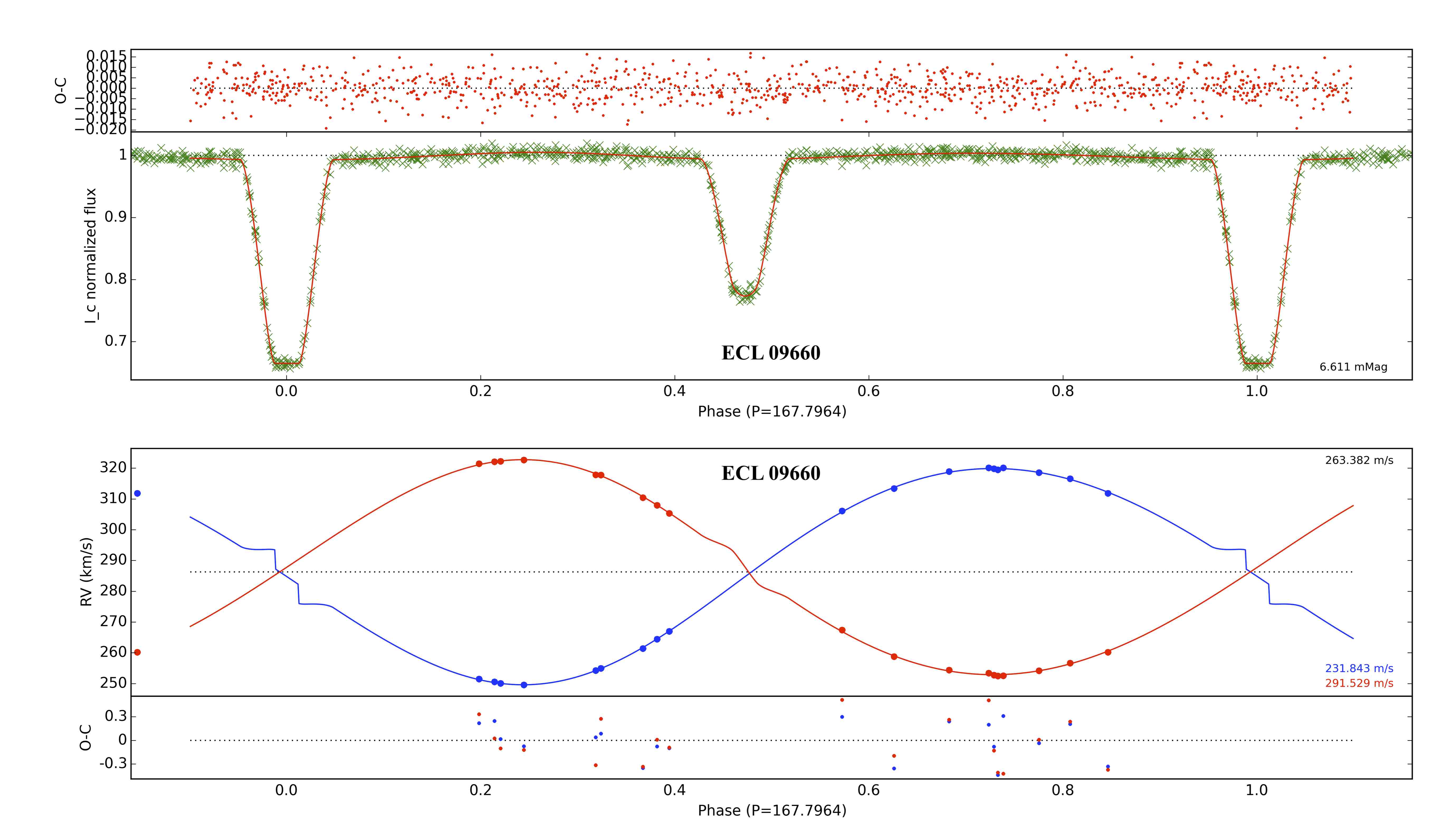} 
\end{minipage}\hfill 
\caption{ The OGLE $I_{\mathrm C}$-band light curve and the radial velocity solutions for OGLE-LMC-ECL-09114 (two upper panels) and OGLE-LMC-ECL-09660  (lower two panels). \label{fig3}}
\end{figure*}

\begin{figure*}
\begin{minipage}[th]{\linewidth}
\includegraphics[angle=0,scale=0.43]{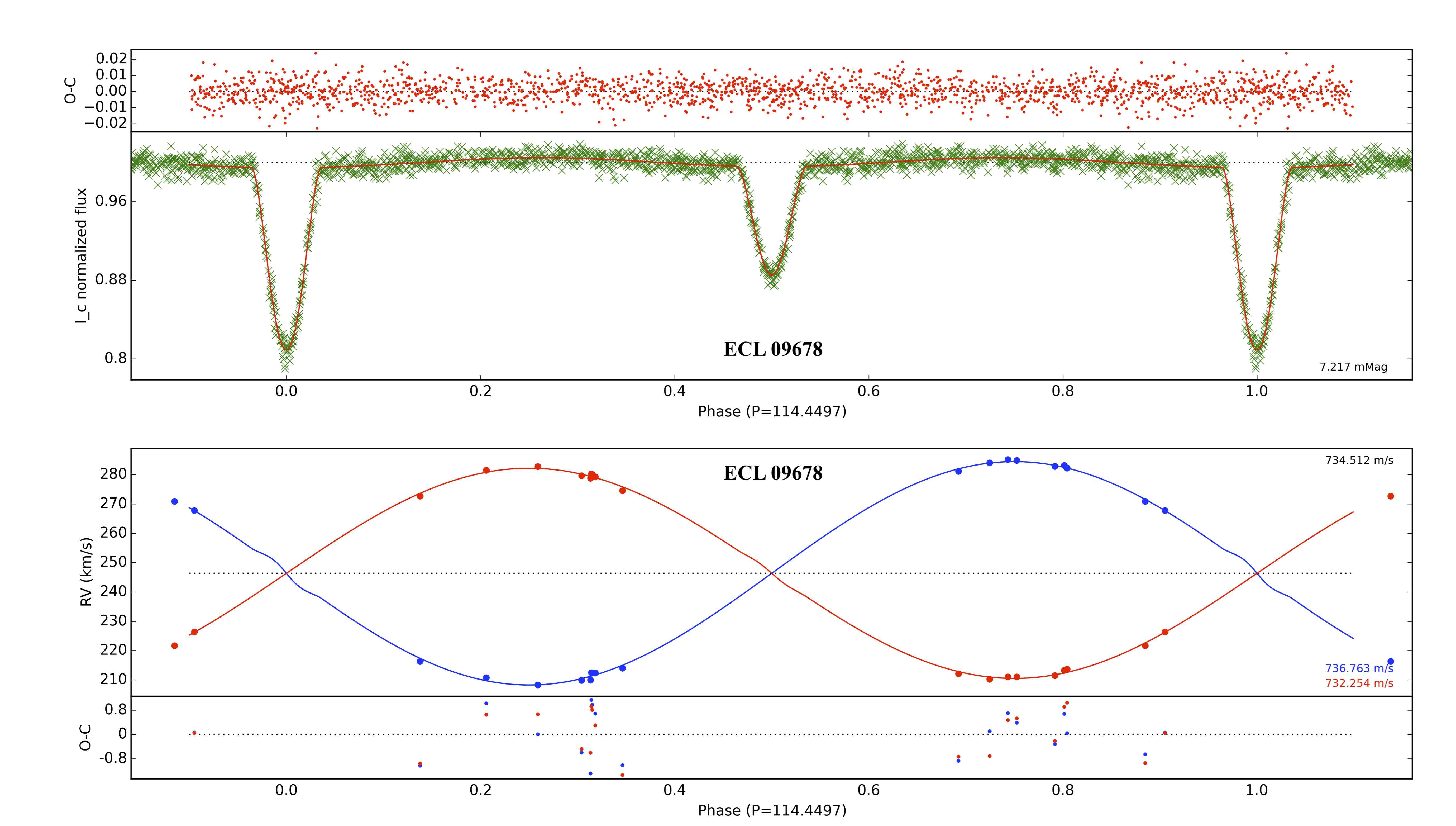}
\\ 
\includegraphics[angle=0,scale=0.43]{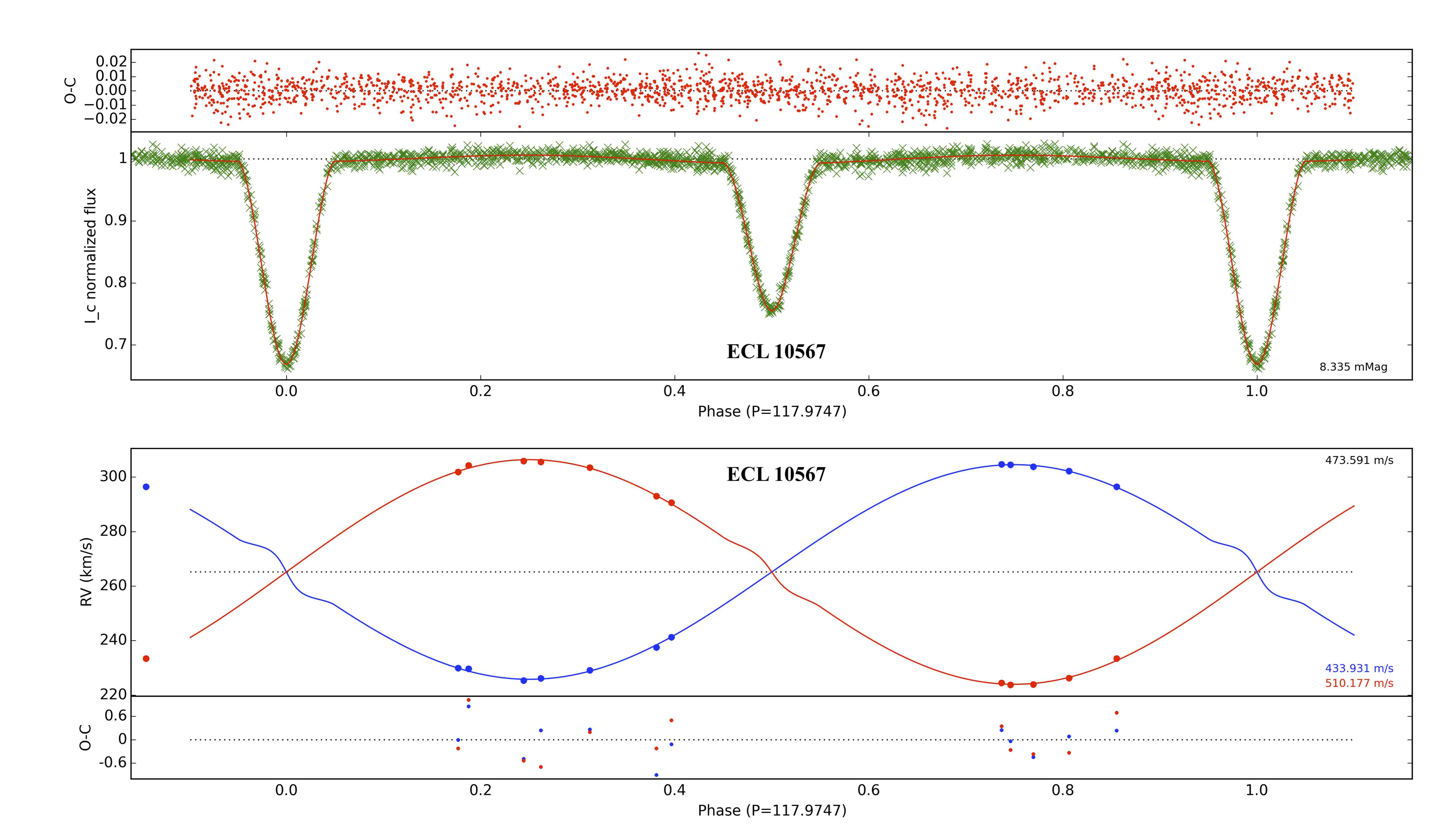} 
\end{minipage}\hfill 
\caption{ The OGLE $I_{\mathrm C}$-band light curve and the radial velocity solutions for OGLE-LMC-ECL-09678 (two upper panels) and OGLE-LMC-ECL-10567  (lower two panels).  \label{fig4}}
\end{figure*}

\begin{figure*}
\begin{minipage}[th]{\linewidth}
\includegraphics[angle=0,scale=0.44]{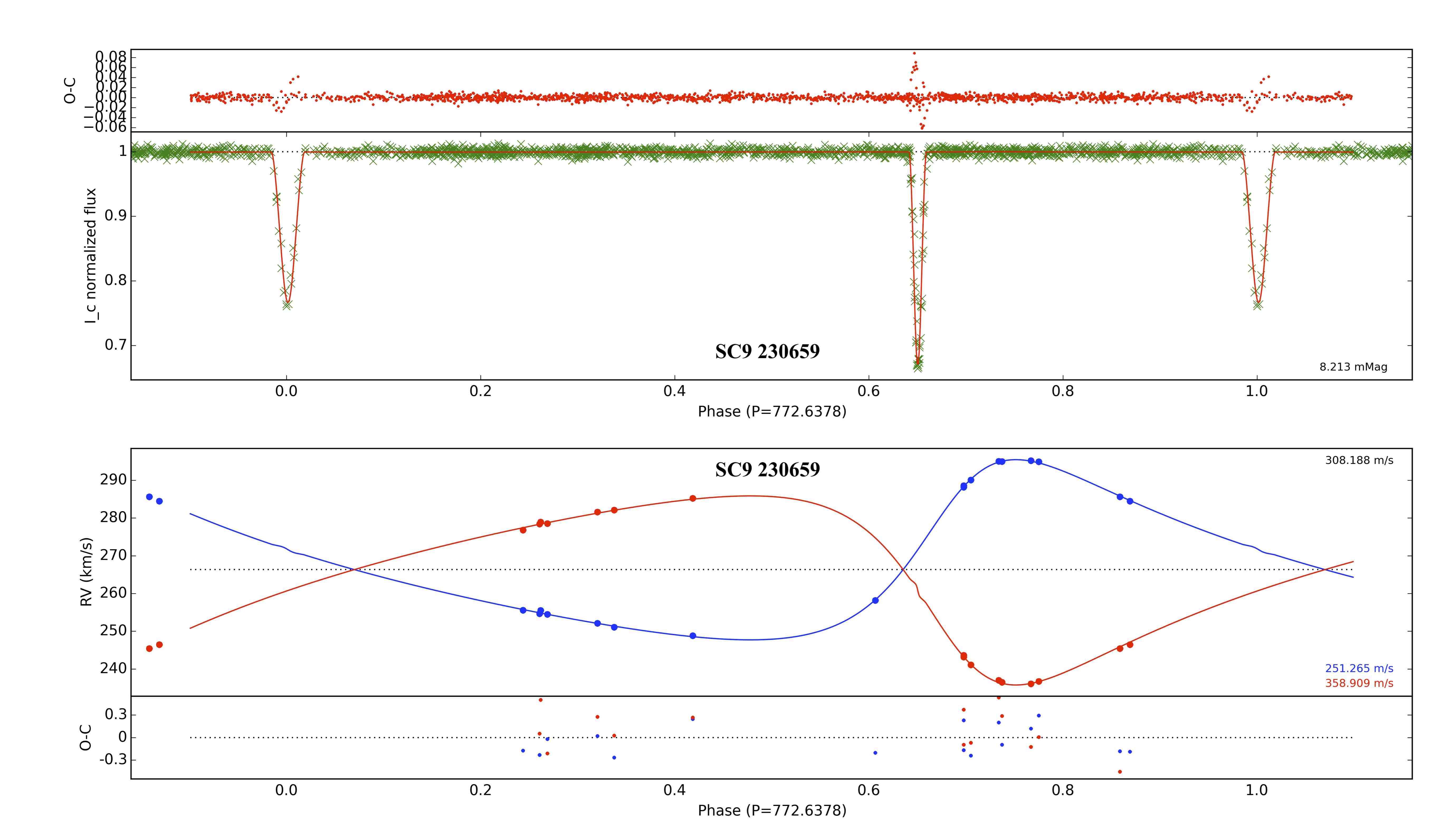}
\\ 
\includegraphics[angle=0,scale=0.44]{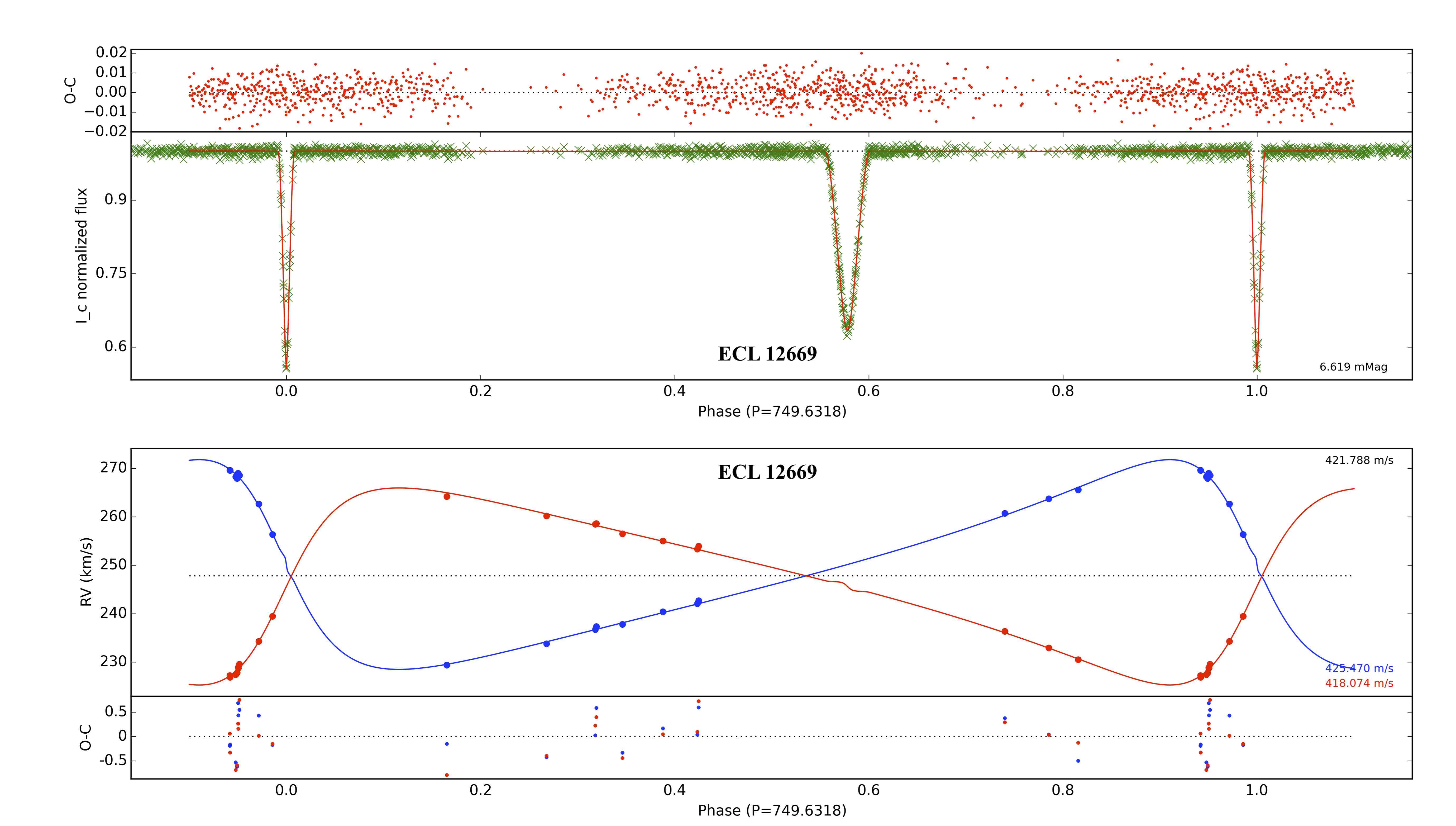} 
\end{minipage}\hfill 
\caption{ The OGLE $I_{\mathrm C}$-band light curve and the radial velocity solutions for LMC SC9\_230659 (two upper panels) and OGLE-LMC-ECL-12669  (lower two panels). Large residua during eclipses of LMC SC9\_230659 are caused by
its periastron movement. \label{fig5}}
\end{figure*}

\begin{figure*}
\begin{minipage}[th]{\linewidth}
\includegraphics[angle=0,scale=0.44]{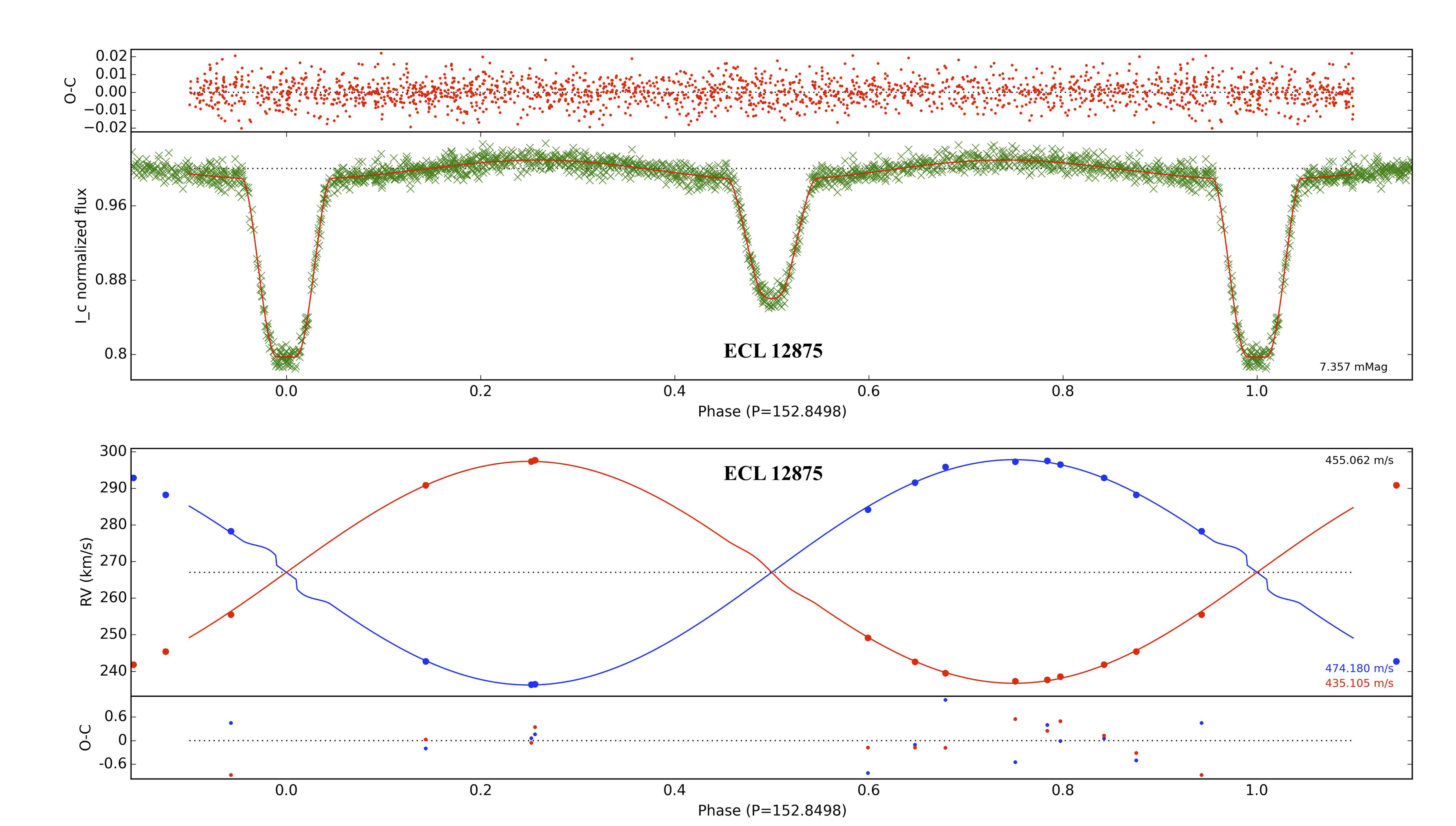}
\\ 
\includegraphics[angle=0,scale=0.44]{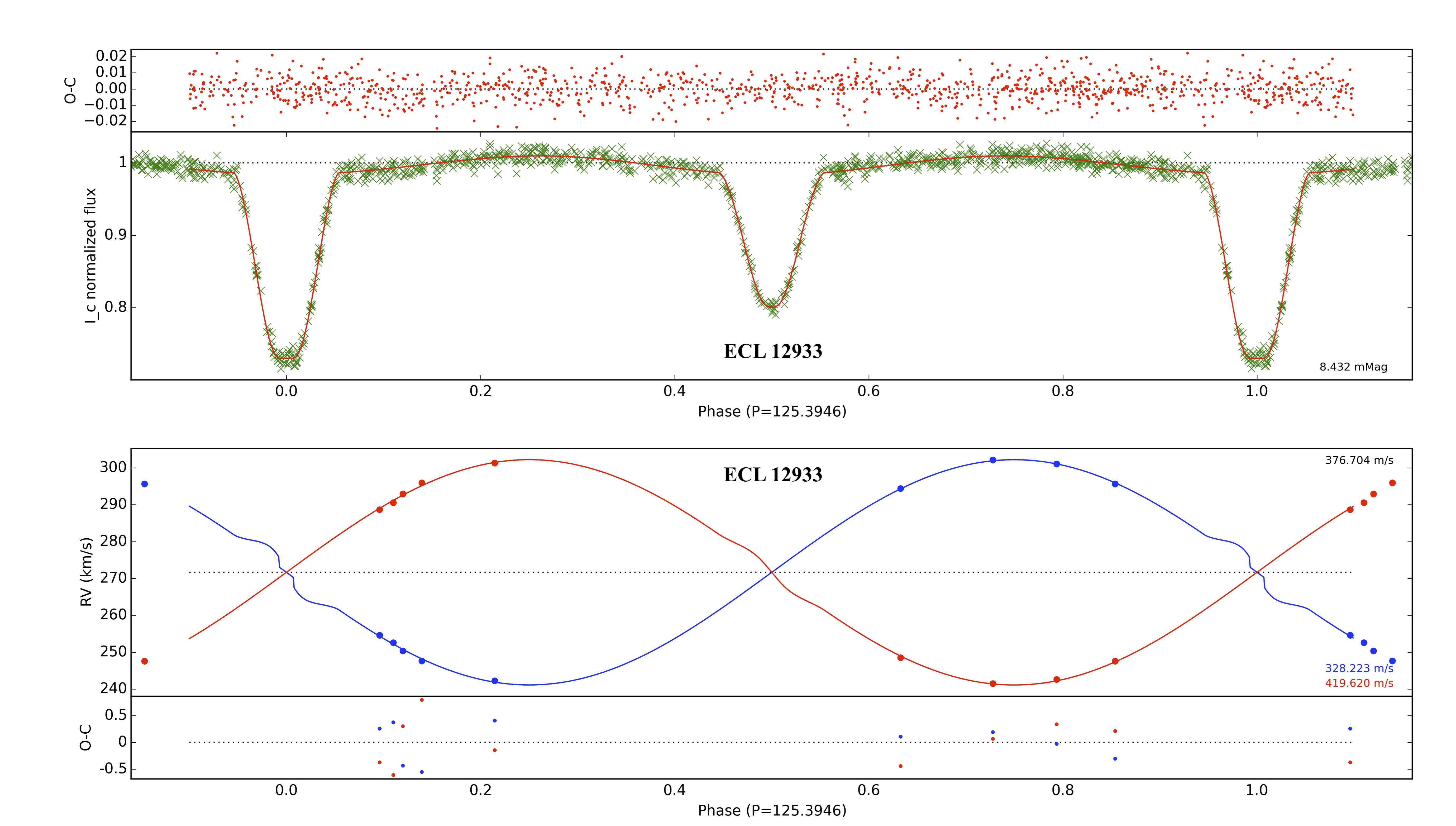} 
\end{minipage}\hfill 
\caption{ The OGLE $I_{\mathrm C}$-band light curve and the radial velocity solutions for OGLE-LMC-ECL-12875 (two upper panels) and OGLE-LMC-ECL-12933  (lower two panels).  \label{fig6}}
\end{figure*}

\begin{figure*}
\begin{minipage}[th]{\linewidth}
\includegraphics[angle=0,scale=0.44]{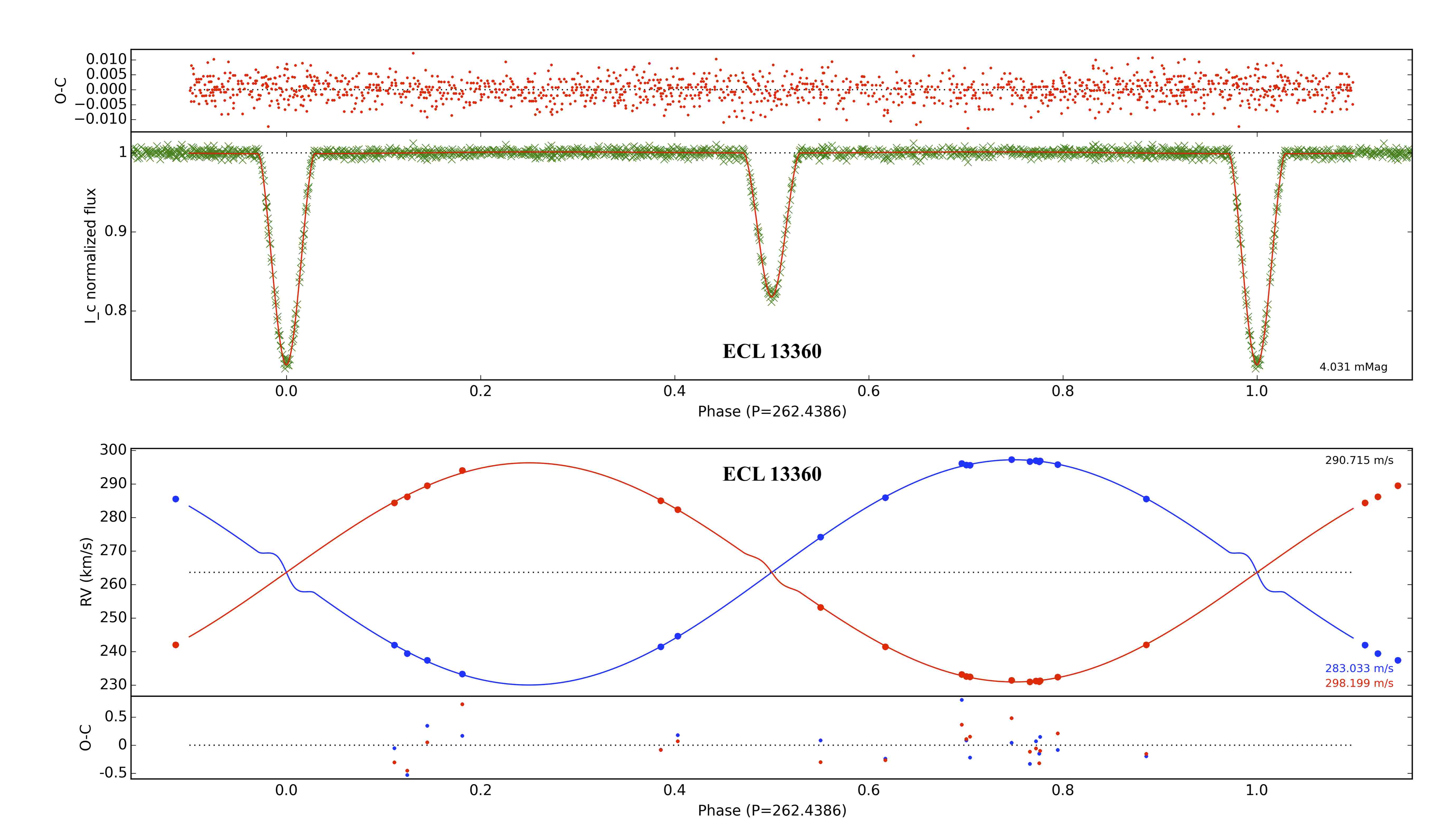}
\\ 
\includegraphics[angle=0,scale=0.44]{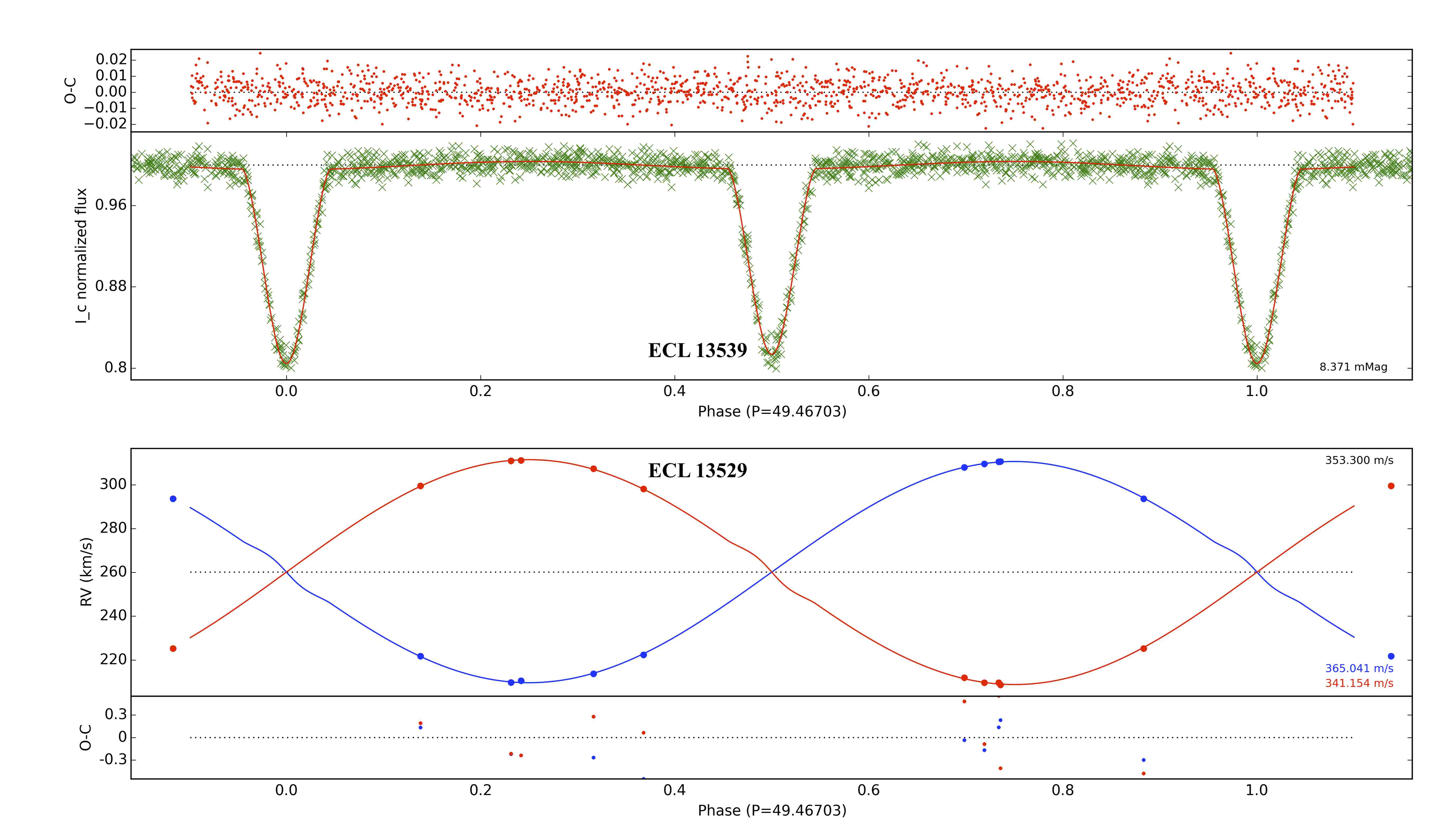} 
\end{minipage}\hfill 
\caption{ The OGLE $I_{\mathrm C}$-band light curve and the radial velocity solutions for OGLE-LMC-ECL-13360 (two upper panels) and OGLE-LMC-ECL-13529  (lower two panels). \label{fig7}}
\end{figure*}

\begin{figure*}
\begin{minipage}[th]{\linewidth}
\includegraphics[angle=0,scale=0.44]{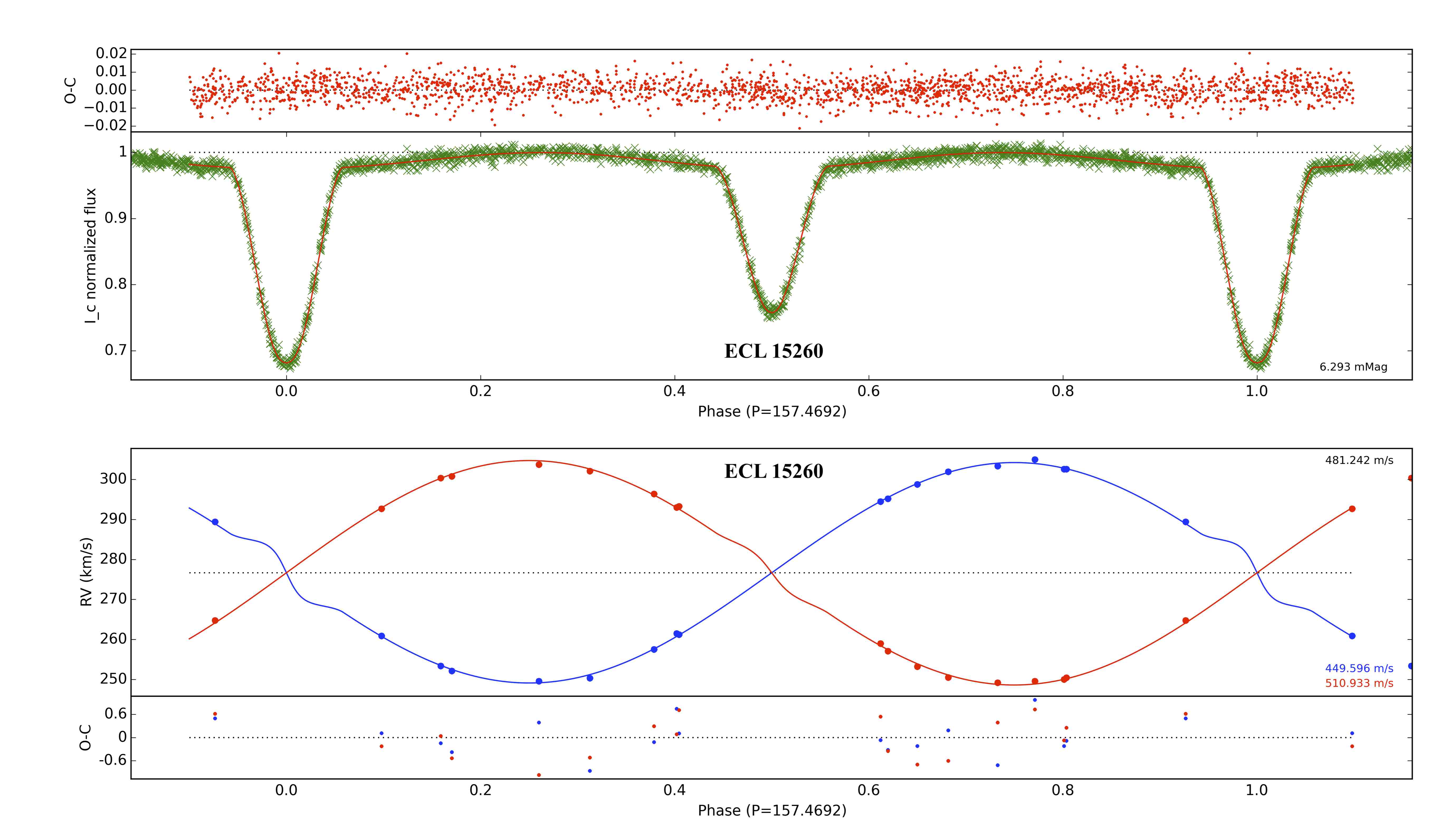}
\\ 
\includegraphics[angle=0,scale=0.44]{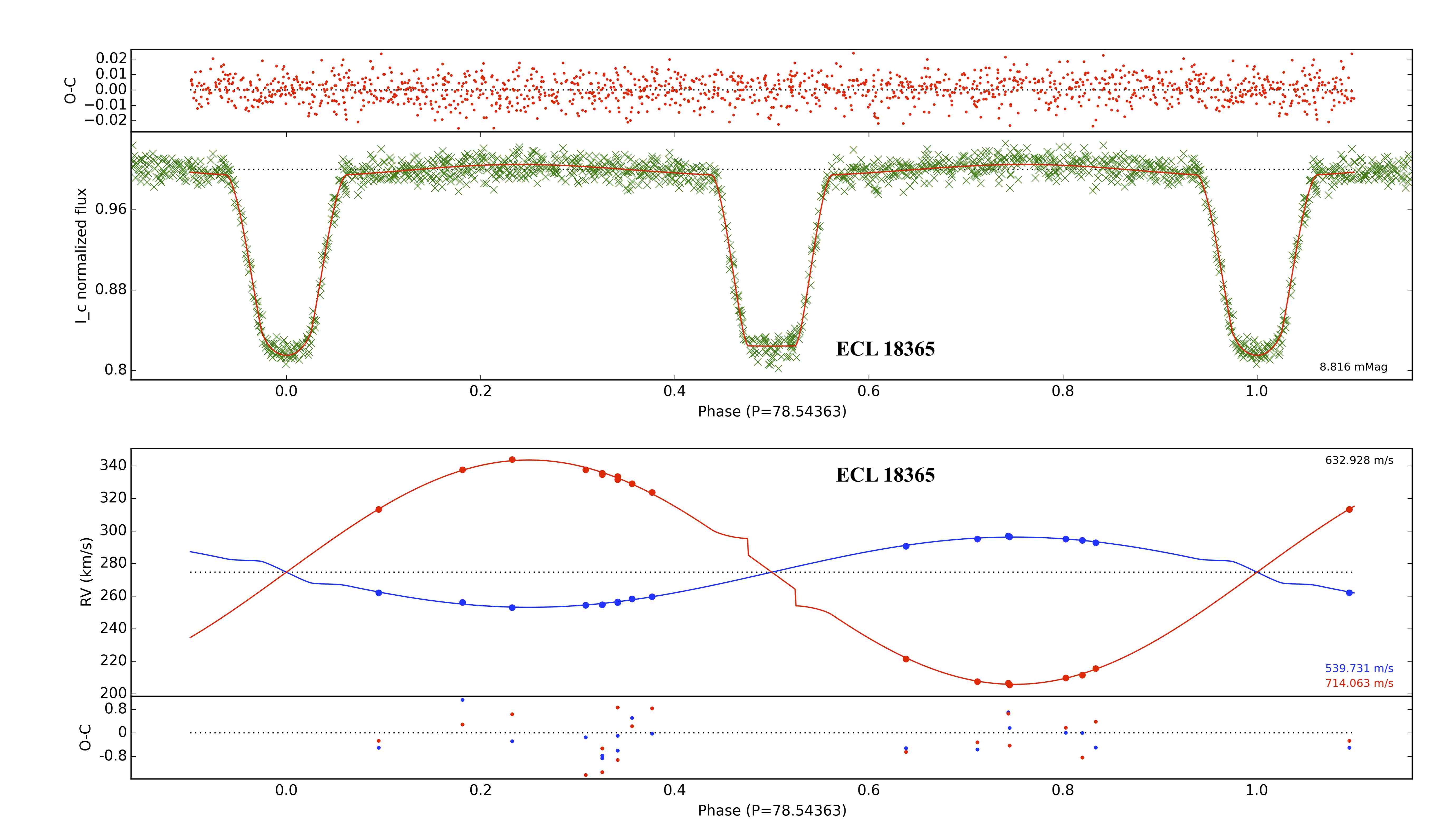} 
\end{minipage}\hfill 
\caption{ The OGLE $I_{\mathrm C}$-band light curve and the radial velocity solutions for OGLE-LMC-ECL-15260 (two upper panels) and OGLE-LMC-ECL-18365  (lower two panels). \label{fig8}}
\end{figure*}

\begin{figure*}
\begin{minipage}[th]{\linewidth}
\includegraphics[angle=0,scale=0.44]{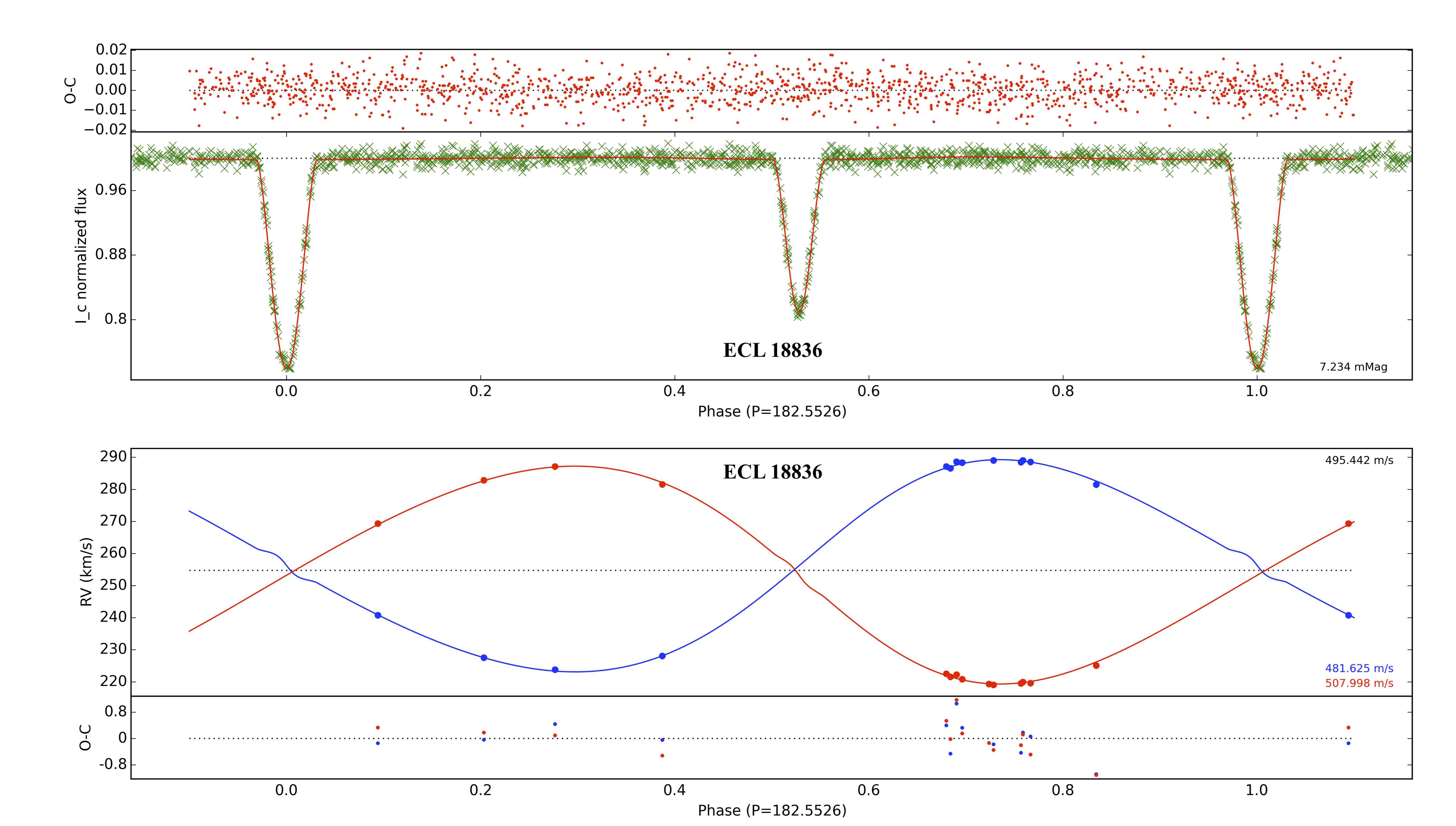}
\\ 
\includegraphics[angle=0,scale=0.44]{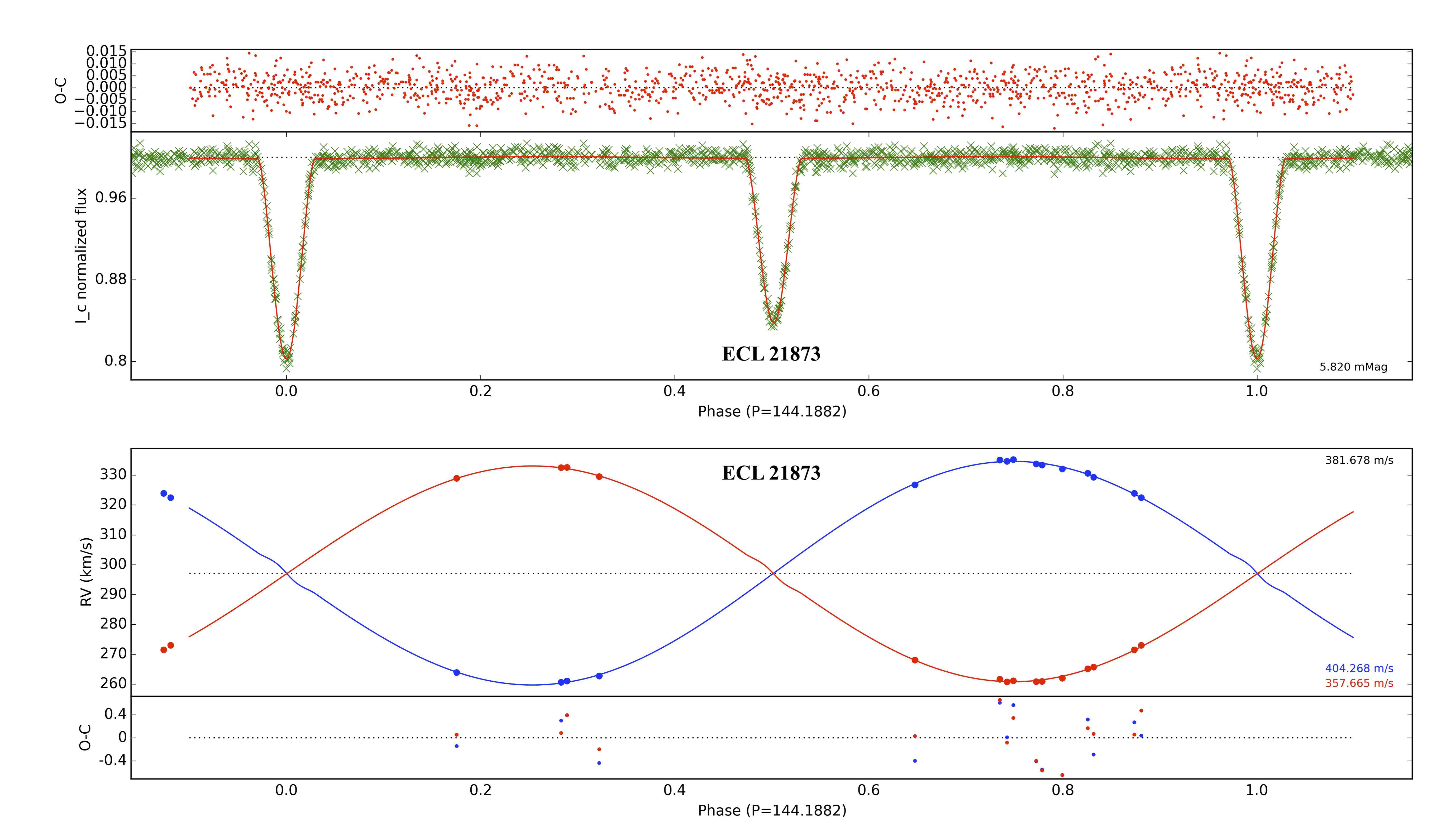} 
\end{minipage}\hfill 
\caption{ The OGLE $I_{\mathrm C}$-band light curve and the radial velocity solutions for OGLE-LMC-ECL-18836 (two upper panels) and OGLE-LMC-ECL-21873  (lower two panels). \label{fig9}}
\end{figure*}

\begin{figure*}
\begin{minipage}[th]{\linewidth}
\includegraphics[angle=0,scale=0.44]{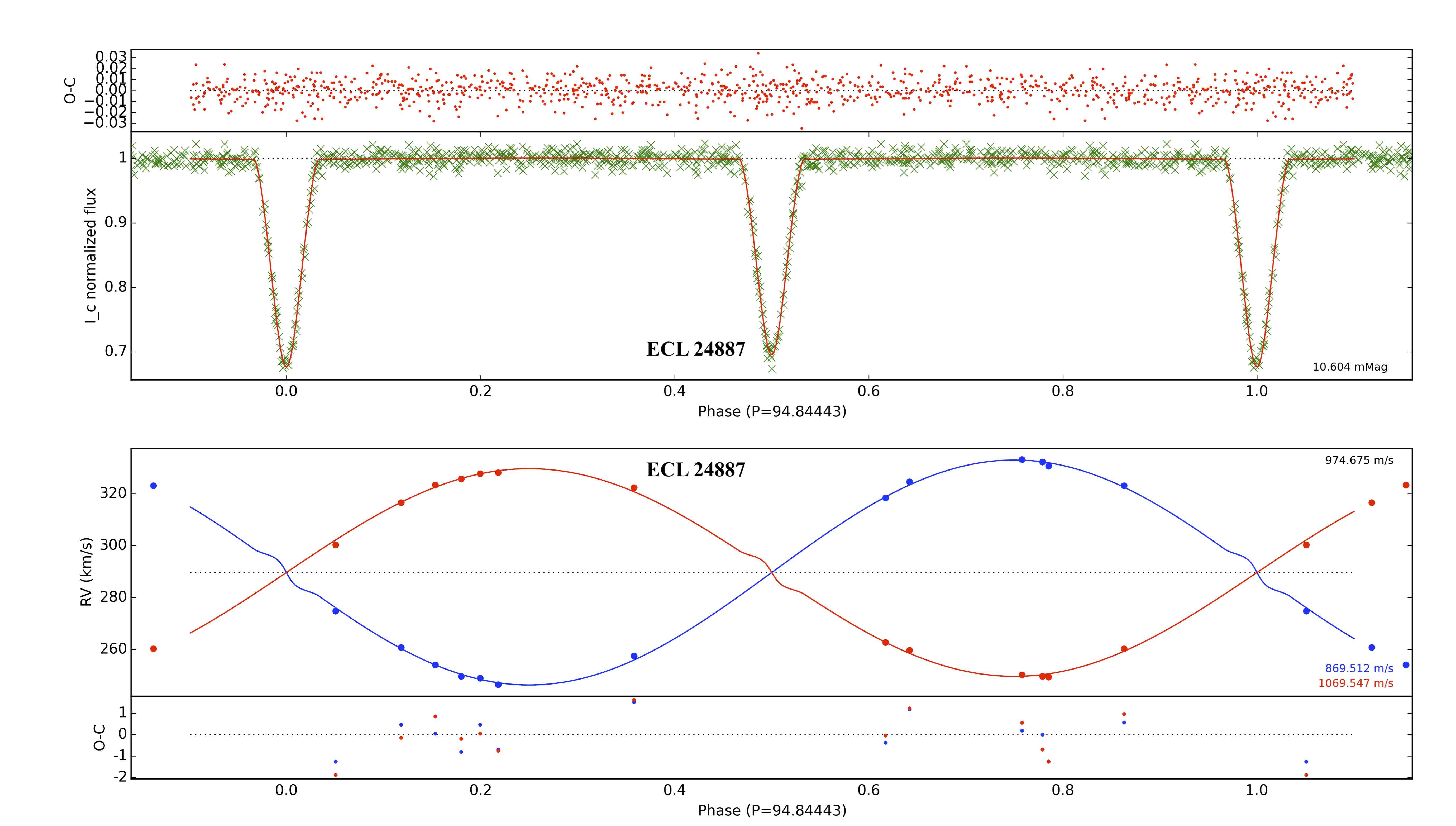}
\\ 
\includegraphics[angle=0,scale=0.44]{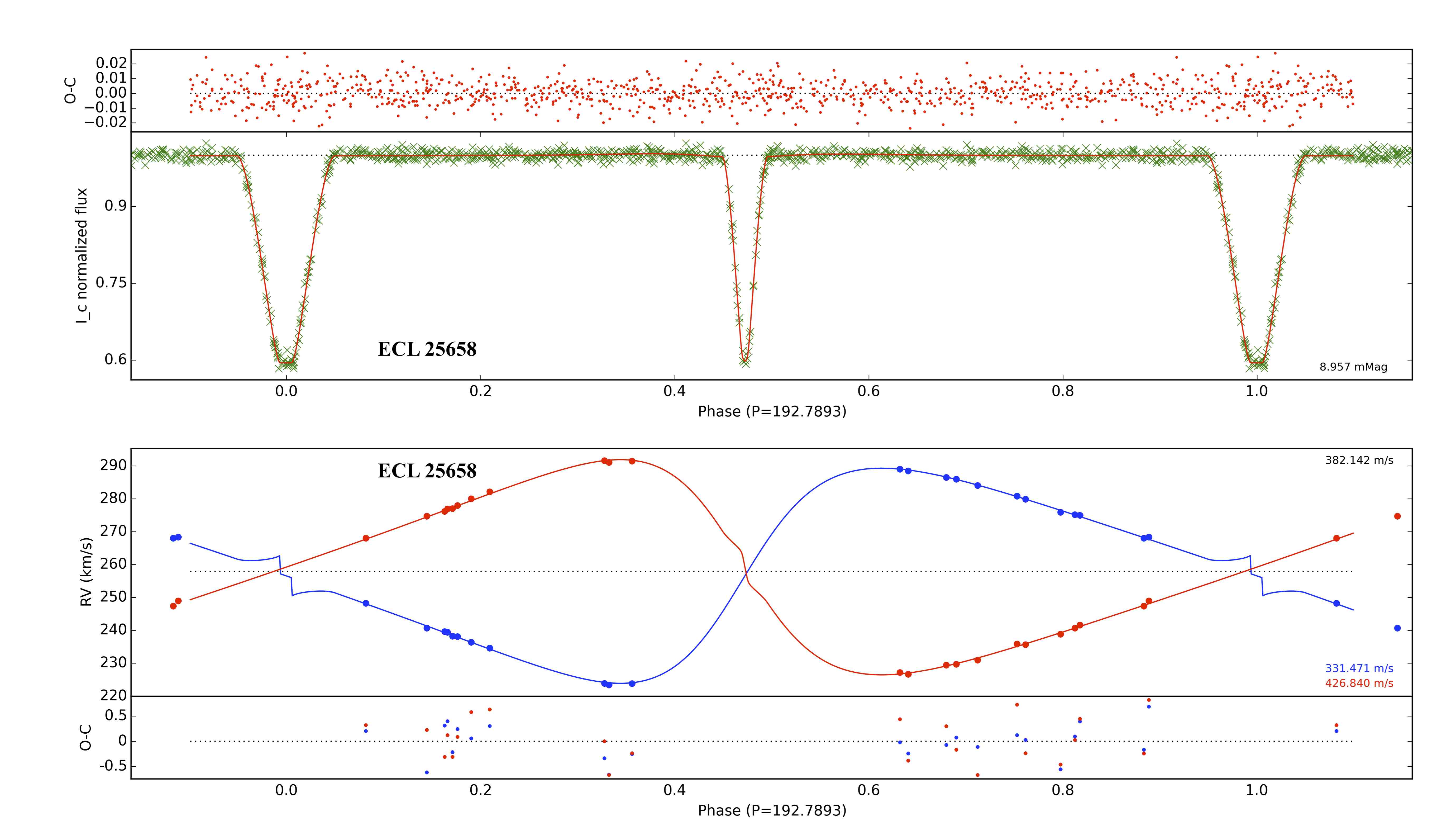} 
\end{minipage}\hfill 
\caption{ The OGLE $I_{\mathrm C}$-band light curve and the radial velocity solutions for OGLE-LMC-ECL-24887 (two upper panels) and OGLE-LMC-ECL-25658  (lower two panels).  \label{fig10}}
\end{figure*}

The OGLE $I_{\mathrm C}$-band light curves and radial velocity curves and solutions are presented in Figures:~\ref{fig1} -- \ref{fig10}. The values given on the upper right or the lower right part of a panel give the root-mean-square of the fit. In case of radial velocities three numbers are given for the {\it rms}, two at lower right (upper for the primary) and at upper right the weighted combined {\it rms}. The light curve panels show normalized photometry (crosses) and the WD model solution is superimposed as a continuous line. For the radial velocity panels the points correspond to measured radial velocities, the lines correspond the WD model solution, and blue and red colors correspond to the primary and the secondary star, respectively. In both radial velocity and light curve panels we do not mark observational errors because during the fitting we used the standard deviation of each curve as the weight.  

\newpage
\section{Comparison with evolutionary models.}
\begin{figure*}
\begin{minipage}[!bh]{0.5\linewidth}
\vspace{-0.2cm}
\includegraphics[angle=0,scale=0.70]{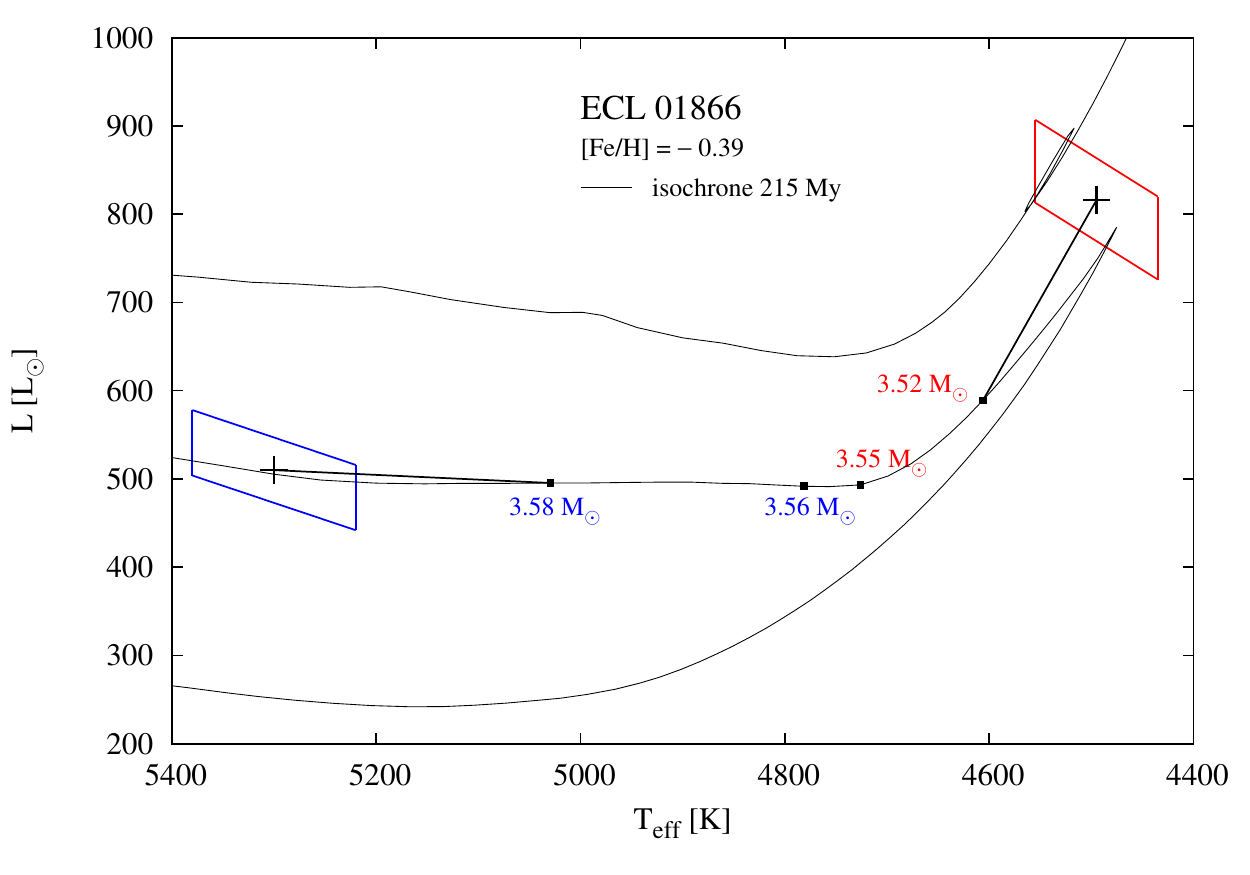} \vspace*{-0.4cm}
\mbox{}\\
\includegraphics[angle=0,scale=0.70]{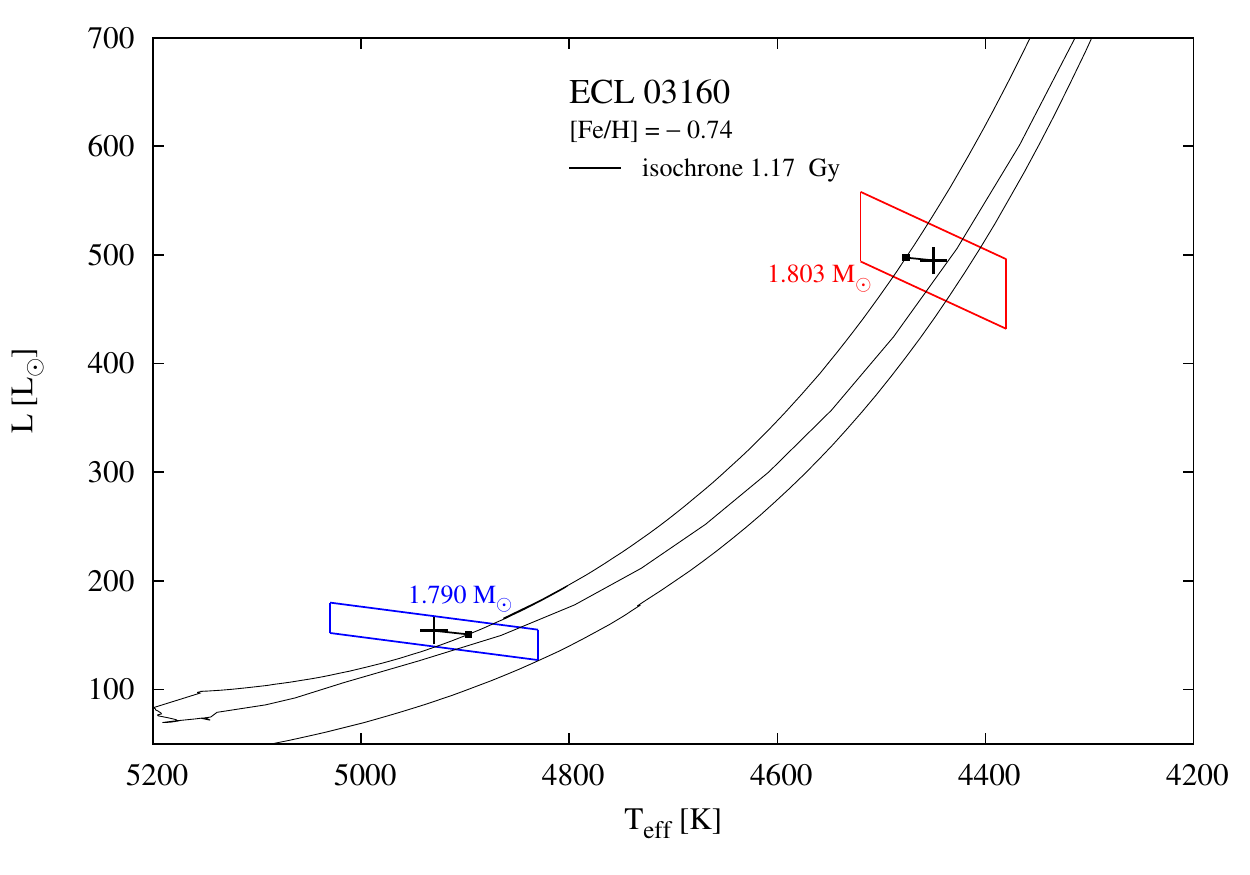}\vspace*{-0.4cm}
\mbox{}\\
\includegraphics[angle=0,scale=0.70]{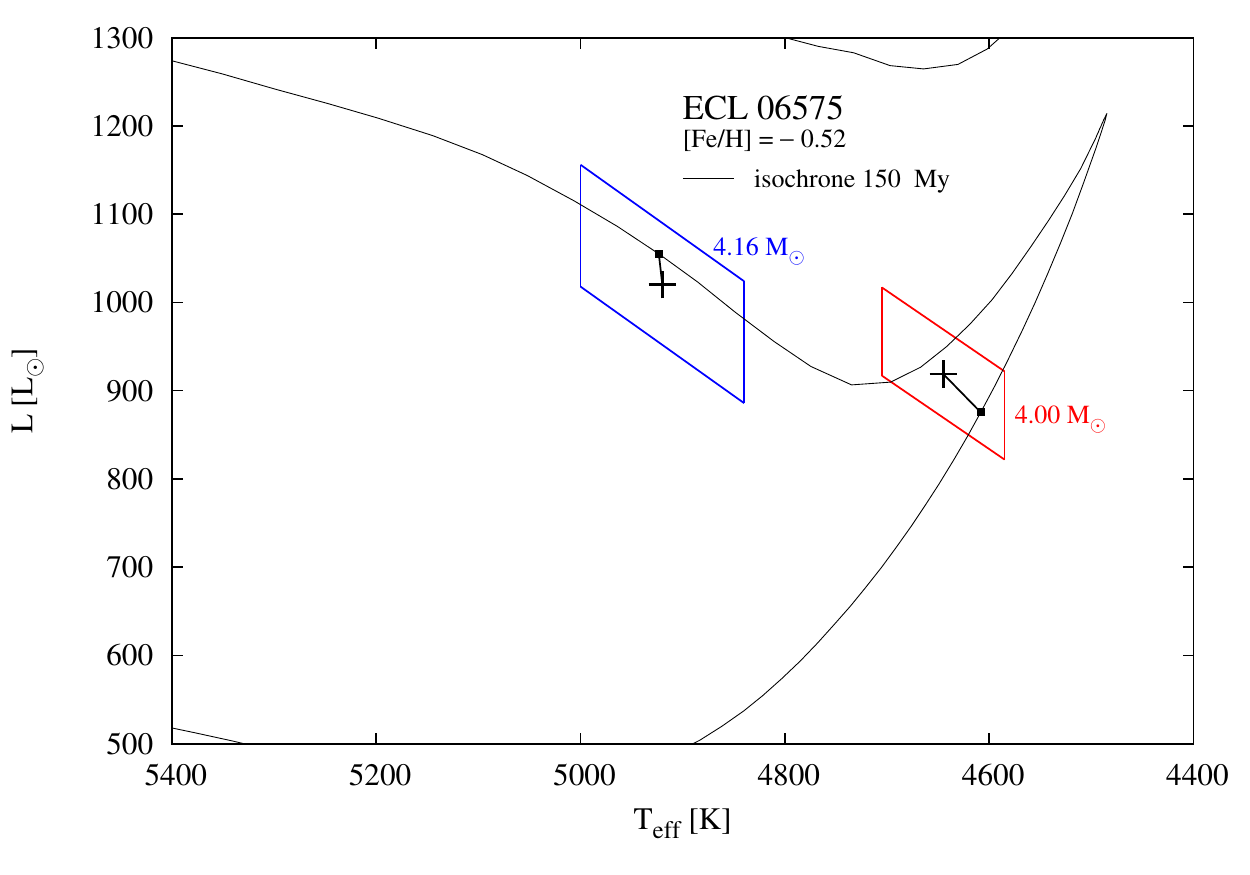}
\end{minipage}\hfill 
\begin{minipage}[!bh]{0.5\linewidth} 
\vspace{-0.2cm}
\includegraphics[angle=0,scale=0.70]{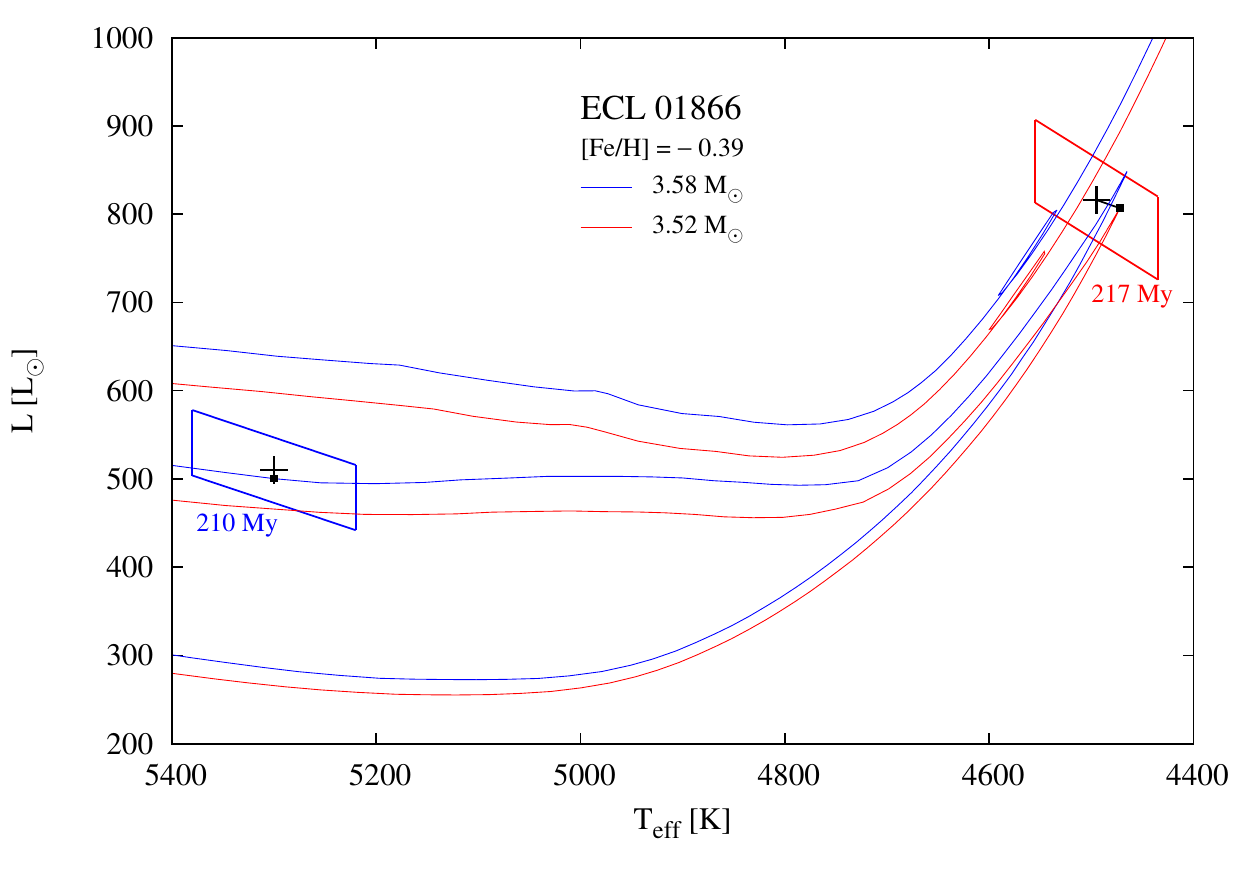}\vspace*{-0.4cm}
\mbox{}\\
\includegraphics[angle=0,scale=0.70]{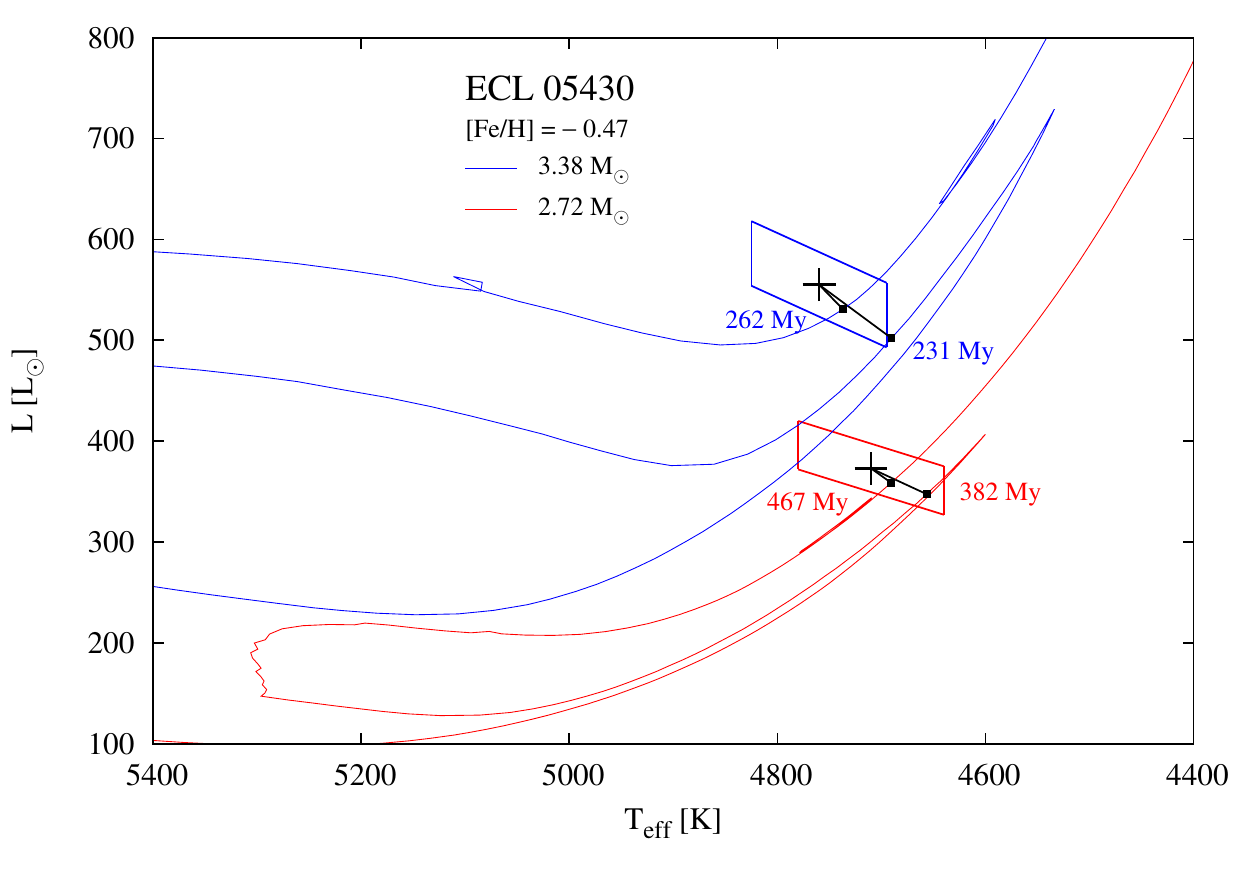}\vspace*{-0.4cm}
\mbox{}\\
\includegraphics[angle=0,scale=0.70]{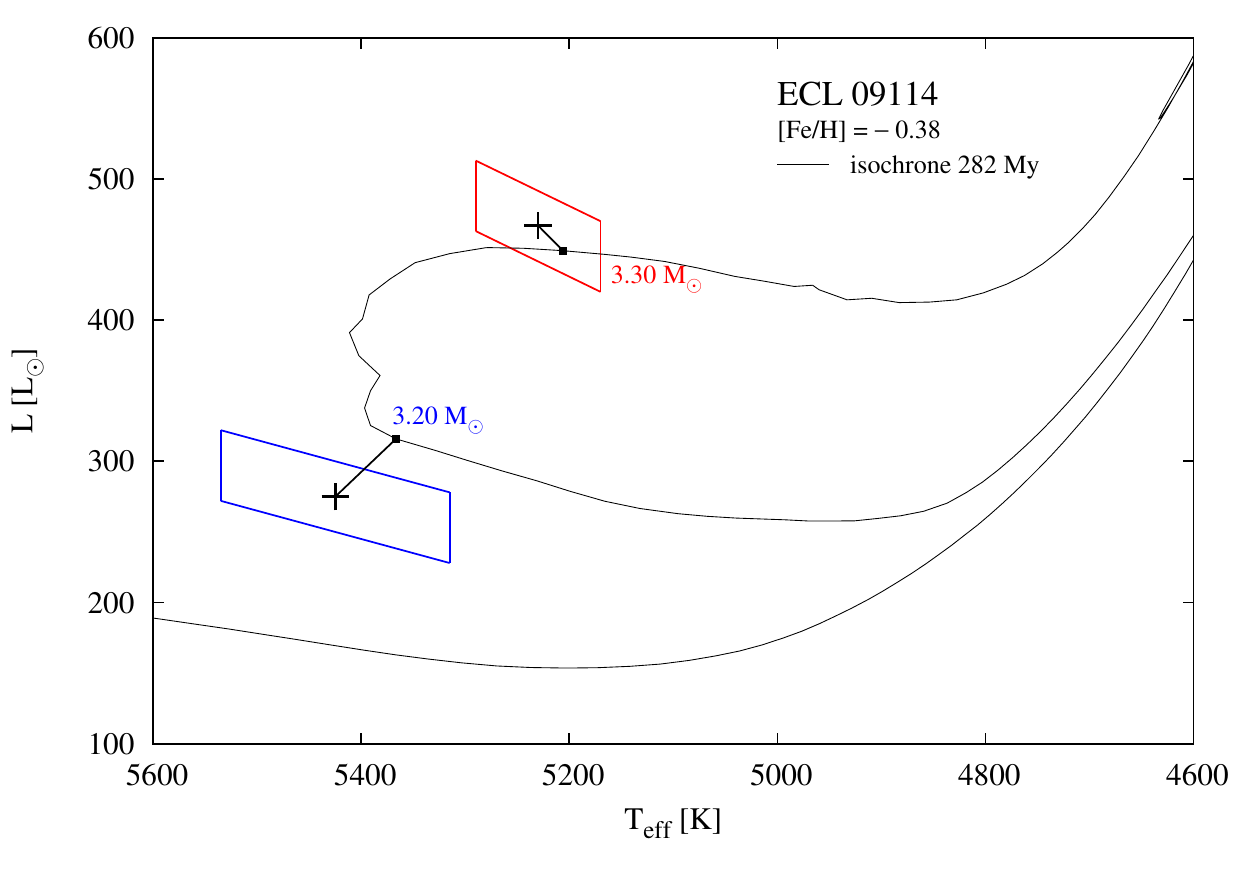}
\end{minipage}\hfill
\caption{ Comparison with the MESA evolutionary code \label{fig11}}
\end{figure*}

\begin{figure*}
\begin{minipage}[th]{0.5\linewidth}
\vspace{-0.2cm}
\includegraphics[angle=0,scale=0.70]{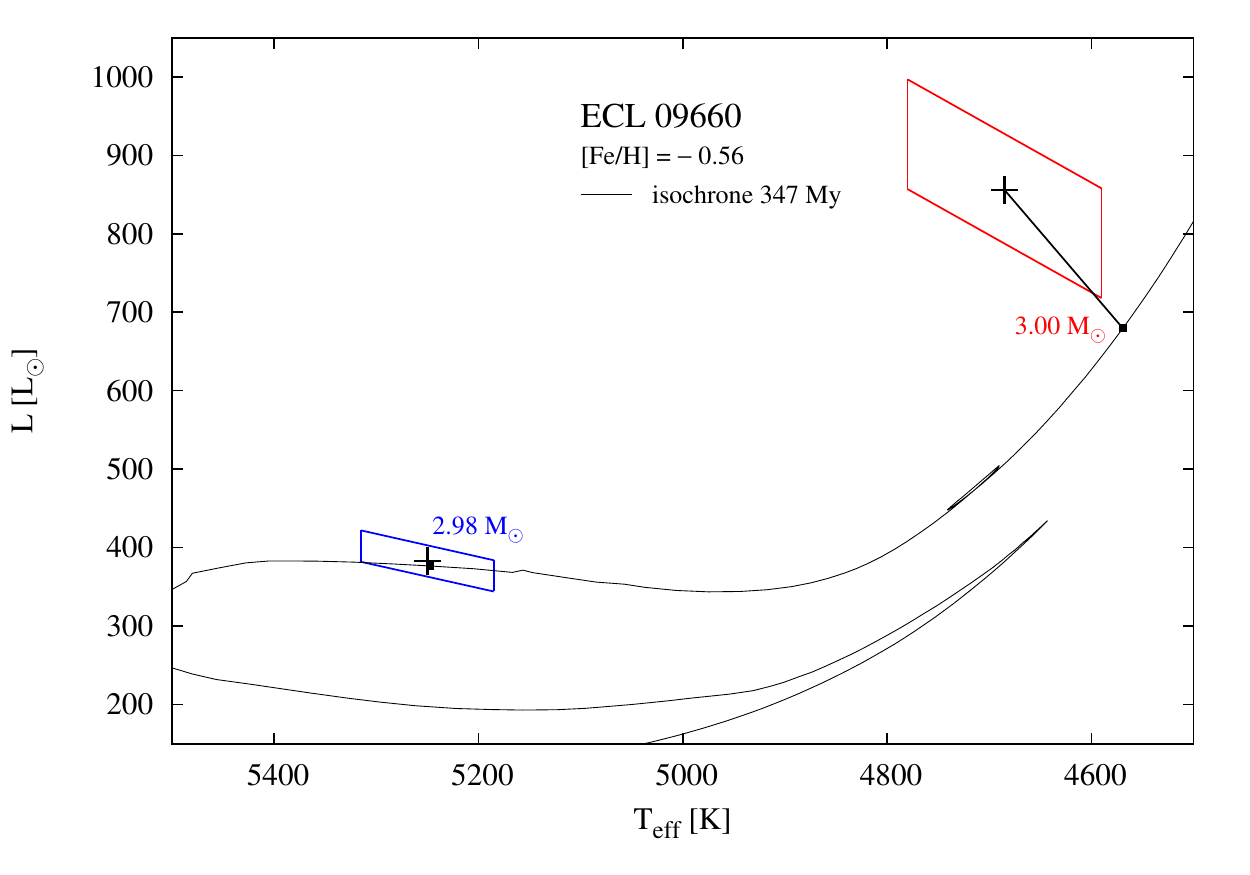} \vspace*{-0.4cm}
\mbox{}
\includegraphics[angle=0,scale=0.70]{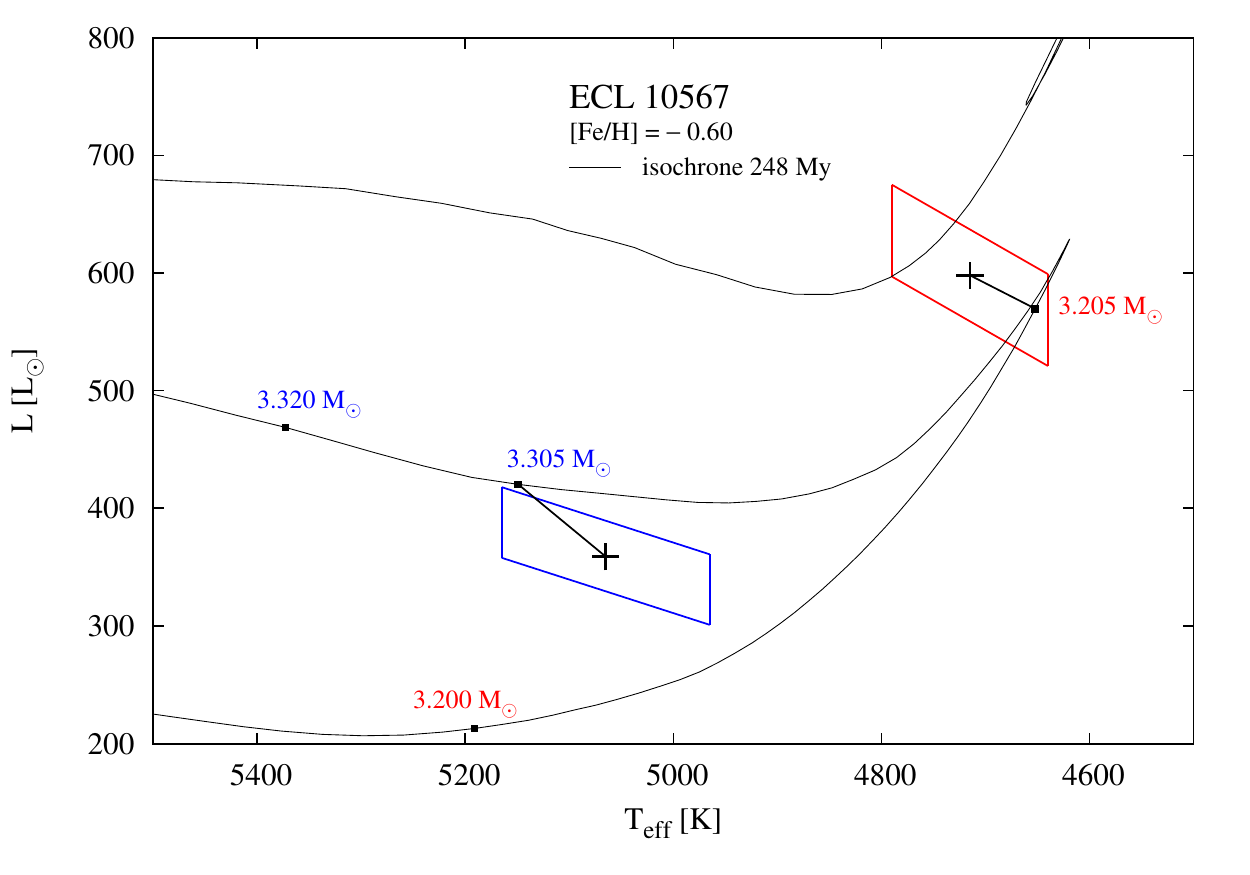} \vspace*{-0.4cm}
\mbox{}\\
\includegraphics[angle=0,scale=0.70]{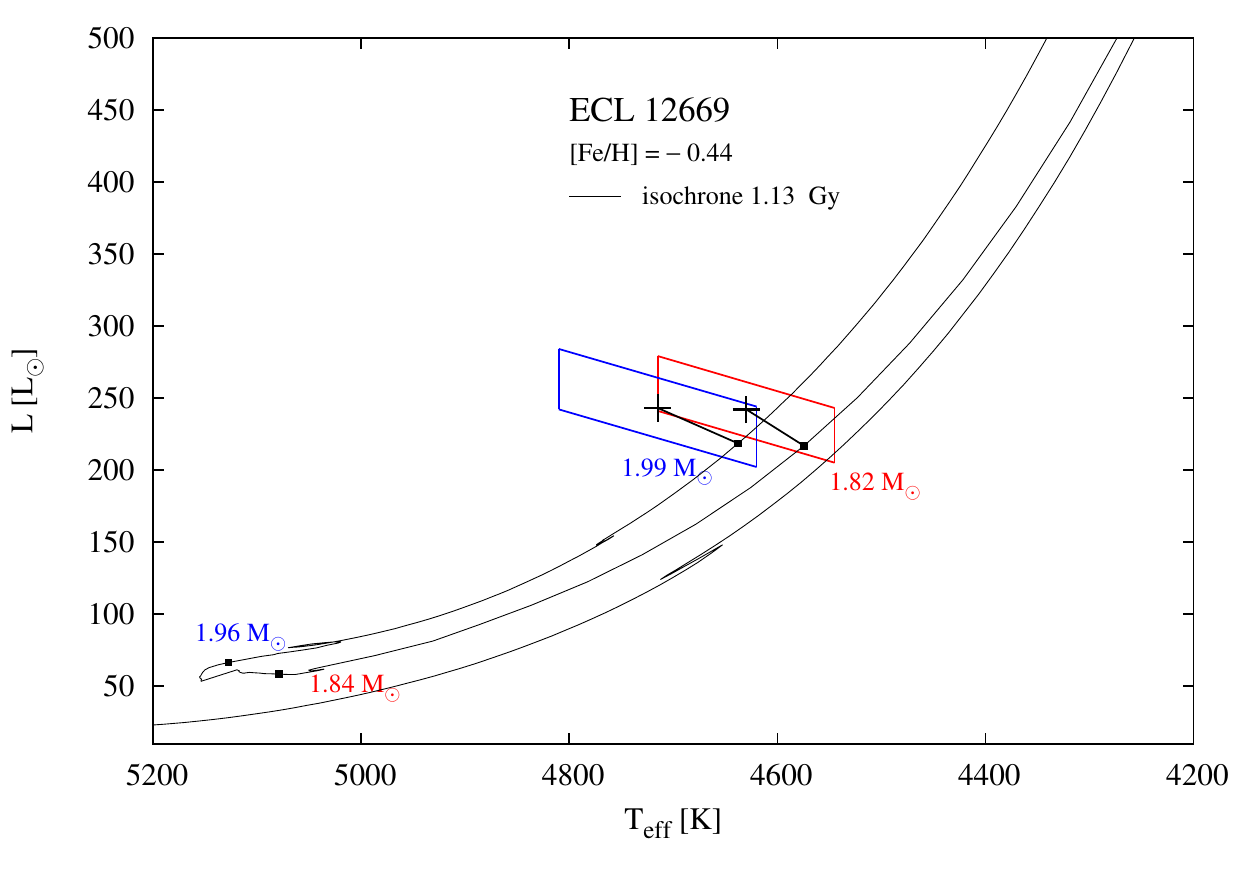}\vspace*{-0.4cm}
\mbox{}\\
\includegraphics[angle=0,scale=0.70]{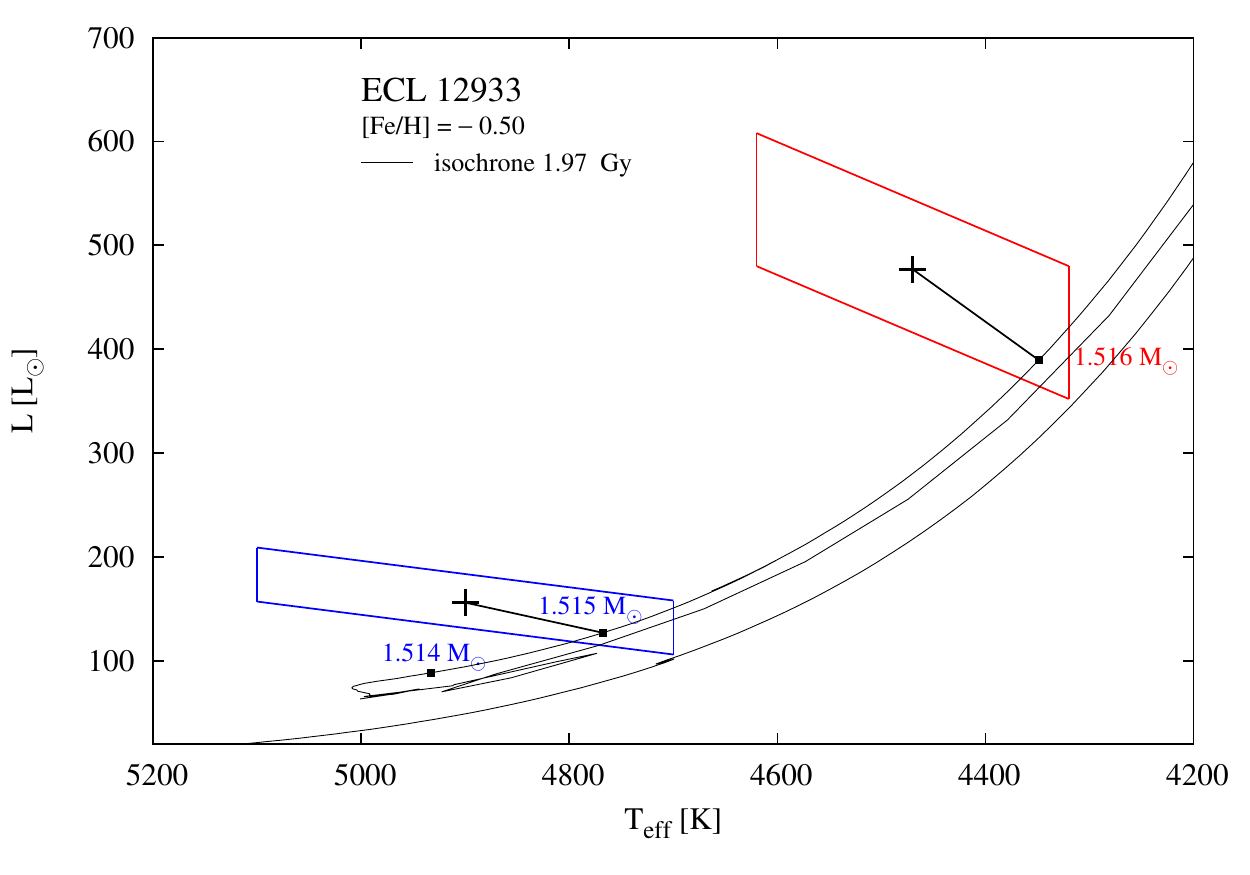}
\end{minipage}\hfill 
\begin{minipage}[th]{0.5\linewidth} 
\vspace{-0.2cm}
\includegraphics[angle=0,scale=0.70]{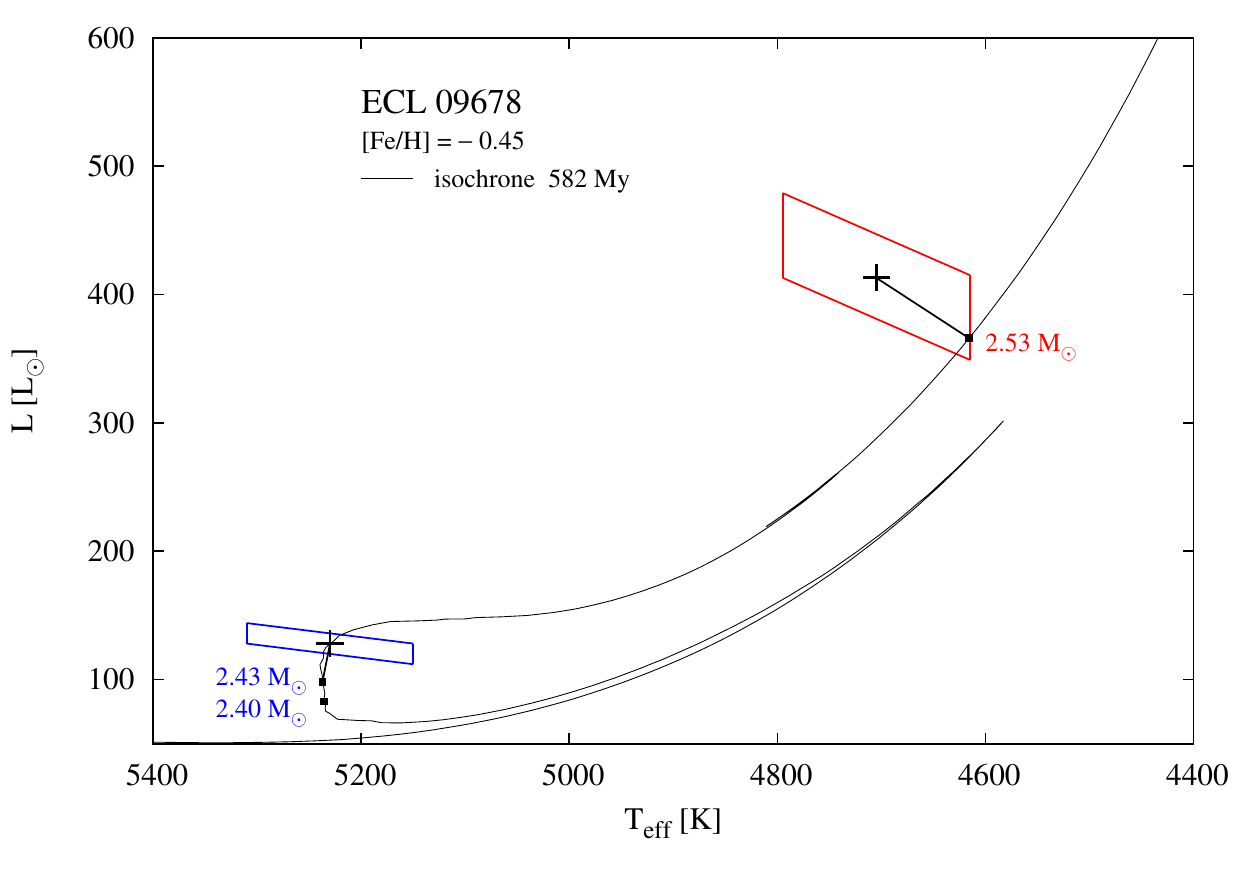}
\mbox{}
\includegraphics[angle=0,scale=0.70]{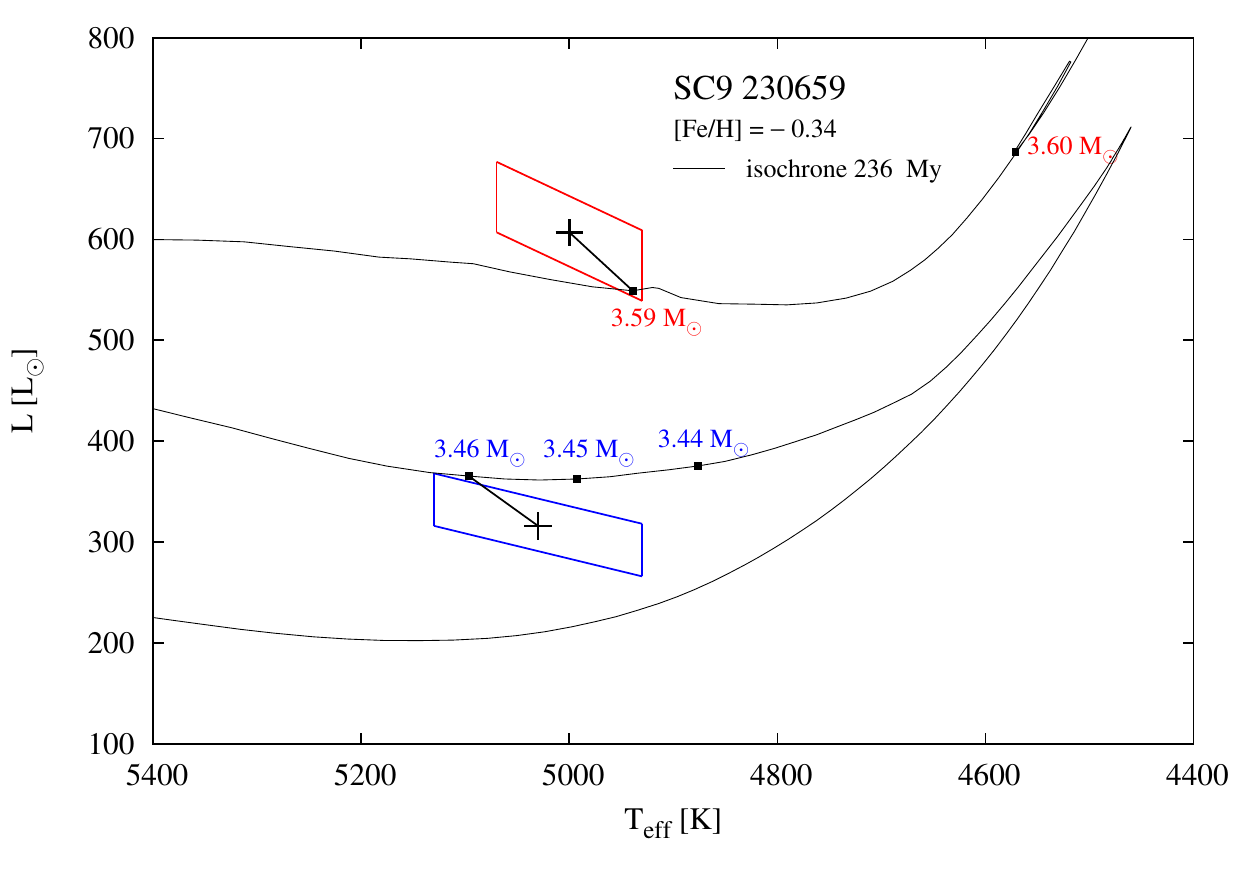}\vspace*{-0.4cm}
\mbox{}\\
\includegraphics[angle=0,scale=0.70]{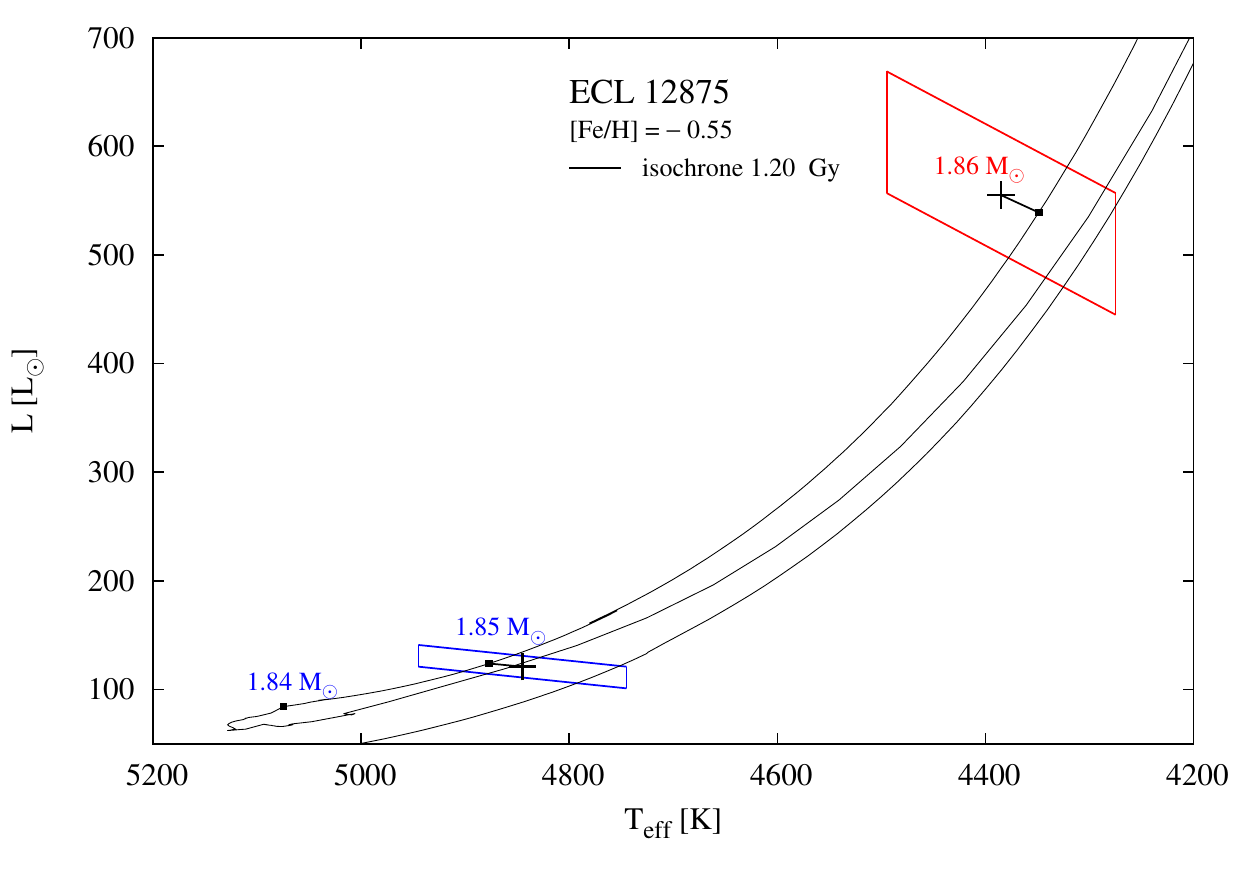}\vspace*{-0.4cm}
\mbox{}\\
\includegraphics[angle=0,scale=0.70]{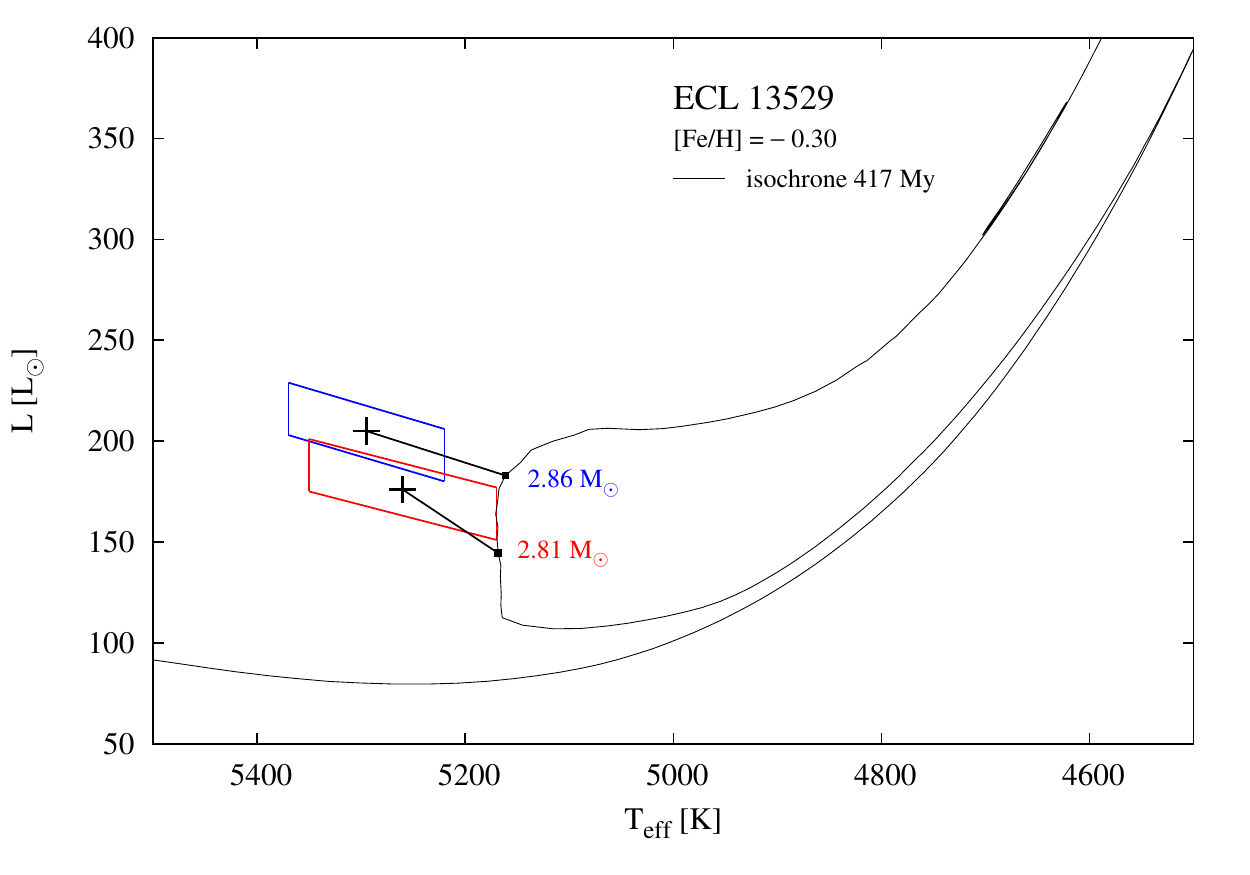}
\end{minipage}\hfill
\caption{ Comparison with the MESA evolutionary code \label{fig12}}
\end{figure*}

\begin{figure*}
\begin{minipage}[th]{0.5\linewidth}
\vspace{-0.2cm}
\includegraphics[angle=0,scale=0.70]{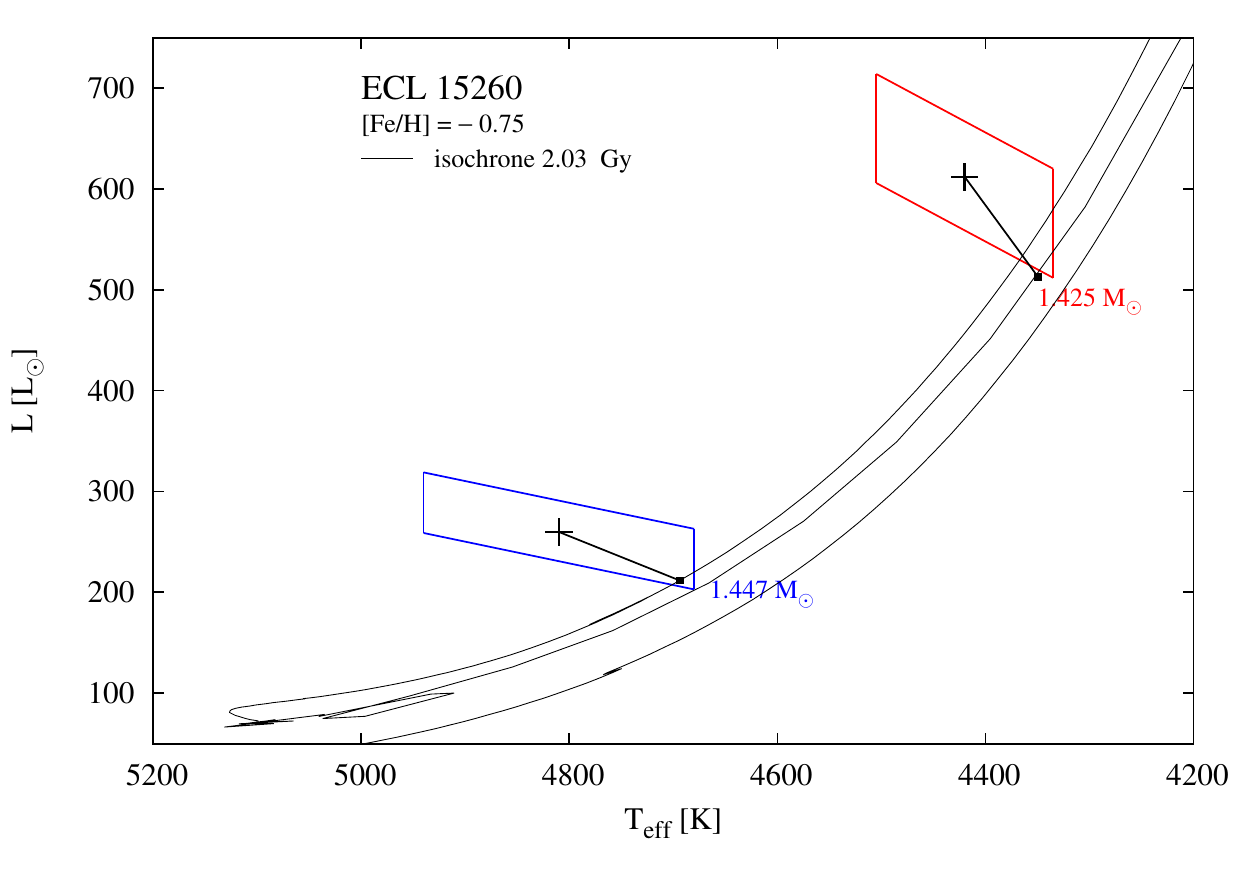}\vspace*{-0.4cm}
\mbox{}
\includegraphics[angle=0,scale=0.70]{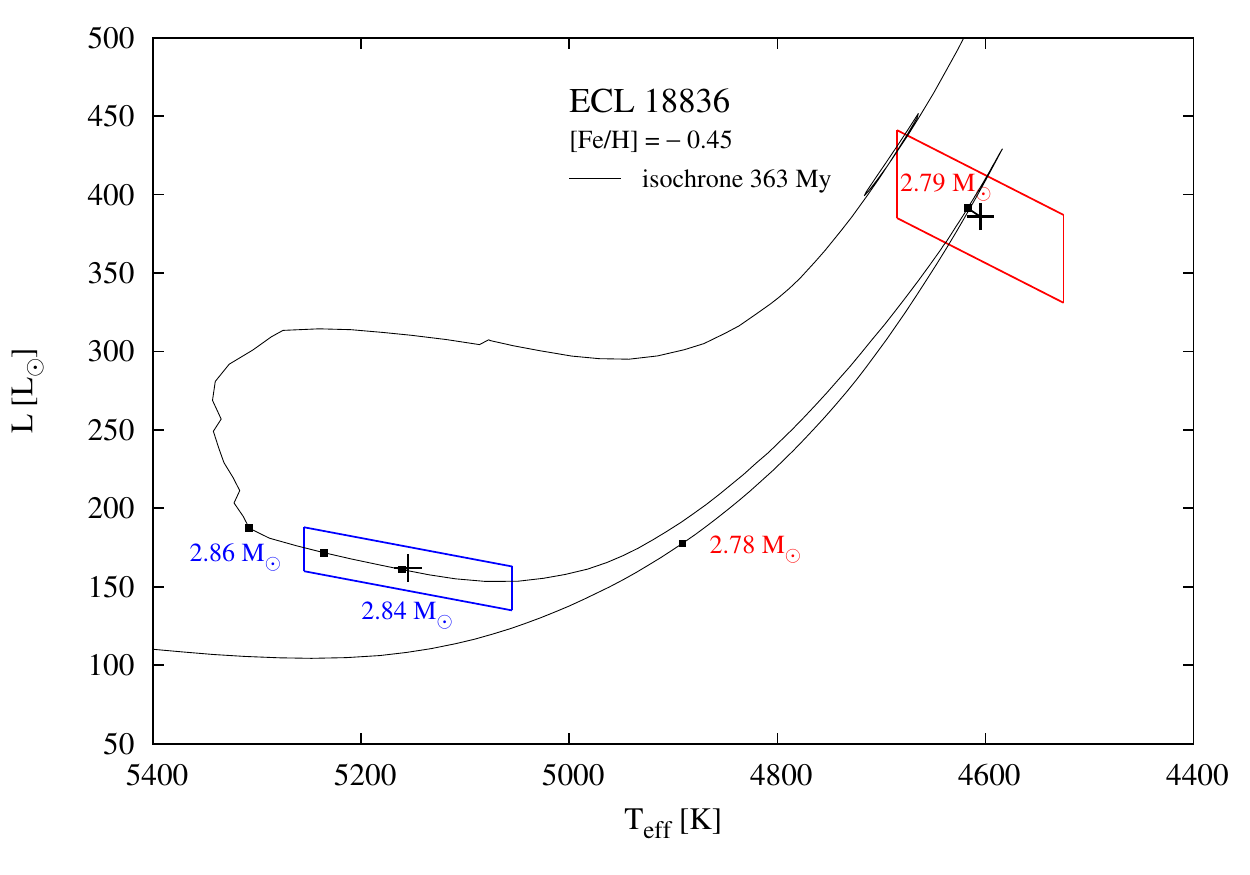}\vspace*{-0.4cm}
\mbox{}\\
\includegraphics[angle=0,scale=0.70]{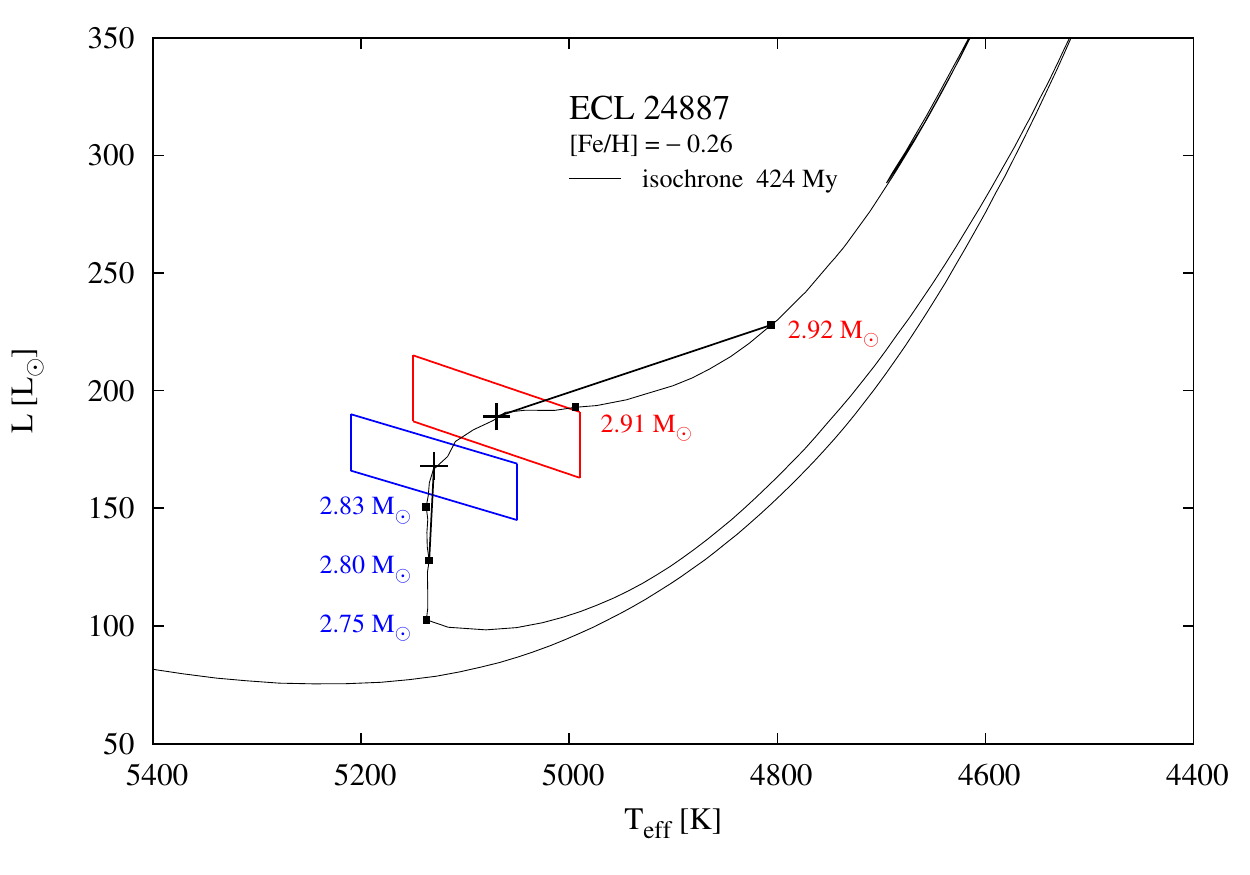}
\end{minipage}\hfill 
\begin{minipage}[th]{0.5\linewidth} 
\vspace{-0.2cm}
\includegraphics[angle=0,scale=0.70]{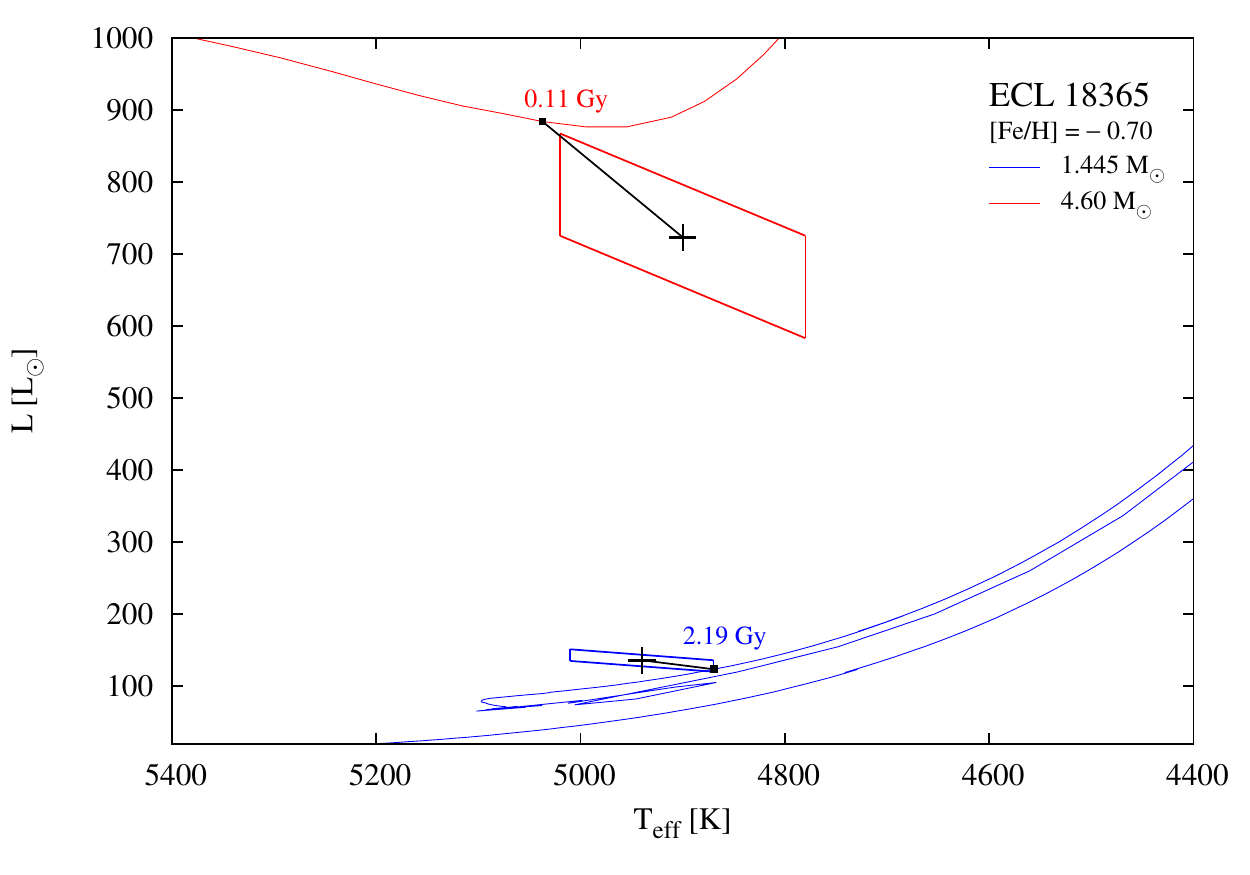}\vspace*{-0.4cm}
\mbox{}
\includegraphics[angle=0,scale=0.70]{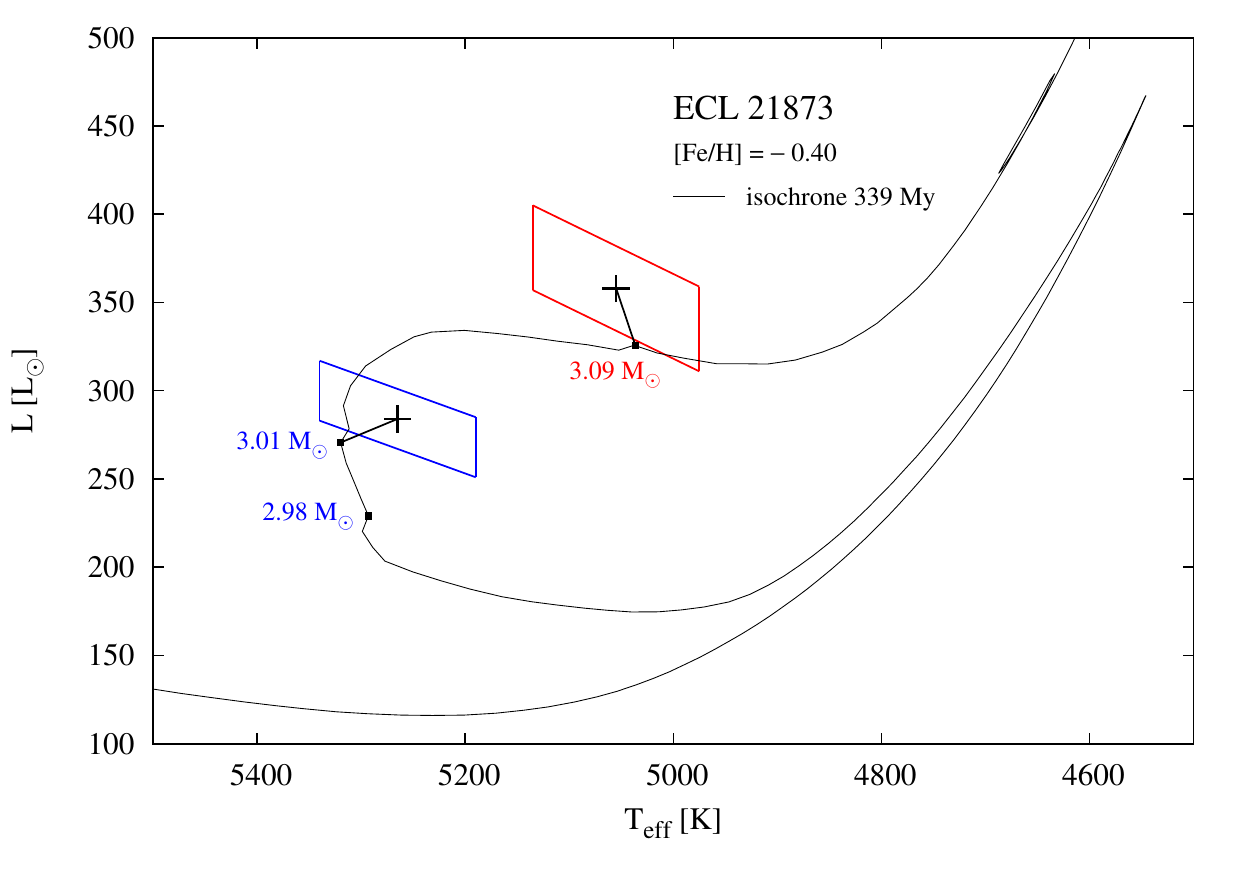}\vspace*{-0.4cm}
\mbox{}\\
\includegraphics[angle=0,scale=0.70]{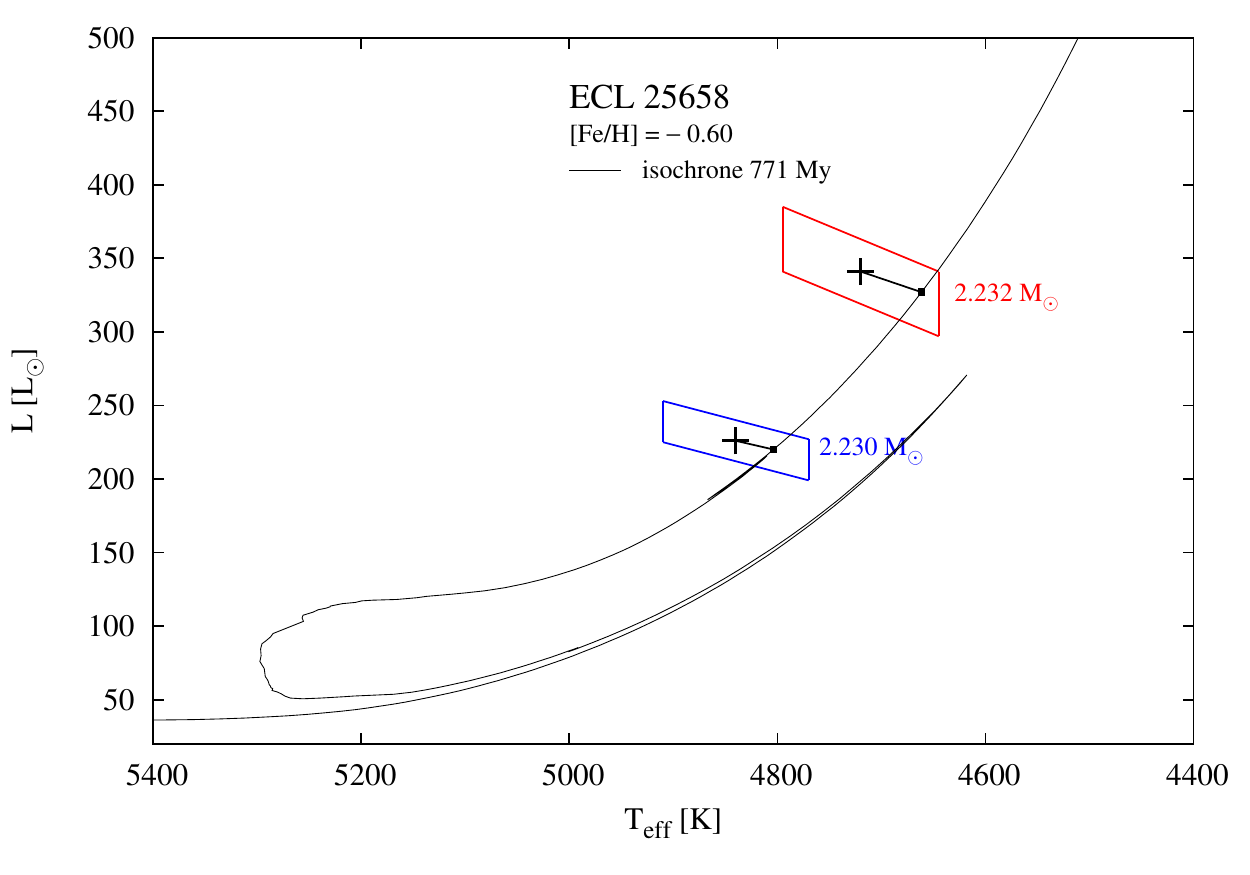}
\end{minipage}\hfill
\caption{ Comparison with the MESA evolutionary code \label{fig13}}
\end{figure*}

The comparison with the MESA evolutionary tracks is shown in Figures~\ref{fig11}, \ref{fig12} and \ref{fig13}. The positions of both components are marked by black crosses, uncertainties on the effective temperature and the luminosity are marked by parallelograms, the blue corresponding to the hotter star and the red corresponding to the cooler component. In most cases only one isochronal fit is shown, assuming the same chemical composition, with a continuous line representing the best isochrone found. The small filled squares on the isochrons are the best-fit position of stars with masses corresponding to masses of components within one sigma. For some systems where components have discrepant ages we  also show individual evolutionary tracks, blue for the hotter star and red for the cooler one. In these cases the small filled square corresponds to the best fitted age of a particular component.    

\end{appendix}
\end{document}